\documentclass[12pt]{article}
\usepackage{amsmath}
\usepackage{graphicx}
\usepackage{amssymb}
\usepackage{enumerate}
\usepackage{natbib}
\usepackage{url} 
\usepackage{amsfonts}
\usepackage{amssymb}
\usepackage{xcolor}

\usepackage{tikz}
\usepackage{bm}

\begin{document}

\def\spacingset#1{\renewcommand{\baselinestretch}%
{#1}\small\normalsize} \spacingset{1}


  \title{\bf Constructing optimal dynamic monitoring and treatment regimes: An application to hypertension care}
  \author{Janie Coulombe 
    \hspace{.2cm}\\
    Department of mathematics and statistics, Université de Montréal\\
    and \\
    Dany El-Riachi \\
    Department of mathematics, Université de Sherbrooke\\
     and \\
   Fanxing Du \\
    College of Pharmacy, University of Florida\\
     and \\
    Tianze Jiao \\
    College of Pharmacy, University of Florida}
  \maketitle

\bigskip
\begin{abstract}
Hypertension is a leading cause of cardiovascular diseases and morbidity, with antihypertensive drugs and blood pressure management strategies having heterogeneous effects on patients. Previous authors exploited this heterogeneity to construct optimal dynamic treatment regimes for hypertension that input patient characteristics and output the best drug or blood pressure management strategy to prescribe. There is, however, a lack of research on optimizing monitoring schedules for these patients. It is unclear whether different monitoring patterns and drug add-on strategies could lower blood pressure differently across patients. We propose a new consistent methodology to develop optimal dynamic monitoring and add-on regimes that is doubly-robust and relies on the theory of Robins' g-methods and dynamic weighted ordinary least squares. We discuss the treatment of longitudinal missing data for that inference. The approach is evaluated in large simulation studies and applied to data from the SPRINT trial in the United States to derive a new optimal rule. This type of rule could be used by patients or physicians to personalize the timing of visit and by physicians to decide whether prescribing an antihypertensive drug is beneficial.
\end{abstract}

\noindent%

\spacingset{1.9} 
\section{Introduction}
Hypertension is a leading cause of cardiovascular diseases and morbidity \citep{gu2010association,wang2020association}. Physicians typically begin treating patients with lifestyle interventions, followed by the administration of antihypertensive drugs. It would be useful to learn about the heterogeneous causal treatment effects of different monitoring and drug add-on strategies for patients with hypertension, such that we can choose the best strategy for a given patient. In statistics, the effect of different treatment strategies can be studied using conditional outcome models, and sequences of decisions can be formalized as dynamic treatment regimes (DTRs). In the case of hypertension, deriving optimal monitoring and treatment regimes would be useful to support clinical decision-making, delay the progression of the hypertension, and avoid unnecessary healthcare utilization. Using the corresponding rules, patients could choose when to visit their physician based on their BP readings and other personal characteristics. The rules could also consider other decisions than monitoring, such as the choice of prescribing antihypertensive drugs or switching drug at a physician visit.

A few authors have developed DTRs using observational data, to adapt the choice of antihypertensive drugs and therapies to patient characteristics with the aim to decrease the burden of hypertension or to target varied systolic BP levels, see, e.g., \cite{kjohnson} and \cite{zhou2022}, who respectively used the g-formula and a nonparametric optimization approach called T-RL \citep{tao2018} to derive the optimal regimes. \cite{jiao2023use} used a dynamic marginal structural model to evaluate the optimal target BP level. Some authors have also proposed and studied different approaches to missing data and irregular visits in DTRs, see e.g., \cite{shortreed2014multiple} and \cite{shen2023estimation} in randomized settings or \cite{coulombesmmr} for the development of a one-stage rule in an observational setting.

Joint optimization of monitoring times and treatments has also been tackled, although much remains to be done in this area. \cite{robins2008estimation} have proposed a theoretical framework for estimation of DTRs under varying monitoring schedules and discussed extrapolation to different monitoring schedules, noting that the causal estimand can vary under different visit processes. Other work in this area focused on static monitoring schedules, i.e., a schedule that cannot be optimized in time as a function of evolving characteristics, see \cite{neug2, neug3}.   \cite{neug3} review existing approaches to jointly modeling treatment and monitoring processes and developing optimal decision rules. They apply some of the previously proposed approaches and review some important considerations with longitudinal data. Them, and \cite{neug1}, focus on time-to-event outcomes, while \cite{neug2} focus on a binary outcome. \cite{robins2008estimation, neug1, neug2}, and \cite{neug3} rely on the \textit{No Direct Effect} (NDE) assumption. This assumption broadly means that visits have no direct effect on covariates (or the outcome), they only have effects on these variables through the treatment choice. In \cite{robins2008estimation}, monitoring a patient refers to performing a diagnostic test that can inform best treatment decisions, such that the NDE assumption is more natural. We do not make this assumption in this work. Instead, we assume that visits can change patient characteristics other than the treatment. For instance, physicians may provide consultations, which motivate patients to modify bad habits, leading them to e.g., stop smoking or improve their diet.  
None of the previously proposed approaches address optimal dynamic monitoring and treatment regimes and most approaches for static regimes assume NDE.

Define a dynamic monitoring and add-on regime (DMAR) as a sequence of rules that dynamically chooses whether to visit, and if so, whether to obtain an additional prescription for an antihypertensive drug. Estimation of an optimal DMAR is challenging in observational studies, as longitudinal data tend to be observed irregularly.  
Times when there is no visit generally do not provide us with additional information on patient characteristics, such that the estimation of visit effect modification by patient characteristics like age and comorbidities must depend on previous values carried forward or imputed. The choice of an approach to address missing data depends on the assumed data generating mechanism (DGM). See e.g., \cite{moreno2018canonical}, for a discussion on the use of causal diagrams to determine whether observed variables are sufficient to adjust for missing data, and if not, the possible sensitivity analyses to assess the impact on inference. 

In this manuscript, we provide, to the best of our knowledge, the first \textcolor{black}{consistent} statistical methodology to estimate optimal DMAR. In addition to being straightforward and not relying on the NDE assumption, the methodology addresses missing data in the time-varying effects modifiers (and could be extended to other missing data, under additional assumptions). In Section 2, we present the notation, causal assumptions, treatment of missing and censored data, and the proposed approach. In Section 3, we demonstrate the approach empirically using large simulation studies. We then apply the proposed method to data from the SPRINT trial in the United States (US), a study on patients with hypertension, to derive an optimal DMAR that consists of a sequence of monthly decision rules, in Section 4. We conclude this work in Section 5.

\section{Methods}
\subsection{Notation}
Suppose a random sample of size $n$ from a larger population, leading to observations on $n$ independent individuals. We use uppercase letters to refer to random variables, lowercase letters to refer to their realizations, and bold notation for matrices and vectors. We assume that individuals are observed at baseline and (possibly) at physician visits occurring at discrete times $t= 1, 2, ..., \tau$ where $\tau$ is an integer and a ``final time" chosen in the study, e.g., one year after cohort entry. In our application to SPRINT data, we use $t=0, 1, ..., 12$ which correspond respectively to the baseline and the 12 subsequent months.

\textcolor{black}{
Each individual has a censoring time denoted by $C_i$. Let $\xi_i(t)$ denotes an indicator of still being in the study at time $t$, i.e., $\xi_i(t):=\mathbb{I}(C_i\ge t)$ where $\mathbb{I}(\cdot)$ is an indicator function. 
We denote the time-dependent drug treatment by $A_i(t)$.  In our application, we focus on an add-on treatment which takes values 1 if there is an add-on, and 0 otherwise, at time $t$.  We are interested in optimizing a final continuous outcome $Y_i(\tau)$, but the approach could be extended to other types of outcome. For each patient $i=1, 2, ..., n$, the potential confounders for the treatment-outcome relationship at time $t$ are denoted by $\mathbf{K}_i(t)$. We assume that the confounders come before the visit and treatment add-on at time $t$, i.e., they are generated before the visit and potential treatment at time $t$. We assume that a set of effect modifiers $\mathbf{Q}_i(t)$ could be used to tailor treatment and monitoring schedule in order to optimize the final outcome. In our application, for instance, we assume that the effect of visiting at each month on the change in BP after one year is different across strata of monthly BP measurements, such that monthly BP acts as an effect modifier. 
}
 
 \textcolor{black}{ We assume that when a visit occurs, variables $\mathbf{Q}_i(t)$ are observed.  
 Effect modifiers are assumed to be observed at those covariate-dependent visit or observation times. The indicator of visiting physician at time $t$ is denoted by $dN_i(t)$, based on a visit counting process $N_i(t):= N_i^{\dagger}(t \wedge C_i)$, where $a \wedge b = min(a,b)$, that counts the number of observation times between times 0 and $t$ for patient $i$. The covariates affecting the observation of the effect modifiers are denoted by the set $\mathbf{V}_i(t)$. That set should contain any variable that affects simultaneously the observation of the effect modifier and the effect modifier itself.}

 \textcolor{black}{For all other processes than the effect modifiers, we assume that covariates are always observed, until censoring occurs. We use the general notation $\mathbf{P}^{\dagger}(t)$ to denote the observed (non-censored) part of the set $\textbf{P}(t)$. In case of censoring before time $\tau$, the final outcome is not observed. We discuss that scenario in Section \ref{miss}. We denote the observed effect modifiers (i.e., non-censored, and observed only when there is a visit) by $\mathbf{Q}^{\dagger}(t)$.}

 \textcolor{black}{
 We further define $R_i(t):=\{dN_i(t), A_i(t)\} \in \{ (0,0), (1,0), (1,1)\}$ the pair of visit and treatment add-on indicators at time $t$, which we will sometimes refer to as the three treatment strategies, noting that there cannot be a treatment add-on, i.e., $A_i(t)=1$, on days or months when there is no visit, i.e., $dN_i(t)=0$. The potential outcome $\widetilde{Y}_i^r(\tau)$ refers to the outcome we would observe at time $\tau$ if patient $i$ had received the regime $\bm{r}=\{r_1, r_2,...,r_{\tau-1} \} \subset \mathcal{R}$ with $\mathcal{R}$ the space of all possible regimes.} 


We use a line over a variable (set) to refer to the \textit{observed} history of the corresponding process, i.e., for an arbitrary process $P_i(t)$,$ \overline{P_i(t)} := \{ P^{\dagger}_i(0) ,  P^{\dagger}_i(1) , ..., P^{\dagger}_i(t) \}$. The full history of variables observed by time $t$ is $\mathcal{H}_i(t):=\{\overline{\xi_i(t)},  \overline{\mathbf{K}_i(t)}, \overline{\mathbf{V}_i(t)},\overline{dN_i(t)}, \overline{\mathbf{Q}_i(t)},\overline{A_i(t)}, \mathbb{I}(t \ge \tau)Y(\tau)\}$.

 A causal diagram depicting the assumed DGM and relations between the variables is given in Web Appendix A.




\subsection{Causal assumptions}\label{ass}
 
The following causal and missingness assumptions are required for identification of the optimal DMAR:
\begin{enumerate}
\item Sequential ignorability \citep{robins2000marginal}, which requires that we have in $\mathbf{K}(t)$ and $\mathbf{V}(t)$ all the variables to block backdoor paths or bias due to conditioning on colliders that could distort the causal effects (in strata of the effect modifiers) of the visit and add-on strategies at time $t$ on the final outcome,
$$\{dN_i(t), A_i(t)\} \perp  \widetilde{Y}_i^r(\tau)  \mid   \mathbf{K}_i(t), \mathbf{V}_i(t), \mathbf{Q}_i(t)=\mathbf{q}_i(t)$$, $\forall \hspace{0.2cm} \mathbf{q}_i(t)$, $t=1, 2, ..., \tau -1$ and all regimes $r \in \mathcal{R}$. In this work, we assume that the visit choice occurs first, and then the add-on choice, if there is a visit. This could be reflected in our confounding adjustment by factorizing the joint distribution of the visit and the add-on treatment variables. This is discussed more in depth in Section \ref{conf};
\item Stable unit treatment value \citep{rubin1980randomization}, which requires that treatment assignment for one individual does not affect the outcome of other individuals, combined with outcome consistency for the final outcome $Y(\tau)$, which means that the final outcome does correspond to one of the potential outcomes under the treatment and visit sequence received by the patient;
\item Positivity, which requires that all regimes in $\mathcal{R}$ have a non-null probability of occurring, considering that at any given time point, patients can receive one of three following ``treatment" strategies: No visit, a visit with no add-on treatment, and a visit with an add-on treatment;
\item \label{hr} Missingness at random (MAR) for the effect modifiers, defined here as
$$ \mathbf{Q}_i(t)  \perp  dN_i(t) \mid \mathbf{V}_i(t);$$
\item Finally, we assume coarsening at random \label{coarsen} \citep{robinsCAR} which assumes that the probability of censoring is independent of the final outcome given the observed history of variables $\mathcal{H}(t)$. Adjustment for informative censoring is further addressed in Section \ref{miss}. 
\end{enumerate}


 
 Note that assumptions \ref{hr} and \ref{coarsen} together mean that we are able to learn about $\mathbf{Q}_i(t)$ on times when there is no observation only by using the information available on $\mathbf{V}_i(t)$ (and other censoring predictors, if censoring is informative). This is key to detecting effect modification due to the visit process, as such effect modification requires that we be able to estimate how the effect of $dN_i(t)$ on $Y_i(\tau)$ varies in different strata of $\mathbf{Q}_i(t)$.
 
\subsection{Formulation of the optimal DMAR} \label{mat}

The causal estimands in this work are the causal effects of different monitoring and add-on treatment regimes assessed within different combinations of strata of patient characteristics $\mathbf{Q}_i(t), t=1,...,\tau-1$. The ``treatment" choice at a given month $t$ is between: No visit at month $t$, i.e., $R_i(t)=(0,0)$; A visit at month $t$ and no change in treatment course, i.e., $R_i(t)=(1,0)$; A visit at month $t$ and a change in treatment course (e.g., an add-on treatment), denoted $R_i(t)=(1,1)$. We present below the formulation of the DMAR in the case where there are no missing data and no informative censoring (such that $\mathbf{P}_i(t)=\mathbf{P}^{\dagger}_i(t)$ generally) and discuss the treatment of missing data in Section \ref{miss}.
 
 First, we separate effect modifiers $\mathbf{Q}(t)$ in two groups that do not need to be mutually exclusive, denoted by $\mathbf{Q}_{V}(t)$ and $\mathbf{Q}_{V,A}(t)$, to emphasize that effect modifiers can be different for the visit and the add-on drug effects. We define two so-called \textit{blip} functions \citep{robins2004optimal} that can be studied to find the optimal treatment for a given individual. The first blip, $\psi_{V,t}\{ r_t,\mathbf{q}_V(t)\}$, \textcolor{black}{is the expected change in potential outcome due to receiving the visit strategy corresponding to $r_t$ at time $t$ (i.e., visiting if $r_t=(1,\cdot)$, and not visiting otherwise) when compared with no visiting, for an individual with characteristics $\mathbf{Q}_V(t)=\mathbf{q}_V(t)$. The second blip, $\psi_{VA, t}\{ r_t,\mathbf{q}_{V,A}(t)\}$, is the additional effect on the potential outcome of receiving the add-on strategy corresponding to $r_t$ at time $t$ (e.g., $r_t=(1,1)$ corresponds to an add-on, and $r_t=(1,0)$ or $(0,0)$ are no add-on) for an individual} with characteristics $\mathbf{Q}_{V,A}(t)=\mathbf{q}_{V,A}(t)$. The linear model is a natural choice for modeling each blip function. In this work, we use the parametric linear models $\psi_{V,t}\{ R(t)=r_t,\mathbf{q}_{V}(t); \bm{\gamma}_t\}= r_t[1] \{ \gamma_{0t} + \bm{\gamma_{1t}}^T\mathbf{q}_V(t)\} $ and $\psi_{VA,t}\{ R(t)=r_t,\mathbf{q}_{V,A}(t); \bm{\gamma_t^*}\}= r_t[1] r_t[2] 
\{ \gamma^*_{0t} + \bm{\gamma^{*T}_{1t}}\mathbf{q}_{V,A}(t) \}$. Note, we use the notation $r_t[1]$ and $r_t[2]$ to refer to the first and second element of $r_t$, which also correspond to the visit and add-on treatment at time $t$, respectively. 
 
 We further assume the following structural model that emphasizes the impact of previous characteristics (including confounders) on the outcome:
 \begin{align}
&E[\widetilde{Y}^r(\tau) \mid  \mathcal{H}(\tau)=\mathbf{h}(\tau); \bm{\beta}, \bm{\gamma} , \bm{\gamma}^*] = \sum_{s=1}^{\tau} f_s\{\mathbf{h}(s); \bm{\beta}_s\} + r_s[1] \{ \gamma_{0s} + \bm{\gamma_{1s}}^T\mathbf{q}_V(s)\}   \nonumber \\  
&\hspace{8cm} +r_s[1] r_s[2] \{ \gamma^*_{0s} + \bm{\gamma^{*T}_{1s}}\mathbf{q}_{V,A}(s) \} , \label{modely}
 \end{align}
 where we call $f\{\mathcal{H}(s)=\mathbf{h}(s); \bm{\beta}\}$ the \textit{treatment-free model}, borrowing the term from \cite{wallace2015doubly}. It is a function of the whole history $\mathbf{h}(s)$ that should minimally contain confounders but could also contain pure predictors of the outcome, i.e., variables that affect the outcome but not the treatment strategy prescribed. Correctly modeling the relationship between the confounders in $\mathcal{H}(t)$ and the outcome via that treatment-free model provides a first attempt at correctly adjusting for confounders. The other attempt is via balancing weights for the confounders, discussed in Section \ref{conf}. 

We define a deterministic decision rule at time $t$, $d_t: \mathbf{q}(t) \rightarrow r_t$ which takes as input patient characteristics $\mathbf{Q}(t)=\mathbf{q}(t)$ and outputs for each time $t$ a decision between the three options listed in the previous Section. The optimal decision rule is then defined as
$$ d_t^{opt}\{ \mathbf{q}(t) \} =  \operatorname*{argmax}_{r_t}  \left\{ \psi_{V,t}\{ r_t,\mathbf{q}_V(t)\} + \psi_{VA, t}\{ r_t,\mathbf{q}_{V,A}(t)\} \right\}.$$

Taking as an example the potential outcome under treatment decision at time $\tau-1$, our approach also corresponds to modeling the difference in potential outcomes using the following nested structural mean models \citep{robins1994correcting}: 
\begin{align}
&E[\widetilde{Y}^{\{r_1,  ..., r_{\tau-2}, r_{\tau-1}\}}(\tau) \mid r_{\tau-1}, \mathbf{Q}(\tau-1)= \mathbf{q}(\tau-1)]- E[\widetilde{Y}^{\{r_1, ..., r_{\tau-2}, (0, 0) \}}(\tau)\mid r_{\tau-1}, \mathbf{Q}(\tau-1)= \mathbf{q}(\tau-1)]   \nonumber \\
  &=E[Y(\tau) \mid r_{\tau-1},   \mathbf{q}(\tau-1)]- E[\widetilde{Y}^{\{r_1, ..., r_{\tau-2}, (0, 0) \}}(\tau)\mid r_{\tau-1},  \mathbf{q}(\tau-1)]  \nonumber \\
& = \psi_{V, \tau-1}\{ r_{\tau-1},\mathbf{q}_{V}(\tau-1); \bm{\gamma}_{\tau-1}\} +\psi_{VA, \tau-1}\{ r_{\tau-1},\mathbf{q}_{V,A}(\tau-1); \bm{\gamma_{\tau-1}^*}\}, \label{first} 
\end{align}
which could be solved using g-estimation.   Under that model, if the blips are correctly specified, the \textit{true} optimal decision at, e.g., time $\tau-1$ is given by:
 \begin{itemize}
     \item  $R(\tau-1)=(0,0)$ (no visit and no add-on drug) if $\psi_{V, \tau-1}\{ r_{\tau-1}=(1, \cdot),\mathbf{q}_{V}(\tau-1); \bm{\gamma}_{\tau-1}\}<0$ and $\psi_{VA, \tau-1}\{ r_{\tau-1}=(1, 1),\mathbf{q}_{V,A}(\tau-1); \bm{\gamma_{\tau-1}^*}\} <0$, or if $\psi_{V, \tau-1}\{ r_{\tau-1}=(1, \cdot),\mathbf{q}_{V}(\tau-1); \bm{\gamma}_{\tau-1}\}<0$ and $\psi_{VA, \tau-1}\{ r_{\tau-1}=(1, 1),\mathbf{q}_{V,A}(\tau-1); \bm{\gamma_{\tau-1}^*}\} >0$ and $\psi_{V, \tau-1}\{ r_{\tau-1}=(1, \cdot),\mathbf{q}_{V}(\tau-1); \bm{\gamma}_{\tau-1}\} + \psi_{VA, \tau-1}\{ r_{\tau-1}=(1, 1),\mathbf{q}_{V,A}(\tau-1); \bm{\gamma_{\tau-1}^*}\} <0$;
     \item  $R(\tau-1)=(1,0)$ (a visit and no add-on drug) if $\psi_{V, \tau-1}\{ r_{\tau-1}=(1, \cdot),\mathbf{q}_{V}(\tau-1); \bm{\gamma}_{\tau-1}\}>0$ and $\psi_{VA, \tau-1}\{ r_{\tau-1}=(1, 1),\mathbf{q}_{V,A}(\tau-1); \bm{\gamma_{\tau-1}^*}\} <0$
     \item  $R(\tau-1)=(1,1)$ (a visit and an add-on drug) if $\psi_{V, \tau-1}\{ r_{\tau-1}=(1, \cdot),\mathbf{q}_{V}(\tau-1); \bm{\gamma}_{\tau-1}\}>0$ and $\psi_{VA, \tau-1}\{ r_{\tau-1}=(1, 1),\mathbf{q}_{V,A}(\tau-1); \bm{\gamma_{\tau-1}^*}\} >0$ or if $\psi_{V, \tau-1}\{ r_{\tau-1}=(1, \cdot),\mathbf{q}_{V}(\tau-1); \bm{\gamma}_{\tau-1}\}<0$ and $\psi_{VA, \tau-1}\{ r_{\tau-1}=(1, 1),\mathbf{q}_{V,A}(\tau-1); \bm{\gamma_{\tau-1}^*}\} >0$ and $\psi_{V, \tau-1}\{ r_{\tau-1}=(1, \cdot),\mathbf{q}_{V}(\tau-1); \bm{\gamma}_{\tau-1}\}+\psi_{VA, \tau-1}\{ r_{\tau-1}=(1, 1),\mathbf{q}_{V,A}(\tau-1); \bm{\gamma_{\tau-1}^*}\} >0$.
 \end{itemize}

 Backward induction, also called dynamic programming, can be used to estimate the optimal DMAR time-dependent parameters $\bm{\gamma}_{t}, \bm{\gamma}^*_{t}$. This is the approach also used to solve g-estimation \citep{robins2004optimal}, dynamic weighted ordinary least squares (dWOLS) blip estimators \citep{wallace2015doubly} and Q-learning (see e.g., \cite{sutton2018reinforcement} who show that it provides optimal rules with Q-learning), and the approach we take in this manuscript. At each step of the backward induction, the parameters are estimated using potentially weighted ordinary least squares estimators. It can be done with the \texttt{lm} package in R, adding as predictors in the outcome regression all the confounders from the treatment-free model, the visit blip variables, and the add-on blip variables, in their right format, and adding a weight argument to incorporate balancing weights. First, a rule for time $\tau-1$ is derived, using $Y(\tau)$ as the outcome in the regression model, and covariates measured by time $\tau-1$ that could act as confounders for the decision at time $\tau-1$, in addition to the blip functions for that time. For subsequent regressions, the outcome is replaced by a pseudo-outcome corresponding to the estimated outcome we would observe if the optimal visit and add-on treatment were received at all future times. Pseudo-outcomes can be computed by removing from the final outcome all the blip functions corresponding to the visit and add-on received each month, and adding the blip functions corresponding to the optimal visit and add-on strategy each month, starting by the end of follow-up and going backward.  
 
 In simulation studies, we assess both a doubly-robust approach in which inverse probability of treatment weights or other balancing weights provide an additional safety net against confounding of the ``treatment" and outcome relationship (leading to weighted least squares estimating equations), which we call WOMA (for Weighted equations for Optimal Monitoring and Add-on) and a singly-robust approach that only uses ordinary least squares to estimate the parameters of both blip functions, which we call QLOMA (for Q-Learning analog equations for Optimal Monitoring and Add-on). These estimators, as opposed to their respective analogs dWOLS and Q-learning  (\cite{wallace2015doubly,murphy2005generalization,chakraborty2011dynamic}), use a new outcome model and new weight specification that tackle specifically the choice of visiting and add-on drug during a visit.

\subsection{Confounding adjustment that results in double robustness}\label{conf}

Some patients are more likely to visit given their personal characteristics, and if they do, some of them are more likely to receive an add-on treatment. The probability of ``receiving" one of the three treatment options is given by $P\{dN_i(t)=d, A_i(t)=a\mid \mathcal{H}_i(t)\}$ for $(d, a)$ pairs $\{(0,0), (1,0), (1,1)\}$. Under some (problem-dependent) assumption of conditional independence, such as $dN_i(t) \perp A_i(t) \mid \mathbf{V}_i(t)$, and the other assumptions listed in Section \ref{ass}, the probability can be factorized as $P\{dN_i(t)=d\mid \mathbf{V}_i(t)\}\times P\{A_i(t)=a\mid dN_i(t)=d, \mathbf{K}_i(t)\}$, a product of two confounding mechanisms. 

To make the WOMA doubly-robust, we propose a double adjustment, with the first adjustment done via a correctly specified \textit{treatment-free} model $f\{\mathbf{h}(t); \bm{\beta}\}$ model in equation \ref{modely}. The second adjustment is done via correctly specified balancing weights \citep{wallace2015doubly, simoneau2020estimating}. We assess two types of balancing weights in simulations. First, under the conditional independence assumption mentioned above, inverse probability of treatment weights are given by the inverse of estimates for $P
\{dN_i(t)=d\mid \mathbf{V}_i(t)\}\times P\{A_i(t)=a\mid dN_i(t)=d, \mathbf{K}_i(t)\}$ where we consider that $P\{A_i(t)=1\mid dN_i(t)=0, \mathbf{K}_i(t)\}$ is 0. Estimates can be obtained by fitting a correctly specified model, e.g., a logistic regression model, to estimate parameters $\bm{\phi}$ in $P\{dN_i(t)=1\mid \mathbf{V}_i(t); \bm{\phi}\}$. Then, only in a subgroup of patients who visited, we model $P\{A_i(t)=1\mid dN_i(t)=1, \mathbf{K}_i(t); \bm{\omega}\}$ by fitting a second logistic regression with the add-on indicator as the outcome.  

The resulting weights are a product of two weights. In practice, they are likely to be highly variable if some patients have near-zero probabilities to receive one of the treatment strategies for some time points. We propose to use overlap weights to reduce the variance of the WOMA. The treatment strategy can take 3 categories at each time, such that the usual overlap weights developed for a binary treatment cannot be used. Instead, we propose to compare the previous, more standard inverse probability weights based on the probability factorization, to a second approach using generalized overlap weights \citep{li2023overlap}, with which we model the visit and add-on jointly. We define these weights as
\begin{align*}
    w_i(t) =  \sum_{r \in \{(0,0), (1,0), (1,1) \}} \left\{ \frac{1}{\sum_{k \in \{(0,0), (1,0), (1,1) \}} \frac{1}{e_{i,k}}} \frac{\mathbb{I}(R_{i}(t)=r)}{e_{i,r}} \right\}
\end{align*}
where $e_{i,k}= P\{dN_i(t)=k[1], A_i(t)=k[2]\mid \mathbf{K}_i(t), \mathbf{V}_i(t)\}.$ These weights consist of the product between standard inverse probability weights and the harmonic mean of the propensity scores. The sum over $r$ above is only non-null for the treatment actually received.

We obtain estimates for the probabilities $e(\cdot)$ by fitting a multinomial regression model as a function of potential confounders of the visit and add-on drug relationships (with the outcome being the 3-category treatment strategy). 
If $\mathbf{K}_i(t) \subset \mathbf{V}_i(t)$, the multinomial model only needs to incorporate covariates $\mathbf{V}_i(t)$. If there are covariates in $\mathbf{K}_i(t)$ that are not included in $\mathbf{V}_i(t)$, the product of overlap weights fitted separately for the visit and the add-on could be used. 

 
 \subsection{Missing data and informative censoring}\label{miss}
 
 A challenge in deriving the optimal DMAR is that variables used in the structural model in equation \ref{modely} are missing at different times not completely at random \citep{rubin1976inference}. In the SPRINT, for instance, the effect modifier BP is not measured at all months.

\textcolor{black}{Inverse intensity of visit weights \citep{lin2004analysis} or inverse probability of missingness weights cannot be used directly to address missing data in our setting because part of the causal effects of interest is the visit effect. Since effect modifiers are only observed when there is a visit, the visit effects would not be identifiable if we merely removed ``non-visit" data and re-weighted the remaining observations. } 

We propose two approaches to address missingness in the effect modifiers: i) Multiple imputations by chained equations (MICE) of effect modifiers, and ii) Last Observation Carried Forward (LOCF). The first approach, MICE, relies on assumption 4 in Section \ref{ass}; 
a causal diagram could be used to determine which predictors of missingness should be incorporated in the imputation models for each variable in the set $\mathbf{Q}(t)$. The LOCF approach, on the other hand, is not guaranteed to provide valid inference even under the assumption of data MCAR. It relies on strong assumptions, such as, the fact that the longitudinal process would not vary after attrition. In addition, that approach does not consider the uncertainty due to the missing values as it only replaces them with the LOCF and treats the new observations as if they were original data \citep{molenberghs2004analyzing}.

We suggest using the method which assumptions align the most with the missingness assumptions one is ready to make about the data. 
If using MICE, we propose the use of a sequential approach to imputation akin to \cite{shortreed2014multiple} 
that aims to preserve the time structure from the original data, but other approaches could work well as long as the missingness assumptions are correct.

The longitudinal data used to develop the optimal DMAR are also likely to be affected by informative censoring that could occur at any patient-dependent time during the study. If it is more realistic to assume that data are censored after a certain point, such that the final outcome $Y(\tau)$ is not observed, a possibility is to use only the patient data from patients who were followed until $\tau$ and re-weight these observations to recreate the full patient sample, or it might be easier to impute all the variables, including the final outcome. If using the first approach, inverse probability of censoring weights (IPCW) \citep{robins2000correcting} can be used to create a pseudo-population that corresponds to the full sample if they were followed until time $\tau$. If some of the weights are extreme, IPCW can be stabilized using a marginal model for the probability of censoring as the numerator. Our approach to address informative censoring is described in Web Appendix B and further assessed in simulations in which we consider two different mechanisms for informative censoring. 

 Note, if using the IPCW approach, patients for whom the final outcome is not observed cannot contribute to our estimating equations for the blip parameters. However, some of their person-time could be used to estimate other nuisance models parameters, such as the parameters in the treatment model leading to inverse or overlap weights. A thorough study on the assumed causal diagram should be done to assess whether using these data for estimating nuisance models can lead to selection bias in the nuisance model parameter estimates \citep{daniel2012using}. In simulations, all patient observations for any patient who was not followed until time $\tau$ were removed and the remaining data were weighted with IPCW.

 The estimation problem for the optimal DMAR, for both the WOMA and QLOMA estimators, corresponds to a series of estimating equations, one for each time point when a decision rule is derived. The estimating equations are shown in Web Appendix C. 
 Demonstration of the consistency of the WOMA when either the treatment-free or the joint visit and treatment models are correctly specified is shown in Web Appendix C, in the case of no missing data. Whether missing data or informative censoring are present, the corresponding missing variable or censoring models must be correctly specified, as well as both blip functions, for consistency of both the WOMA and QLOMA estimators. We also review in Web Appendix D the asymptotic theory and variance estimation of these estimators.


\section{Simulation studies}

We assessed the WOMA and QLOMA in large simulation studies, under different scenarios for the DGM. We focused on a 2-stage optimal DMAR, i.e, a sequence of 2 decision rules to choose between the 3 treatment options, no visit, visit with no add-on drug, and visit with an add-on drug, to optimize an outcome observed at a third time point.

We assessed scenarios in which either the set of effect modifiers was always observed, or it contained MAR missing values. Two approaches to address missing data were evaluated: imputation with MICE, or LOCF. For MICE, we used a sequential approach in which the baseline data were imputed first, once, then month 1 data were imputed using the imputed baseline data as predictors and any time-dependent information available at month 1, and so on, always adding variables measured previously in time as predictors in the future imputation models. We repeated this procedure 25 times to obtain 25 completed datasets and obtained each final blip estimate by averaging the corresponding 25 estimates from the 25 imputed datasets. 
In additional simulation studies, we induced informative censoring that depended either on baseline covariates or time-dependent covariates, respectively. In all scenarios above, we assessed the following 6 estimators for the optimal DMAR:
\begin{enumerate}
\item[--] The QLOMA that adjusts correctly ($\hat{\gamma}_{Oc}$ for ``Outcome" and ``Correct") or wrongly ($\hat{\gamma}_{On}$) for confounders and predictors of visits via a treatment-free model in the outcome model;
\item[--] The WOMA that incorporates correctly specified balancing weights and incorporates a wrongly specified treatment-free model in the outcome model, $\hat{\gamma}_{On-Wc}$ (which stands for not correct outcome model and correct weights);
\item[--] The WOMA that incorporates correctly specified treatment-free model in the outcome model and wrongly specified balancing weights, $\hat{\gamma}_{Oc-Wn}$;
\item[--] The WOMA that incorporates correctly specified balancing weights and correctly specified treatment-free model in the outcome model, $\hat{\gamma}_{Oc-Wc}$.
\item[--] The WOMA that incorporates wrongly specified balancing weights and wrongly specified treatment-free model in the outcome model, $\hat{\gamma}_{On-Wn}$.

\end{enumerate}

We now describe the DGM, removing the patient index $i$ to simplify notation. Note that the notation $\mathcal{N}(a,b)$ refers to the Normal distribution with mean $a$ and standard deviation $b$. First, we simulated two confounders at baseline as $K_1(0) \sim \mathcal{N}(4, 3)$ and $K_2(0) \sim \mathcal{N}(5, 1.4)$, a blood pressure measurement $Y(0) \sim \mathcal{N}(120, 13)$, and a randomized drug treatment at baseline $A(0)\sim \text{Bernoulli}(0.5)$.

Data at month 2 were simulated as follows (data at month 1 were simulated in a similar manner except that $Y(1)$ depended on less previous variables): $K_1(2) \sim \mathcal{N}(\mu_{k12}, 3)$ with $\mu_{k12}= -26 + 0.8 \hspace{0.1cm} K_1(1) +0.2\hspace{0.1cm} Y(1)+0.1\hspace{0.1cm}A(1)+0.1\hspace{0.1cm}dN(1)$; $K_2(2) \sim \mathcal{N}(\mu_{k22}, 3)$, with $\mu_{k22}= -43 + 1 \hspace{0.1cm} K_2(1) +0.2\hspace{0.1cm} Y(1)+0.1\hspace{0.1cm}A(1) - 0.1\hspace{0.1cm}dN(1)$;  $Y(2) \sim \mathcal{N}(\mu_{y2}, 3)$ with $\mu_{y2}= 122  +  0.05 \hspace{0.1cm} A(0)  -0.005 Y(0) + 0.02\hspace{0.1cm} K_1(0) +   0.02 \hspace{0.1cm} A(0) \times X(0)   + 0.004  \hspace{0.1cm}A(0) \times Y(0)   +  0.02  \hspace{0.1cm} Y(0) \times K_1(0) +  0.02 \hspace{.1cm} K_1(1) - 1.4 \hspace{.1cm} A(1)\times dN(1) -0.0005 +1\hspace{0.1cm} K_1(1)  \times dN(1) +1 \hspace{0.1cm} Y(1) +  0.002 \hspace{0.1cm} dN(1)\times Y(1) + 0.1\hspace{0.1cm} A(1)\times dN(1)\times K_1(1) + 0.04\hspace{0.1cm} A(1)\times dN(1)\times Y(1)$; $dN(2) \sim \text{Bernoulli}(p_{dN2})$ with $\text{logit}(p_{dN2})=-18 + 0.3 \hspace{0.1cm} K_1(1)-0.8 \hspace{0.1cm} K_2(1) + 0.1 \hspace{0.1cm} A(1) + 0.02 \hspace{0.1cm} Y(1)$ and $A(2) \mid (dN(2)=1)  \sim \text{Bernoulli}(p_{A2})$ with  $\text{logit}(p_{A2})=0.4\hspace{0.1cm} K_1(1)  -0.05\hspace{0.1cm} K_2(1) +  0.2\hspace{0.1cm} A(1) -0.04 \hspace{0.1cm} Y(1) $ and $A(2) \mid (dN(2)=0)  = 0$. The outcome at month 3 was simulated as  $Y(3) \sim \mathcal{N}(\mu_{y3}, 3)$ with $\mu_{y3}=  
 134  +  0.05 \hspace{0.1cm} A(0)  -0.005 Y(0) + 0.02\hspace{0.1cm} K_1(0) -0.6 \hspace{0.1cm} K_2(0) +   0.02 \hspace{0.1cm} A(0) \times K_1(0)   + 0.004  \hspace{0.1cm}A(0) \times Y(0)   +  0.02  \hspace{0.1cm} Y(0) \times K_1(0)
 +  0.02 \hspace{.1cm} K_1(1) -1.5 \hspace{0.1cm} K_2(1)  - 1.4 \hspace{.1cm} A(1)\times dN(1) -0.005 \hspace{0.1cm} Y(1)  + 0.18 \hspace{0.1cm} K_1(1)  \times dN(1)  +  0.002 \hspace{0.1cm} dN(1)\times Y(1) + 0.1\hspace{0.1cm} A(1)\times dN(1)\times K_1(1)+ 0.04\hspace{0.1cm} A(1)\times dN(1)\times Y(1)  + 1.5\hspace{0.1cm} A(2)\times dN(2) -0.005 \hspace{0.1cm} Y(2) + 0.02 \hspace{0.1cm} K_1(2)  + 0.02 \hspace{0.1cm} dN(2)    -1.5 \hspace{0.1cm} K_2(2) +                           1 \hspace{0.1cm} dN(2) \times K_1(2)+   0.01\hspace{0.1cm} dN(2)\times Y(2) -1.2 \hspace{0.1cm} A(2)\times dN(2) \times K_1(2) +  0.01 \hspace{0.1cm} dN(2)\times A(2) \times Y(2). $

We used two criteria to evaluate the proposed estimators. First, using 1000 simulations and a sample size of 50000, we computed the empirical bias and standard error of the blip parameters for the last decision rule, i.e., the rule at the last time point when a decision need be made (i.e., $\tau-1$ in general, here $t=2$). In the results, we denote by $\gamma_0=1$, $\gamma_K=1$, $\gamma_Y=0.01$, $\gamma_0^*=1.5$, $\gamma_K^*=-1.2$, and $\gamma_Y^*=0.01$ the parameters corresponding to the two blips for decision at time 2, in the model for $Y(3)$. Second, the WOMA and QLOMA were evaluated in terms of difference in value function, with the value function defined as the expected outcome under a treatment strategy. We computed the difference between the value function under the treatment actually received and under the estimated optimal treatment decisions (found from the WOMA and 
 QLOMA rules estimated with simulations on samples of 50000 patients from the previous step) in a new population of 1 million patients. This quantity indicates how the outcome would be higher, or lower, on average, under the optimal treatment, i.e., it represents the gain from optimizing the treatment over the two time points.

 In the scenario with missing data in the effect modifiers, we merely removed data on the tailoring variables on any time $t$ when $dN(t)=0$. In the scenarios with informative censoring, censoring was respectively generated as follows: $\xi(2) \sim \text{Bern}(1, p_{c2})$ with $p_{c2}=\text{expit}\{8 + 0.2 \hspace{0.1cm} A(0) - 0.1 \hspace{0.1cm} Y(0) + 0.2\hspace{0.1cm} K_1(0)\}$ and $\xi(1) \sim \text{Bern}(1, p_{c1})$ with $p_{c1}=\text{expit}\{10 - 0.2 \hspace{0.1cm} A(0) - 0.1 \hspace{0.1cm} Y(0) + 0.1\hspace{0.1cm} K_1(0)\}$, which together led to an approximate 25\% rate of censoring; and $\xi(t) \sim \text{Bern}(1, p_{ctd})$ with $p_{ctd}=\text{expit}\{10 - 0.2 \hspace{0.1cm} A(t) - 0.1 \hspace{0.1cm} Y(t) + 0.1\hspace{0.1cm} K_1(t)\}$ in the case of time-dependent censoring, which led to a censoring proportion around 7\% (i.e., 7\% were loss to follow-up before the measurement of the final outcome).

 More details on simulation studies, including how data were generated at time 1 and how we parameterized the wrong nuisance models, are given in Web Appendix E. The causal diagram corresponding to the simulation DGM is also presented in Web Appendix E. In the manuscript, we present the results for the 6 estimators, corresponding to different versions of the WOMA and QLOMA, for the scenarios with no missing data or mith MAR data addressed using MICE. We present in Web Appendix F the results when using LOCF instead. In the main analysis above, we assessed generalized overlap weights to adjust for confounding. In Web Appendix G, we show the same analyses using inverse probability of treatment weights.


\subsection{Results}\label{poftr}

\begin{figure}[hbt!]
\includegraphics[width=1.05\textwidth]{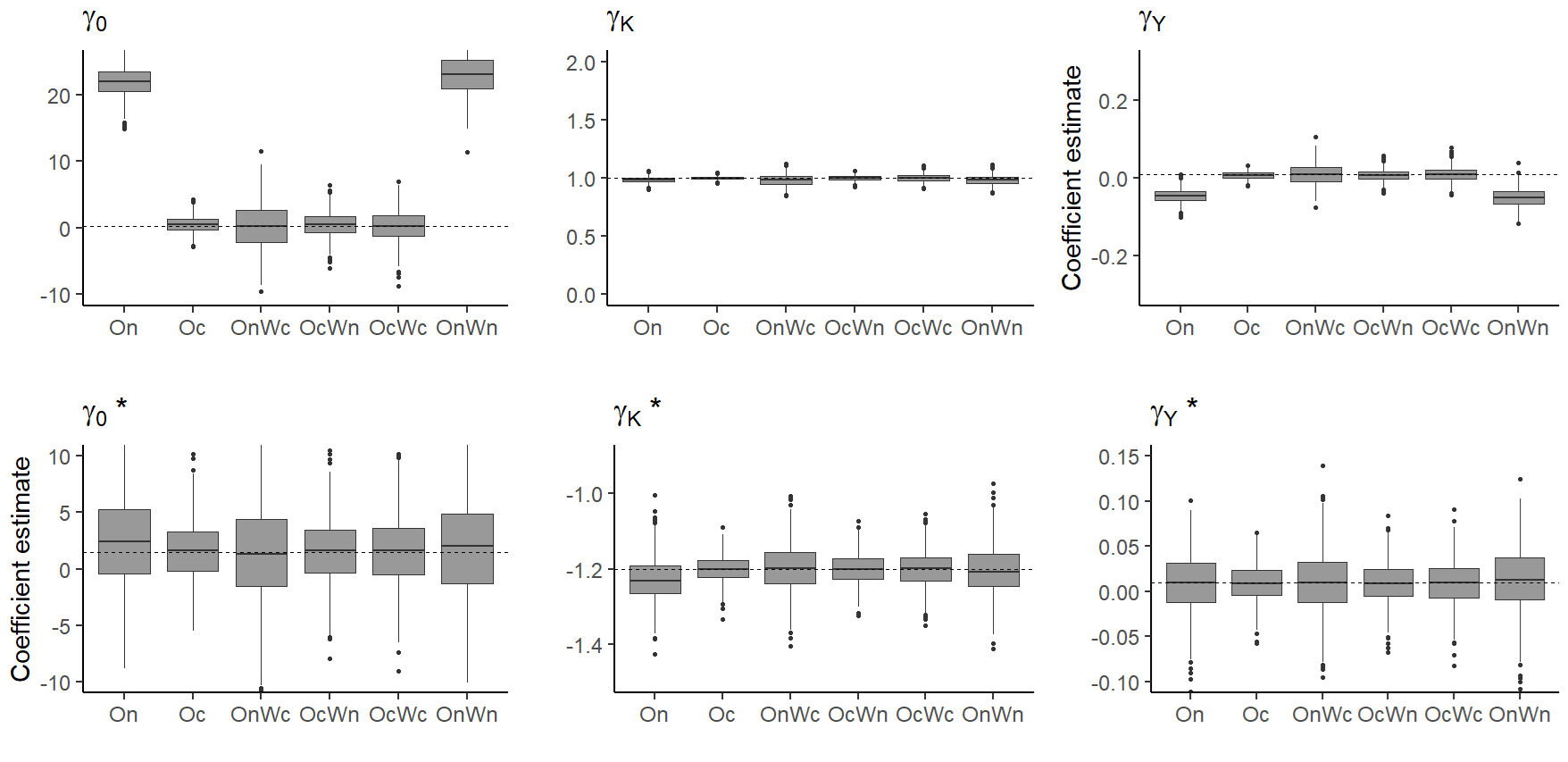}
\caption{Results under \textbf{no missing data} in the effect modifiers, with the weighted estimator using \textbf{overlap weights}. Each boxplot represents the distribution of 1000 point estimates. Sample size is 50000. The dashed line represents the true effect. On and Oc: Not correct and correct outcome models; Wn and Wc: Not correct and correct weight models; The top coefficients are those from the visit blip function (intercept, interaction between visit and confounder, and interaction between visit and previous outcome) while the bottom coefficients are the analogs from the treatment blip function. The estimators $O_c$ and $O_n$ correspond to the QLOMA, and all other estimators to the WOMA.}  \label{figove1}
\end{figure}

\begin{figure}[hbt!]
\includegraphics[width=1.\textwidth]{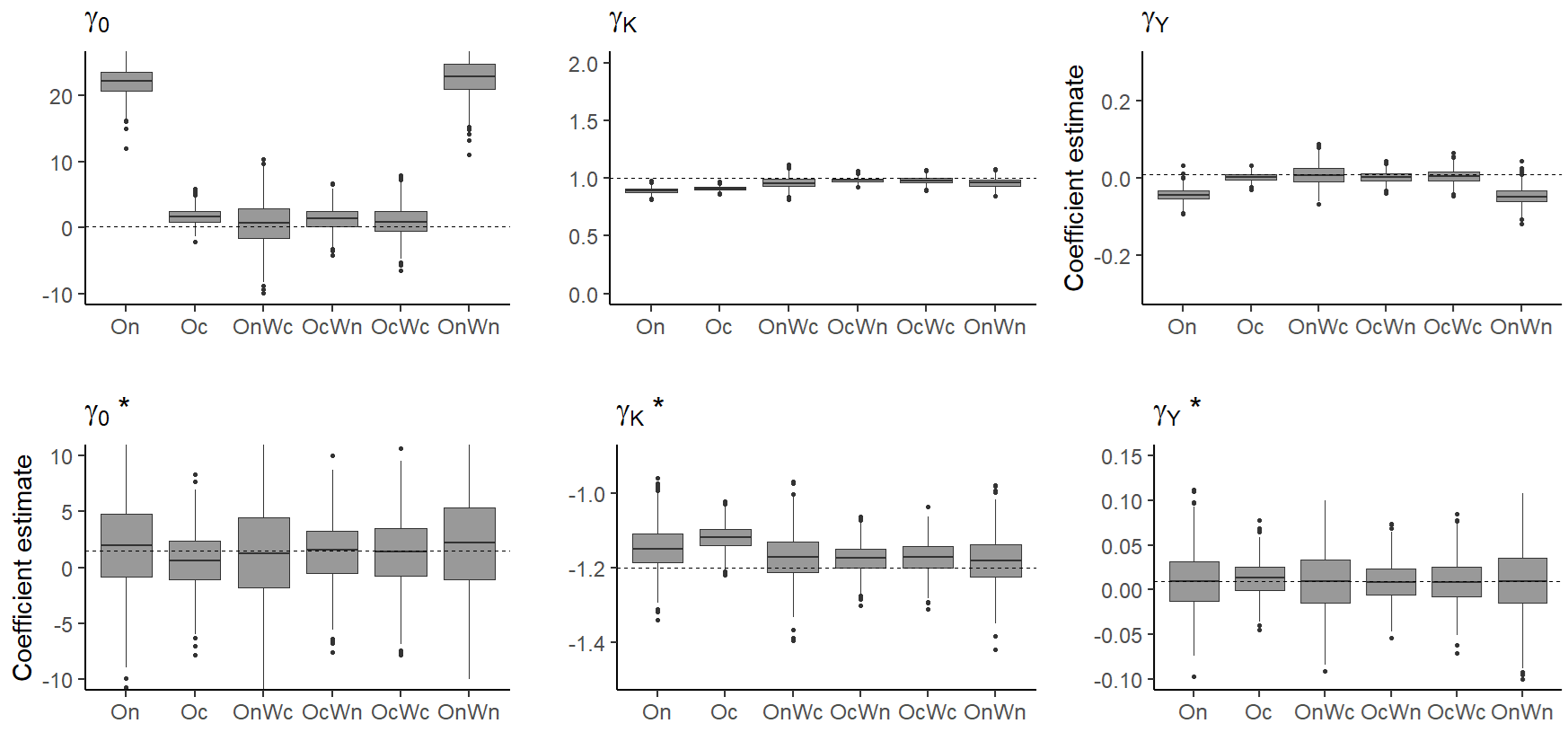}
\caption{ Results under \textbf{missing data and using MICE}, with the weighted estimator using \textbf{overlap weights}. Each boxplot represents the distribution of 1000 point estimates (averaged over 25 imputed datasets). Sample size is 50000. The dashed line represents the true effect.   On and Oc: Not correct and correct outcome models; Wn and Wc: Not correct and correct weight models; The top coefficients are those from the visit blip function (intercept, interaction between visit and confounder, and interaction between visit and previous outcome) while the bottom coefficients are the analogs from the treatment blip function. The estimators $O_c$ and $O_n$ correspond to the QLOMA, and all other estimators to the WOMA.}\label{figove1mice}
\end{figure}

 In the comparison of the blip coefficients, the overlap weights led to a much smaller variance of the coefficients than using inverse probability of treatment weights (Figure \ref{figove1} compared with Web Figure 4 in Web Appendix G). As expected, the blip parameter estimates all converged to the truth when using the QLOMA with a correctly specified treatment-free model, or the WOMA with at least one of the two nuisance models (weight or treatment-free) correctly specified. The bias was important for the coefficient $\gamma_0$ representing the (non-modified) causal effect of visit on the outcome, in settings in which the estimators were not consistent. Interestingly, LOCF performed poorly in simulation studies (Web Figure 3 in Web Appendix F) but the WOMA seems to be protected against the violation of the LOCF assumptions. The QLOMA combined with LOCF performed the most poorly with highly biased coefficients corresponding to the (non-modified) effect of treatment (coefficient $\gamma_0^*$) and effect of visit ($\gamma_0$). The coefficient corresponding to the interaction term between the visit indicator, the treatment and the previous outcome ($\gamma_Y^*$) was also importantly biased when using the QLOMA (Web Figure 3), while we did not have this issue with MICE or when we did not incorporate missing data in the effect modifiers (Figures 1 and \ref{figove1mice}).

 The adjustment for informative censoring via inverse probability of censoring weights led to consistent estimators when the censoring predictors were either time-fixed or time-dependent (Web Appendix H).

 In the simulated cohort of 1 million patients, the value function under the treatment received was 209.2. Under the optimal DMAR using the new estimators, the value was 10.2 points higher (219.4), for both the QLOMA and WOMA and independently of whether the nuisance models were correctly specified or not. The benefit was similar when using either overlap weights or inverse probability weights for the adjustment of confounding (less than 0.01 difference).
 
The empirical bias and standard deviation for all blip coefficients in each scenario assessed in simulation studies are presented in Web Tables 1-3 in Web Appendix I.

\section{Application}

We applied the WOMA and QLOMA to data from the \textit{Systolic Blood Pressure Intervention Trial} (SPRINT) in the United States (US). The SPRINT is a longitudinal study aimed at comparing two BP control strategies and their effect on different cardiovascular outcomes. The two strategies it compares are defined as either intensive, \textit{Adjusting treatment if the systolic BP measured at a given visit is higher than 120 mmHg}, or standard, \textit{Adjusting treatment if the systolic BP measured at a given visit is higher than 140 mmHg} \citep{ambrosius2014design,sprint2015randomized}. We had access to data on 9361 patients aged over 50 followed in the SPRINT and randomized to one of the two BP management strategies. Enrolled patients were diagnosed with hypertension with a systolic BP in the range of 130-180 mmHg at baseline. While the control strategy (i.e., choice between 120 and 140 mmHg as the threshold) was randomly assigned to patients, antihypertensive drugs that patients received during follow-up were prescribed based on their evolving health condition. 

The observation scheme in SPRINT is complex. A flow chart showing the planned and unplanned observation times is available with its protocol \citep{sprint2016systolic}. 
Routine visits were scheduled monthly for the first 3 months, and quarterly thereafter until the occurrence of outcome, loss-to-follow-up, or end of study, whichever occurred first. When physicians titrated or prescribed additional antihypertensive drugs, they requested to have an unscheduled visit in one month. Missing visits could occasionally occur at scheduled visits, yet more frequently at unscheduled visits. We assumed that missing data in between planned visit times and those missing at planned visit times are MAR or MCAR, in which cases our approaches are appropriate. 

We restricted the analysis to the first year of follow-up, during which, according to the protocol for this randomized controlled trial, patients were expected to be observed at baseline, months 1, 2, 3, 6, 9, and 12, see e.g., \cite{sprint2016systolic}.  Other covariate-dependent visits could occur in between those times. 

\subsection{Confounders, treatment, effect modifiers and outcome definitions}

The SPRINT data contains demographics, time-dependent BP measurements, clinical outcomes (such as cardiovascular events), report of adverse events, medication use, including information on different antihypertensive drugs prescribed, qualify of life (QOL) measures, and laboratory tests. In this analysis, we did not use data on QOL and laboratory tests, nor did we use (the rare) report of adverse events.

Similary to the simulation study, we used as the exposure the pairs of indicators $\{dN_i(t), A_i(t)\}$ where $A_i(t)=1$ corresponds to a drug add-on on at month $t$ and $dN_i(t)=1$ means there was a physician visit any day of that month (with e.g., month 1 corresponding to days 0 to 30; month 2 corresponding to days 31 to 60; and so on). We defined the add-on as any antihypertensive drug class switch or any additional antihypertensive drug class on that month when compared with the previous month. The 20 different antihypertensive drug classes that were considered are listed at the beginning of Web Appendix J. To define continuous use of a drug, and because prescription duration was not available for most prescriptions, we allowed a 120 days grace period from each prescription start date. We chose 120 days because this is the expected duration for antihypertensive drugs in this dataset and the median duration for the prescriptions that did have an end date, in the database. To summarize the add-on definition and what it implies for the causal contrasts, each month $t$, the subgroup $A(t)=0$ was a combination of patients who never switched from their original antihypertensive drug, or patients who had switched before but continued the same treatment course that month (when compared to the pevious month). On the other hand, the group $A(t)=1$ contained patients who had a treatment switch or add-on at time $t$. The estimated causal effects each month therefore corresponded to those of visiting and the additional causal effect of switching treatment or adding-on, no matter whether the patient was still on the original treatment or a new treatment since time 0. 

The final outcome was defined as the \textit{absolute reduction in systolic BP} at time $\tau=12$ from cohort entry ($t=0$). The baseline confounders included the indicator of being in the intensive BP management group (as opposed to standard), age (continuous), sex assigned at birth, race (white, black, hispanic, and other), systolic BP, high-density lipoprotein (aka total cholesterol) in mg/dL, body mass index in $kg/m^2$, coronary vascular disease (yes/no), indicator of ever smoker, alcohol abuse (yes/no), aspirin use (yes/no), and statin use (yes/no). The time-dependent confounders included the indicator of visit at the previous month, the indicator of add-on at the previous month, and the systolic BP both at the previous month and the current month. 

The potential effect modifiers included the indicator of being in the intensive BP management group, age, female sex, race, BMI, indicator of coronary vascular disease, indicator of ever smoker, alcohol abuse (yes/no), aspirin use (yes/no), statin use (yes/no),  high-density lipoprotein, as well as the baseline systolic BP and the current (time-dependent) systolic BP measure (e.g., BP at month 11 for the rule at month 11).

\subsection{Addressing missing data} 
To address missing data, with most missing data being in the time-dependent systolic BP, we used a sequential imputation approach 
combined with MICE to estimate 25 completed datasets. In each conditional specification for the models used with MICE, we included all the other variables available for the analysis, observed before or by that time, including confounders, effect modifiers, and the previous add-on variables. The visit indicator was never included in the conditional specifications to avoid any convergence issue with the MICE algorithm, noting that missing data are mostly (or always) missing at times when $dN_i(t)=0,$ and vice-versa.

On each imputed dataset, monthly overlap weights were estimated by fitting separate monthly multinomial models. In the first imputed dataset, for month 11, we assessed their performance in terms of recovered balance in patient characteristics across the 3 treatment strategies by comparing mean (standard deviation, SD) for continuous, or percentages for categorical confounders, before and after weighting with overlap weights. 

Then, on each of the 25 datasets, we compared the WOMA and QLOMA by estimating with each estimator 10 optimal rules, one for each month from 2 to 11. We did not optimize decision at times 0 (baseline), 1, or 12, because no treatment add-on or switch was possible at time 0, no patient in the SPRINT data received an add-on at time 1, and that we assumed the outcome at month 12 would only be affected by any treatment given by time 11 or before. 

For each approach and for each month, results were pooled by taking the mean blip parameter estimate across the 25 estimates (from 25 imputed datasets), for each parameter. Finally, results from the two approaches were compared in terms of value function under the treatment actually received and the optimal DMAR, in terms of blip parameter estimates, and, only for the WOMA, in terms of the frequency of statistically significant effect modifiers across 25 imputed datasets, for each effect modifier separately.

\textcolor{black}{This study protocol has been approved by the Ethics committee for research in science and in health of Université de Montréal (project no. 2024-5845-1).}


\subsection{Results}

The cohort included 4683 patients in the standard BP management group and 4678 patients in the intensive group. At baseline, the average systolic BP was 139.7 mmHg (SD 15.39) and 139.7 mmHg (SD 15.77) in the standard and intensive groups, respectively. Each month, we found a few observations ($2\%$ average across all months) for which there was an antihypertensive add-on ($A_i(t)=1)$ even if there was no visit ($dN_i(t)=0$). This is due to our definition of visit, where a visit means that systolic BP was measured. Indeed, there could be visits at which systolic BP was not measured. On those few visits, if there was an add-on, we forced $dN_i(t)=1$ to consider properly the causal effect of the add-on on the final outcome. 

Web Table 4 shows in Web Appendix J the balance in patient characteristics at month 11, across those who did not visit, those who visited and received no add-on treatment, and those who visited with an add-on treatment, before and after using the overlap weights. This is after imputation and using the first imputed dataset. Before weighting, we find the most important differences across the three groups in the baseline systolic BP, the randomized treatment at baseline (intensive/standard), age, race, systolic BP at month 10, having received an add-on at month 10, and having visited or not at month 10 (the variables for which the standardized mean differences, SMD, not shown, were greater than 0.1). The overlap weights led to good balance in characteristics across the three groups (all SMD $<0.02$ after weighting).

The rate of observation on each month 1, 2, 3, 6, 9, and 12 were respectively 96, 94, 95, 93, 91, and 51\%. In between the planned monthly visits, there were also 13,756 unplanned visits in the first year of follow-up, with the average visit occurring at 185.5 days into follow-up (SD 92.77). These unplanned visits could provide additional information on BP measurements. Data were coarsened at the monthly level, taking the average of any systolic BP measurements during that month. We assumed no visit if systolic BP was not measured on a month. We did similarly for the add-on treatment and coarsened any monthly information on changes of drug classes. After coarsening, the months with no planned visit in the physician guidelines (i.e., months 4, 5, 7, 8, 10, and 11), had BP observation rates of 14, 23, 11, 17, 8, and 14\%, respectively. 

Before imputation, the average BP change from baseline to time $\tau=12$ was -10.9 mmHg. After imputation, in the first imputed dataset, it was equal to -11.1 mmHg. Over the 12 months of follow-up, 6\% of the person-time (person-month) was categorized as receiving an add-on or having a treatment switch.
 
 The value function shows an additional reduction of systolic BP between baseline and month 12 of 16.8 mmHg with the WOMA decisions and 15 mmHg with the QLOMA decisions, when compared with the treatment actually received (Table \ref{tabvalue}); these two values are the average benefit across all 25 imputed datasets. Several effect modifiers were found to be statistically significant in the monthly decision rules, for both the WOMA and QLOMA (Web Appendix J, Web Table 5 to 14). The effect modifiers that were the most often statistically significant (i.e., more than 12 times out of 25 imputed datasets) in the visit blip were the control group (intensive/standard), body mass index, high-density lipoprotein, systolic BP at baseline, and current systolic BP. In particular, the control group (intensive/standard) was a significant effect modifier more than 12 times out of 25 at months 2, 5, 8, 10, and 11, in the visit blip; the current systolic BP was also significant more than 12 times out of 25 at months 2, 5, 7, 8, 10, and 11. Those that were the most often statistically significant in the add-on blip were body mass index, current systolic BP, and race. This is not surprising, as body mass index and BP are key factors in the clinical decision making process for patients with hypertension. The effect modifier that was important in the add-on blip for the most months (i.e., months 2, 9, and 10) was race. 
 We also find in Web Table 15 the monthly contingency tables of the received vs the optimal strategy in the first imputed dataset, which show that every month, several patients did not receive the optimal treatment and visit strategy according to our rules.

\begin{table}
  \begin{center}
    \caption{Value function under the treatment received and the optimal treatment decisions, with optimal DMAR estimated using the QLMOA or the WOMA estimators, SPRINT trial, years 2010-2013. \vspace{0.01cm}\\ }  \label{tabvalue}
    \begin{tabular}{|l|c|c| c|}
    \hline 
               Approach & Treatment received, & Optimal treatment, & Difference in   \\ 
               & reduction since baseline & reduction since baseline & reduction, mmHg \\ \hline  
     QLOMA, mean (SD)  & 11.1 ($<$0.1) & 26.2 (0.9) & 15.0 (0.9) \\   \hline
        WOMA, mean (SD)  & 11.1 ($<$0.1) & 27.9 (1.1) & 16.8 (1.1) \\   \hline
         \end{tabular}
  \end{center}
\end{table}

\section{Conclusion}

It can be challenging to draw causal inferences in longitudinal data with missing visits. However, longitudinal data with irregular observation times provide a unique opportunity to study the effect of visits on clinical outcomes of interest. Visits can not only allow treatment add-on if needed, but they can also lead to health benefits or pitfalls by changing patient habits and other patient exposures or by reducing stress. 

In this work, we proposed the first doubly-robust approach, the WOMA estimator, to derive optimal DMAR. The method requires only one of the overlap weights or the treatment-free model to be correctly specified for consistency, along with any model for the missing effect modifiers and censoring if there are missing data or informative censoring. The doubly-robust approach was also compared to a novel singly-robust approach, QLOMA, in simulation studies. In addition to not relying on the NDE assumption, the WOMA relies on weighted least squares (WLS) theory, which is understood by most analysts and easy to implement. Theoretical properties of WLS-type estimators can be derived using standard WLS asymptotic theory and variance be derived using e.g., sandwich or two-step estimators \citep{huber1967behavior,newey1994series}. In addition, we have found that overlap weights perform better to develop optimal DMAR than inverse probability weights, and that MICE is a convenient approach to derive optimal DMAR when the effect modifiers are missing at some points in time, under a MAR assumption. LOCF has shown to perform poorly, especially when using the QLOMA, while the WOMA seems to confer some protection against the violation of LOCF assumptions. 

Most interestingly, in an application to the SPRINT trial, several effect modifiers were found to be statistically significant in the monthly outcome models, and the proposed approach resulted in an estimated additional 16.8 mmHg reduction in systolic BP compared with the treatment and visit schedule received, on average.

Our approach relies on strong, but standard causal assumptions, such as consistency, conditional exchangeability, and positivity. The positivity assumption is probably the one most at risk of being violated in practice, especially when the number of time points over which to derive optimal DMAR. Yet, we have obtained consistent estimates of both blip functions' parameters for the last stage decision in our simulation studies. Moreover, we have obtained higher value functions under the derived optimal DMAR in simulations and application when compared with the treatment strategies that patients had received. In our application to SPRINT, a limitation of our analysis was the potential for unmeasured confounding. Most patient characteristics were not updated frequently during follow-up, except for the visit indicator, add-on indicator, and systolic BP measurements. We also observed a rapid decline of observation rate between months 1 and 12 from 91\% to 51\% such that our approach relies heavily on the imputation models. 

In practice, researchers can choose to coarsen data more (e.g., to estimate a DMAR with decision rules every 3 months) such that positivity is less likely to be an issue for inference. \cite{ferreira2020impact} have discussed data coarsening in time, in the context of estimating longitudinal, marginal structural models parameters with targeted maximum likelihood estimators \citep{van2012targeted}. They found that discretization can change the target parameter and can induce bias in the estimation due to an inappropriate adjustment for time-dependent confounders. It will be interesting to further study the impact of positivity violation and time discretization on our approach in future work. 

Our proposed approach is particularly interesting in settings, such as the one in our application, in which patients have all the information needed to apply the rules. Patients often have information on their previous diagnoses and comorbidities, as well as medication usage. 
Given that BMI, race, target systolic BP and systolic BP measured at the current visit are key confounders for both visit and add-on models, it is possible to provide personalized scheduling when physicians see patients, based on the optimal DMAR. Patients could update these values with their physicians, for example one week before the scheduled visit, to determine whether the visit is necessary. Our approach could also be extended along the lines of Partially Adaptive Treatment Strategies \citep{talbot2014partially} if patients only have information about a subset of potential effect modifiers at home. 
Importantly, the proposed approach can participate in encouraging patients to be actively involved in their health management, which has multiple benefits and could help improving treatment adherence \citep{nunes2009medicines}.






\section{Acknowledgements}

This work was supported by an award from The Banting Discovery Foundation and the 
Canadian Statistical Sciences Institute (CANSSI) to author JC. 

 \section*{Supplementary Material}

The supplementary material at the end of this document  includes Web Appendices A to J which respectively contain the causal diagram assumed in this work; a demonstration of how informative censoring was addressed;  a demonstration of consistency of the blip estimators; the asymptotic variance estimator for a one-stage DMAR;  more details about the simulation study and the corresponding causal diagram; the simulation results when using LOCF for the missing data; the results when using inverse probability of treatment weights instead of overlap weights; the simulation results under informative censoring; tables of empirical bias and standard deviation of the blip parameter estimates for all simulations; a table of characteristics before and after weighting, the blip parameter estimates in the application to the SPRINT, and contingency tables for the optimal vs the observed treatment strategies each month.
 

 \bibliographystyle{chicago}
\bibliography{sample.bib}

\newpage
\noindent\textbf{ Supplementary Material for ``Constructing optimal dynamic monitoring and treatment regimes: An application to hypertension care''}

 \textbf{Web Appendix A: The causal diagram assumed in this work}
 
  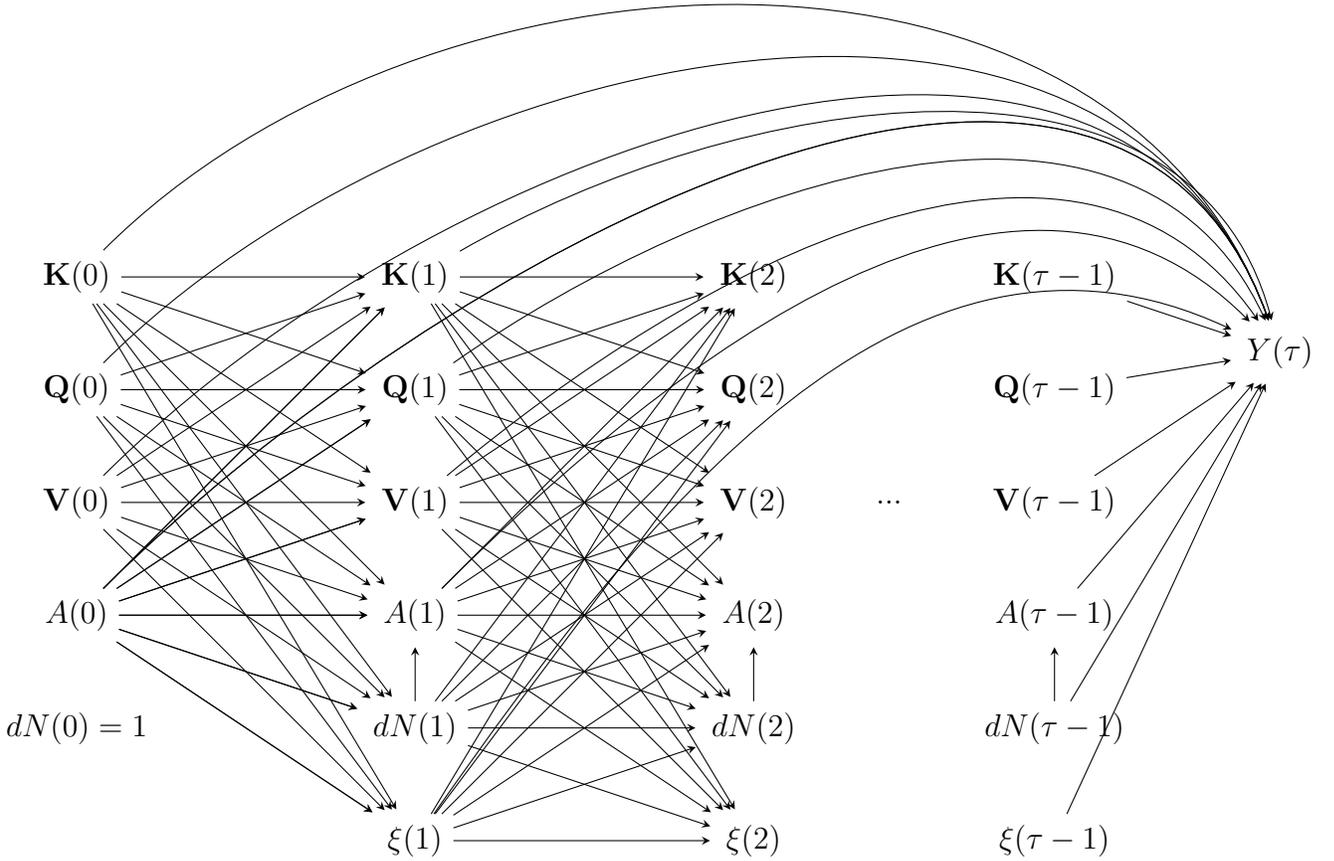
\begin{figure}[hbt!]
 
\begin{tikzpicture}[%
->,
shorten >=2pt,
>=stealth,
node distance=1cm,
pil/.style={
->,
thick,
shorten =2pt,}
]


\node(K0) at (0,0){$\mathbf{K}(0)$};
\node(Q0) at (0,-1.5){$\mathbf{Q}(0)$};
\node(V0) at (0,-3){$\mathbf{V}(0)$};
\node(A0) at (0,-4.5){$A(0)$};
\node(dN0) at (0,-6){$dN(0)=1$};

\node(K1) at (4.5,0){$\mathbf{K}(1)$};
\node(Q1) at (4.5,-1.5){$\mathbf{Q}(1)$};
\node(V1) at (4.5,-3){$\mathbf{V}(1)$};
\node(A1) at (4.5,-4.5){$A(1)$};
\node(dN1) at (4.5,-6){$dN(1)$};
\node(C1) at (4.5,-7.5){$\xi(1)$};

\node(K2) at (9,0){$\mathbf{K}(2)$};
\node(Q2) at (9,-1.5){$\mathbf{Q}(2)$};
\node(V2) at (9,-3){$\mathbf{V}(2)$};
\node(A2) at (9,-4.5){$A(2)$};
\node(dN2) at (9,-6){$dN(2)$};
\node(C2) at (9,-7.5){$\xi(2)$};

\node(dot) at (10.8, -3){...};

\node(Kt1) at (13,0){$\mathbf{K}(\tau-1)$};
\node(Qt1) at (13,-1.5){$\mathbf{Q}(\tau-1)$};
\node(Vt1) at (13,-3){$\mathbf{V}(\tau-1)$};
\node(At1) at (13,-4.5){$A(\tau-1)$};
\node(dNt1) at (13,-6){$dN(\tau-1)$};
\node(Ct1) at (13,-7.5){$\xi(\tau-1)$};


 \node(yt) at (16,-1){$Y(\tau)$};

 \draw[->] (K0) to (K1);
 \draw[->] (K0) to [out=45, in=110] (yt);
 \draw[->] (K0) to (Q1);
 \draw[->] (K0) to (V1);
 \draw[->] (K0) to (A1);
 \draw[->] (K0) to (dN1);
 \draw[->] (K0) to (C1); 
 \draw[->] (Q0) to (K1);
 \draw[->] (Q0) to [out=45, in=110] (yt);
 \draw[->] (Q0) to (Q1);
 \draw[->] (Q0) to (V1);
 \draw[->] (Q0) to (A1);
 \draw[->] (Q0) to (dN1);
 \draw[->] (Q0) to (C1);

  \draw[->] (V0) to (K1);
 \draw[->] (V0) to [out=45, in=110] (yt);
 \draw[->] (V0) to (Q1);
 \draw[->] (V0) to (V1);
 \draw[->] (V0) to (A1);
 \draw[->] (V0) to (dN1);
 \draw[->] (V0) to (C1);

  \draw[->] (A0) to (K1);
 \draw[->] (A0) to [out=45, in=110] (yt);
 \draw[->] (A0) to (Q1);
 \draw[->] (A0) to (V1);
 \draw[->] (A0) to (A1);
 \draw[->] (A0) to (dN1);
 \draw[->] (A0) to (C1);

  \draw[->] (A1) to [out=45, in=135] (yt);
 \draw[->] (K1) to [out=30, in=105] (yt);
 \draw[->] (Q1) to [out=35, in=115](yt);
 \draw[->] (V1) to [out=40, in=125](yt);
 \draw[->] (C1) to [out=55, in=155] (yt);
 
  \draw[->] (At1) to (yt);
 \draw[->] (Kt1) to  (yt);
 \draw[->] (Qt1) to (yt);
 \draw[->] (Vt1) to (yt);
\draw[->] (dNt1) to (yt);
 \draw[->] (Ct1) to (yt);

 \draw[->] (K1) to (K2);
  \draw[->] (K1) to (Q2);
 \draw[->] (K1) to (V2);
 \draw[->] (K1) to (A2);
 \draw[->] (K1) to (dN2);
 \draw[->] (K1) to (C2);

  \draw[->] (Q1) to (K2);
  \draw[->] (Q1) to (Q2);
 \draw[->] (Q1) to (V2);
 \draw[->] (Q1) to (A2);
 \draw[->] (Q1) to (dN2);
 \draw[->] (Q1) to (C2);

\draw[->] (dN2) to (A2);
\draw[->] (dNt1) to (At1);
\draw[->] (dN1) to (A1);

  \draw[->] (V1) to (K2);
  \draw[->] (V1) to (Q2);
 \draw[->] (V1) to (V2);
 \draw[->] (V1) to (A2);
 \draw[->] (V1) to (dN2);
 \draw[->] (V1) to (C2);

  \draw[->] (A1) to (K2);
  \draw[->] (A1) to (Q2);
 \draw[->] (A1) to (V2);
 \draw[->] (A1) to (A2);
 \draw[->] (A1) to (dN2);
 \draw[->] (A1) to (C2);

  \draw[->] (dN1) to (K2);
  \draw[->] (dN1) to (Q2);
 \draw[->] (dN1) to (V2);
 \draw[->] (dN1) to (A2);
 \draw[->] (dN1) to (dN2);
 \draw[->] (dN1) to (C2);

  \draw[->] (C1) to (K2);
  \draw[->] (C1) to (Q2);
 \draw[->] (C1) to (V2);
 \draw[->] (C1) to (A2);
 \draw[->] (C1) to (dN2);
 \draw[->] (C1) to (C2);

  \draw[->] (A0) to (K1);
 \draw[->] (A0) to [out=45, in=110] (yt);
 \draw[->] (A0) to (Q1);
 \draw[->] (A0) to (V1);
 \draw[->] (A0) to (A1);
 \draw[->] (A0) to (dN1);
 \draw[->] (A0) to (C1);

 
\end{tikzpicture}

\caption{Causal diagram representing the data generating mechanism we assume for the observations used in developing the optimal DMAR. We do not represent any causal effect occurring after time 2 and before time $\tau-1$ in this diagram. There is no assumed direct effect from each visit indicator to the final outcome (except for the last month, $\tau-1$ after which there is no covariate updated before the outcome is measured) but there is indirect effects through covariates and treatments. In this diagram, covariates $\mathbf{V}(t)$ represent any variables, other than confounders, previous add-on, etc., that simultaneously affect the future visit indicator and the final outcome. Using the manuscript notation, therefore, the set $\mathbf{V}(t)$ would include not only the sets $\mathbf{V}(t)$ from the diagram but also all other variables that simultaneously cause the visit indicator and final outcome.}\label{DAGfig}
\end{figure}

\newpage
\textbf{Web Appendix B: Method used to address informative censoring  }

To address censoring (Results presented in Web Appendix H), we used inverse probability of censoring weights proposed originally by Robins (1993); see also Robins and Finkelstein (2000).

Time is discrete in our simulation studies, so we fitted a pooled logistic regression model that combined all time points. In our simulations, the outcome is measured at ``month 3" and patients could have their follow-up censored at ``month 2" or ``month 3" (further referred to as $m2$ and $m3$, respectively).
In the case of time-fixed predictors of censoring (measured at baseline, i.e., time 0), we fitted the following censoring model:

$$P(C_i>m3) = \{1-P(C_i=m2)\}\times\{1-P(C_i=m3)\},$$

with $C_i$ the censoring time of individual $i$. In simulations, time-fixed censoring depended on covariates $A(0), Y(0), X(0)$ which respectively corresponded to the add-on treatment, effect modifier/BP, and confounder at baseline. These covariates affected censoring both at times $C_2$ and $C_3$ (under two different mechanisms) but one logistic regression model could be used to estimate the following probability:
$$P(C_i>m3 \mid A(0), Y(0), X(0)).$$
   In that model, the outcome was the indicator of being censored by month 3 or at month 3 (such that the final outcome at month 3 would not be observed).

 In the case of time-dependent predictors of censoring, we generated informative censoring using the same mechanism at times $m2$ and $m3$ (i.e., same coefficients in the censoring generating mechanism for each covariate, at both times) but the censoring at both times depended on evolving characteristics (i.e., it depended on $A(t)$, $Y(t)$ and $X(t)$ for month time $t$). In that case, we needed cumulated inverse probability of censoring weights and to transform the dataset into long format to fit a pooled logistic regression and obtain an estimator for
 
\begin{align}
   & P(C_i>m3 \mid A(2), Y(2), X(2), A(3), Y(3), X(3)) = \nonumber \\
   & \{1-P(C_i=m2 \mid  A(2), Y(2), X(2))\}\times\{1-P(C_i=m3\mid A(3), Y(3), X(3))\}. \nonumber
\end{align}   

\noindent \textbf{References}\vspace{0.05cm}\\

\noindent Robins, J. M. (1993). ``Information recovery and bias adjustment in proportional hazards regression analysis of randomized trials using surrogate markers." In \textit{Proceedings of the Biopharmaceutical Section, American Statistical Association}, 24(3), pp. 24-33.\vspace{0.05cm}\\

\noindent Robins, J. M., and Finkelstein, D. M. (2000). ``Correcting for noncompliance and dependent censoring in an AIDS clinical trial with inverse probability of censoring weighted (IPCW) log‐rank tests." \textit{Biometrics}, 56(3), pp. 779-788.
\newpage

\textbf{Web Appendix C: Identifiability and proof of consistency of the proposed estimating equations } \vspace{0.1cm}\\
We demonstrate the consistency of the estimator for a one-stage decision rule; we assume in these derivations that there is no missing data but otherwise, missing data can be replaced with multiple imputations by chained equations or LOCF depending on our assumptions about the missingness mechanism. Similar arguments can be used for multiple-stage rules. We are interested in the estimand  
\begin{align*}
&E[\widetilde{Y}^{r_{1}}(\tau) \mid \mathbf{Q}(1)=\mathbf{q}(1)]- E[\widetilde{Y}^{ (0, 0) }(\tau)\mid \mathbf{Q}(1)=\mathbf{q}(1)]  
\end{align*}
which represents the ``treatment effect" for strategy $r_1$ compared to $(0,0)$ (no visit, no add-on) in the stratum of characteristics $\mathbf{Q}(1)=\mathbf{q}(1)$. As in the manuscript, we assume the following structural nested mean model (updated for a one-stage rule):
 \begin{align}
&E[\widetilde{Y}^{r_1}(\tau) \mid  \mathcal{H}(1)=\mathbf{h}(1); \bm{\beta}_1, \bm{\gamma}_1 , \bm{\gamma}_1^*] =   f_1(\mathbf{h}(1); \bm{\beta}_1) + r_1[1] \{ \gamma_{0,1} + \bm{\gamma_{1,1}}^T\mathbf{q}_V(1)\}   \nonumber \\  
&\hspace{8cm} +r_1[1] r_1[2] \{ \gamma^*_{0,1} + \bm{\gamma^{*T}_{1,1}}\mathbf{q}_{V,A}(1) \}.   \label{eq0}
 \end{align}

\noindent \textit{1. Identifiability} \vspace{0.2cm}\\
Under the causal assumptions mentioned in the main manuscript and the modeling assumptions above, we have
\begin{align}
&E[\widetilde{Y}^{r_{1}}(\tau) \mid \mathbf{Q}(1)=\mathbf{q}(1)]- E[\widetilde{Y}^{ (0, 0) }(\tau)\mid \mathbf{Q}(1)=\mathbf{q}(1)]  \nonumber \\
& \hspace{1cm} =E_{\mathbf{X}(1)\setminus \mathbf{Q}(1)}\left\{ E[\widetilde{Y}^{r_{1}}(\tau) \mid \mathbf{Q}(1)=\mathbf{q}(1), \mathbf{X}(1)]- E[\widetilde{Y}^{ (0, 0) }(\tau)\mid \mathbf{Q}(1)=\mathbf{q}(1), \mathbf{X}(1)] \right\} \label{eq1} \\
& \hspace{1cm}  =E_{\mathbf{X}(1)\setminus \mathbf{Q}(1)}\left\{ E[\widetilde{Y}^{r_{1}}(\tau) \mid  dN_i(1)=r_1[1], A_i(1)=r_1[2],   \mathbf{q}(1), \mathbf{X}(1)]\right\}- \nonumber \\
& \hspace{3cm}E_{\mathbf{X}(1)\setminus \mathbf{Q}(1)}\left\{ E[\widetilde{Y}^{ (0, 0) }(\tau)\mid  dN_i(1)=0,  A_i(1)=0,      \mathbf{q}(1), \mathbf{X}(1)]  \right\}  \label{eq2} \\
& \hspace{1cm} =    E_{\mathbf{X}(1)\setminus \mathbf{Q}(1)  }\left\{  \psi_{V, 1}( r_{ 1},\mathbf{q}( 1); \bm{\gamma}_{1}) +\psi_{VA,  1}( r_{ 1},\mathbf{q}( 1); \bm{\gamma_{ 1}^*})  \right\}  \label{eq3}\\
&\hspace{1cm} =      \psi_{V, 1}( r_{ 1},\mathbf{q}( 1); \bm{\gamma}_{1}) +\psi_{VA,  1}( r_{ 1},\mathbf{q}( 1); \bm{\gamma_{ 1}^*}).   \label{eq4}  
 \end{align}
where (\ref{eq1}) is due to taking the iterated expectation with respect to covariates that are confounders but not effect modifiers; (\ref{eq2}) is due to conditional exchangeability; and (\ref{eq4}) is due to the fact that the blip functions only depend on effect modifiers so taking the expectation over other confounders (excluding these effect modifiers) does not affect them. Note that, by the consistency assumption, the potential outcomes in e.g., equation (\ref{eq0}) above equal the observed outcomes so these blips can be estimated with the observed outcomes. Under these assumptions, we have
\begin{align*}
&E[\widetilde{Y}^{r_{1}}(\tau) \mid \mathbf{Q}(1)=\mathbf{q}(1)]- E[\widetilde{Y}^{ (0, 0) }(\tau)\mid \mathbf{Q}(1)=\mathbf{q}(1)] =  
 \psi_{V, 1}( r_{ 1},\mathbf{q}( 1); \bm{\gamma}_{1}) +\psi_{VA,  1}( r_{ 1},\mathbf{q}( 1); \bm{\gamma_{ 1}^*}) 
\end{align*}
which can be estimated to obtain effect modification estimates.\vspace{0.1cm}\\

\noindent \textit{2. Estimation} \vspace{0.2cm}\\
We assume parametric linear models for the blip functions, as follows
$$\psi_{V,1}( r_1,\mathbf{q}_{V}(1); \bm{\gamma}_1)= r_1[1] \{ \gamma_{0,1} + \bm{\gamma_{1,1}}^T\mathbf{q}_V(1)\}$$
$$\psi_{VA,1}( r_1,\mathbf{q}_{V,A}(1); \bm{\gamma_1^*})= r_1[1] r_1[2] 
\{ \gamma^*_{0,1} + \bm{\gamma^{*T}_{1,1}}\mathbf{q}_{V,A}(1) \}.$$ 

\noindent Using an approach analogous with Q-learning and assuming  a linear function for $f(\cdot)$ such that the partial derivative of $f_1(\mathbf{h}_i(1); \bm{\beta}_1)$ is given by $\mathbf{h}_i(1),$ the parameters in the blip functions can be estimated using the set of estimating equations
\begin{align*}
& \sum_{i=1}^n \left( \begin{matrix} \mathbf{h}_i(1) \\
r_{i1}[1] \mathbf{q}_{iV}(1) 
\\
r_{i1}[1] r_{i1}[2] \mathbf{q}_{iV,A}(1)
\end{matrix} \right) \left[  Y_i^{r_1}(\tau) - f_1(\mathbf{h}_i(1); \bm{\beta}_1) - r_{i1}[1] \{ \gamma_{0,1} + \bm{\gamma_{1,1}}^T\mathbf{q}_{iV}(1)\}\right] +\\
&\sum_{i=1}^n \left( \begin{matrix} \mathbf{h}_i(1) \\
r_{i1}[1] \mathbf{q}_{iV}(1) 
\\
r_{i1}[1] r_{i1}[2] \mathbf{q}_{iV,A}(1)
\end{matrix} \right) \left[ -r_{i1}[1] r_{i1}[2] 
\{ \gamma^*_{0,1} + \bm{\gamma^{*T}_{1,1}}\mathbf{q}_{iV,A}(1) \} \right] =\bm{0}
\end{align*}
which is equivalent to fitting a least squares regression with the outcome $Y_i^{r_1}(\tau)$ and the set of independent variables that includes an intercept, the confounders, effect modifiers and their interaction with the treatment strategies. \vspace{0.1cm}\\

If there is censoring and that the outcome is only observed in patients who are followed until (at least) time $\tau$, the equations can be modified by adding the non-censoring indicator, as follows:

\begin{align*}
& \sum_{i=1}^n \xi_i(\tau) \left( \begin{matrix} \mathbf{h}_i(1) \\
r_{i1}[1] \mathbf{q}_{iV}(1) 
\\
r_{i1}[1] r_{i1}[2] \mathbf{q}_{iV,A}(1)
\end{matrix} \right) \left[  Y_i^{r_1}(\tau) - f_1(\mathbf{h}_i(1); \bm{\beta}_1) - r_{i1}[1] \{ \gamma_{0,1} + \bm{\gamma_{1,1}}^T\mathbf{q}_{iV}(1)\}  \right] + \\
& \sum_{i=1}^n \xi_i(\tau) \left( \begin{matrix} \mathbf{h}_i(1) \\
r_{i1}[1] \mathbf{q}_{iV}(1) 
\\
r_{i1}[1] r_{i1}[2] \mathbf{q}_{iV,A}(1)
\end{matrix} \right) \left[ -r_{i1}[1] r_{i1}[2] 
\{ \gamma^*_{0,1} + \bm{\gamma^{*T}_{1,1}}\mathbf{q}_{iV,A}(1) \}  \right] =\bm{0}.
\end{align*}
An approach to address informative censoring is discussed in Web Appendix B. The following derivations assume no censoring and that missing data have already been replaced.\vspace{0.2cm}\\

\noindent \textit{3. Double robustness}\vspace{0.2cm}\\
The estimating equations can be made doubly robust, as in Wallace and Moodie (2015) and Simoneau et al. (2020), using balancing weights. We propose the use of overlap weights as they have been shown, in the context of estimating causal effects, to be less variable than inverse probability of treatment weights for treatment effect  estimation (Li, Thomas, and Li, 2019). Using these weights denoted by $w_i(t)$ for patient $i$ at time $t$ leads to the updated equations for our one-stage rule:
\begin{align*}
& \sum_{i=1}^n w_i(1)\left( \begin{matrix} \mathbf{h}(1) \\
r_1[1] \mathbf{q}_V(1) 
\\
r_1[1] r_1[2] \mathbf{q}_{V,A}(1)
\end{matrix} \right) \left[  Y^{r_1}(\tau) - f_1(\mathbf{h}(1); \bm{\beta}_1) - r_1[1] \{ \gamma_{0,1} + \bm{\gamma_{1,1}}^T\mathbf{q}_V(1)\} \right]+ \\
&  \sum_{i=1}^n w_i(1)\left( \begin{matrix} \mathbf{h}(1) \\
r_1[1] \mathbf{q}_V(1) 
\\
r_1[1] r_1[2] \mathbf{q}_{V,A}(1)
\end{matrix} \right) \left[ -r_1[1] r_1[2] 
\{ \gamma^*_{0,1} + \bm{\gamma^{*T}_{1,1}}\mathbf{q}_{V,A}(1) \}  \right]=\bm{0}.  
\end{align*}
where 


 \begin{align*}
    w_i(t) =  \sum_{r \in \{(0,0), (1,0), (1,1) \}} \left\{ \frac{1}{\sum_{k \in \{(0,0), (1,0), (1,1) \}} \frac{1}{e_{i,k}(t)}} \frac{\mathbb{I}(R_{i}(t)=r)}{e_{i,r}(t)} \right\}
\end{align*}
and
$$e_{i,r}(t)=P(R_i(t)=r \mid \mathbf{h_i(t)};\bm{\delta}).$$
In simulation studies, we modeled $P(R_i(t)=r \mid \mathbf{h(t)})$ using a multinomial regression model to obtain generalized overlap weights (Li and Li, 2019). The equations are doubly robust in the sense that they are consistent for the blip parameters if either the overlap weights are correctly specified, or if the treatment-free model is correctly specified. Consistency of the proposed estimator always requires the blip functions to be correctly specified.

In the demonstration below, we denote the correct models using the symbol $\dagger$ onto the parameters and the functional format. For instance, the correct treatment-free model is given by $f^\dagger_1(\mathbf{h}(1); \bm{\beta}_1^\dagger)$ and the correct overlap weights (based on the correct multinomial model) are denoted by $ w_i^\dagger(1).$ Because the blip functions must always be correctly specified for the proposed estimator to be consistent, we add the same symbol over the parameters of the blip functions. The proof of consistency follows. \vspace{0.1cm}\\
 
\noindent \textit{a. Consistency when only the treatment-free model is correctly specified} \vspace{0.15cm}\\

\noindent We denote \begin{align*}
 \mathbf{q}(1)= \left[ \begin{matrix}  \mathbf{q}_{V}(1)\\
\mathbf{q}_{V,A}(1) 
\end{matrix} \right].
  \end{align*}
  Then,
 \begin{align*}
& E \left[  \sum_{i=1}^n  w_i(t) \left( \begin{matrix} \mathbf{h}(1) \\
r_1[1] \mathbf{q}_V(1) 
\\
r_1[1] r_1[2] \mathbf{q}_{V,A}(1)
\end{matrix} \right) \left[  Y^{r_1}(\tau) - f_1^\dagger(\mathbf{h}(1); \bm{\beta}_1^\dagger) - r_1[1] \{ \gamma_{0,1}^\dagger + \bm{\gamma_{1,1}}^{\dagger T}\mathbf{q}_V(1)\} \right] \right] \\
& \hspace{1.5cm}  - E \left[  \sum_{i=1}^n  w_i(t) \left( \begin{matrix} \mathbf{h}(1) \\
r_1[1] \mathbf{q}_V(1) 
\\
r_1[1] r_1[2] \mathbf{q}_{V,A}(1)
\end{matrix} \right)   r_1[1] r_1[2] 
\{ \gamma^{*,\dagger}_{0,1} + \bm{\gamma^{*,\dagger T}_{1,1}}\mathbf{q}_{V,A}(1) \}   \right] \\
&= E_{\mathbf{h}(1),\mathbf{q}(1)} E \left[  \sum_{i=1}^n  w_i(t) \left. \left( \begin{matrix} \mathbf{h}(1) \\
r_1[1] \mathbf{q}_V(1) 
\\
r_1[1] r_1[2] \mathbf{q}_{V,A}(1)
\end{matrix} \right) \left[  Y^{r_1}(\tau) - f_1^\dagger(\mathbf{h}(1); \bm{\beta}_1^\dagger) - r_1[1] \{ \gamma_{0,1}^\dagger + \bm{\gamma_{1,1}}^{\dagger T}\mathbf{q}_V(1)\} \right] \right\vert \mathbf{h}(1),\mathbf{q}(1)\right] \\ 
& \hspace{1.5cm}  - E_{\mathbf{h}(1), \mathbf{q}(1)} E \left[  \sum_{i=1}^n  w_i(t) \left. \left( \begin{matrix} \mathbf{h}(1) \\
r_1[1] \mathbf{q}_V(1) 
\\
r_1[1] r_1[2] \mathbf{q}_{V,A}(1)
\end{matrix} \right)   r_1[1] r_1[2] 
\{ \gamma^{*,\dagger}_{0,1} +  \bm{\gamma^{*,\dagger T}_{1,1}}\mathbf{q}_{V,A}(1) \}   \right\vert \mathbf{h}(1),\mathbf{q}(1)\right] \\
&= E_{\mathbf{h}(1),\mathbf{q}(1)} \left[ \sum_{i=1}^n  w_i(t)   \left( \begin{matrix} \mathbf{h}(1) \\
r_1[1] \mathbf{q}_V(1) 
\\
r_1[1] r_1[2] \mathbf{q}_{V,A}(1)
\end{matrix} \right) \right] \\
&  \times  E \left[   Y^{r_1}(\tau) - f_1^\dagger(\mathbf{h}(1); \bm{\beta}_1^\dagger) - r_1[1] \{ \gamma_{0,1}^\dagger + \bm{\gamma_{1,1}}^{\dagger T}\mathbf{q}_V(1)\} -r_1[1] r_1[2] 
\{ \gamma^{*,\dagger}_{0,1} +  \bm{\gamma^{*,\dagger T}_{1,1}}\mathbf{q}_{V,A}(1) \}  \vert \mathbf{h}(1),\mathbf{q}(1) \right]  \\ 
&=0,
\end{align*}
where the last equality comes from the fact that under a correctly specified outcome model, the second part of the equation (two rows above) cancels out:
\begin{align*}
&E \left. \left[   Y^{r_1}(\tau) - f_1^\dagger(\mathbf{h}(1); \bm{\beta}_1^\dagger) - r_1[1] \{ \gamma_{0,1}^\dagger + \bm{\gamma_{1,1}}^{\dagger T}\mathbf{q}_V(1)\} -r_1[1] r_1[2] 
\{ \gamma^{*,\dagger}_{0,1} +  \bm{\gamma^{*,\dagger T}_{1,1}}\mathbf{q}_{V,A}(1) \}  \right\vert \mathbf{h}(1),\mathbf{q}(1)\right]\\
&=E \left. \left[   Y^{r_1}(\tau) \right\vert \mathbf{h}(1),\mathbf{q}(1) \right] -  f_1^\dagger(\mathbf{h}(1); \bm{\beta}_1^\dagger) - r_1[1] \{ \gamma_{0,1}^\dagger + \bm{\gamma_{1,1}}^{\dagger T}\mathbf{q}_V(1)\} -r_1[1] r_1[2] 
\{ \gamma^{*,\dagger}_{0,1} +  \bm{\gamma^{*,\dagger T}_{1,1}}\mathbf{q}_{V,A}(1) \}  \\
&= 0. \vspace{0.1cm}\\
\end{align*}
  \textit{b. Consistency when only the overlap weights are correctly specified}\vspace{0.2cm}\\
Our demonstration is inspired by both Supplementary Materials from Wallace and Moodie (2015) and Simoneau et al. (2020). First, note that using weights that will make the treatment strategy (e.g., $R(1)$ in our one-stage rule) independent of the potential confounders in $\mathbf{h}(1)$ will lead to consistent estimators for the blip parameters if the blip functions are correctly specified. Thus, we are after weights $w_0\{\mathbf{h}(1)\}, w_1\{\mathbf{h}(1)\}$ and $w_2\{\mathbf{h}(1)\}$ with the indices 0, 1 and 2 respectively corresponding to strategies $(0,0), (1,0), (1,1)$ for $\{dN(t), A(t)\},$ that will make the exposure and confounders independent. Denote
\begin{align*}
&\mathbf{H}_W(1) = \mathbb{I}\{R(1)=(0,0)\} w_0\{\mathbf{h}(1)\} \mathbf{H}(1)+\mathbb{I}\{R(1)=(1,0)\} w_1\{\mathbf{h}(1)\}\mathbf{H}(1)+\mathbb{I}\{R(1)=(1,1)\} w_2\{\mathbf{h}(1)\} \mathbf{H}(1),\\
& R_{W}(1) = \mathbb{I}\{R(1)=(0,0)\} w_0\{\mathbf{h}(1)\} R(1)+\mathbb{I}\{R(1)=(1,0)\} w_1\{\mathbf{h}(1)\} R(1)+\mathbb{I}\{R(1)=(1,1)\} w_2\{\mathbf{h}(1)\} R(1).
\end{align*}

\noindent Further denote
\begin{align*} 
&\pi_1\{\mathbf{h}(1)\} = P\{R(1)=(1,0)\mid  \mathbf{H}(1)= \mathbf{h}(1)\}, \hspace{0.2cm}\pi_{W,1}\{\mathbf{h}_W(1)\} = P\{R_{W}(1)=(1,0)\mid  \mathbf{H}_W(1)=\mathbf{h}_W(1)\}\\
&\pi_2\{\mathbf{h}(1)\} = P\{R(1)=(1,1)\mid  \mathbf{H}(1)= \mathbf{h}(1)\}, \hspace{0.2cm}\pi_{W,2}\{\mathbf{h}_W(1)\} = P\{R_{W}(1)=(1,1)\mid  \mathbf{H}_W(1)=\mathbf{h}_W(1)\}\\
&\pi_0\{\mathbf{h}(1)\}=1-\pi_1\{\mathbf{h}(1)\}-\pi_2\{\mathbf{h}(1)\}, \hspace{0.2cm} \pi_{W,0}\{\mathbf{h}(1)\}=1-\pi_{W,1}\{\mathbf{h}(1) \}-\pi_{W,2}\{\mathbf{h}(1)\}.
\end{align*}

\noindent Independence of the exposure and confounders after weighting means that:
\begin{align*}
E[\mathbf{H}_W(1) \mid R_{W}(1)=(0,0)]=E[\mathbf{H}_W(1) \mid R_{W}(1)=(1,0)]=E[\mathbf{H}_W(1) \mid R_{W}(1)=(1,1)].
\end{align*} 
\noindent One way to obtain such an equality is by choosing $w_0\{\mathbf{h}(1)\}, w_1\{\mathbf{h}(1)\}$ and $w_2\{\mathbf{h}(1)\}$ above such that

\begin{align}
\frac{ P\{R_{W}(1)=(0,0) \mid  \mathbf{H}_W(1)=\mathbf{h}_W(1)\}}{P_{0}} =&
\frac{P\{R_{W}(1)=(1,0)\mid  \mathbf{H}_W(1)=\mathbf{h}_W(1)\}}{P_{1}}  \nonumber \\
&= \frac{P\{R_{W}(1)=(1,1)\mid  \mathbf{H}_W(1)=\mathbf{h}_W(1)\}}{P_2}, \label{ho}
\end{align}
\noindent where $P_0 =P
\{R_{W}(1)=(0,0)\} , P_1= P\{R_{W}(1)=(1,1)\}, P_2 = P\{R_{W}(1)=(1,1)\}$ are the marginal probabilities of the different treatment strategies at time 1. One solution to this system of equations is that in which all numerators are equal and all denominators are equal in (\ref{ho}).\vspace{0.2cm}\\

\textbf{All numerators being equal} means that in the weighted data, the probabilities of receiving all three treatment strategies given characteristics $\mathbf{h}_W(1)$ is the same (i.e., all 1/3). Returning to the unweighted data, denote
$k=  w_0\{\mathbf{h}(1)\} \pi_0\{\mathbf{h}(1)\}+  w_1\{\mathbf{h}(1)\}\pi_1\{\mathbf{h}(1)\} + w_2\{\mathbf{h}(1)\}\pi_2\{\mathbf{h}(1)\}.$  Each numerator in (\ref{ho}) can then be expressed as
\begin{align*}
& P\{R_{W}(1)=(0,0)\mid  \mathbf{H}_W(1)=\mathbf{h}_W(1)\}= 1/k \hspace{0.1cm} w_0\{\mathbf{h}(1)\} \pi_0\{\mathbf{h}(1)\},\\
&P\{R_{W}(1)=(1,0)\mid  \mathbf{H}_W(1)=\mathbf{h}_W(1)\}=1/k\hspace{0.1cm} w_1\{\mathbf{h}(1)\} \pi_1\{\mathbf{h}(1)\},\\
&P\{R_{W}(1)=(1,1)\mid  \mathbf{H}_W(1)=\mathbf{h}_W(1)\}= 1/k\hspace{0.1cm}  w_2\{\mathbf{h}(1)\} \pi_2\{\mathbf{h}(1)\}. 
\end{align*}
    
\noindent The left-hand side of each of the three equations above equals 1/3 and we can show that the right-hand side will also equal 1/3 for our three possible treatment strategies if we use balancing weights of the form
$w_0\{\mathbf{h}(1)\} \pi_0\{\mathbf{h}(1)\} =  w_1\{\mathbf{h}(1)\}\pi_1\{\mathbf{h}(1)\}= w_2\{\mathbf{h}(1)\}\pi_2\{\mathbf{h}(1)\}.$ Taking as an example the case $R(1)=(1,1),$ we have:
$$P\{R_{W}(1)=(1,1) \mid \mathbf{h}_W(1)\} = 1/k\hspace{0.1cm} P\{R(1)=(1,1)\mid \mathbf{h}(1)\} w_2\{\mathbf{h}(1)\} = 1/k \hspace{0.1cm}\pi_2\{\mathbf{h}(1)\}w_2\{\mathbf{h}(1)\}.$$
If the weights satisfy the balancing condition \textit{(generalized overlap weights are members of the group of balancing weights, as discussed in Li 2023)}, then
\begin{align*}
w_0\{\mathbf{h}(1)\}= \frac{w_2\{\mathbf{h}(1)\}\pi_2\{\mathbf{h}(1)\}}{\pi_0\{\mathbf{h}(1)\}} , \hspace{0.5cm} w_1\{\mathbf{h}(1)\}= \frac{w_2\{\mathbf{h}(1)\}\pi_2\{\mathbf{h}(1)\}}{\pi_1\{\mathbf{h}(1)\}}.
\end{align*}
such that $k=   w_0\{\mathbf{h}(1)\} \pi_0\{\mathbf{h}(1)\}+  w_1\{\mathbf{h}(1)\}\pi_1\{\mathbf{h}(1)\} + w_2\{\mathbf{h}(1)\}\pi_2\{\mathbf{h}(1)\} = 3 w_2\{\mathbf{h}(1)\}\pi_2\{\mathbf{h}(1)\}$
and
$ P\{R_{W}(1)=(1,1) \mid \mathbf{h}_W(1)\} =  1/k \hspace{0.1cm}\pi_2\{ \mathbf{h}(1)\} w_2\{\mathbf{h}(1)\}= 1/3 $ as expected.\vspace{0.2cm}\\

\textbf{ The other condition for consistency is for all denominators in (\ref{ho}) to be equal}. This is more straightforward to demonstrate under the balancing condition, by comparing the different probabilities. First, denote $l=\int f_X(x) k dx$. For instance, for the strategies (0,0) and (1,0), we have $P\{R_W(1)=(0,0)\} = \frac{1}{l}\int_{\mathbf{h}(1)} f_{\mathbf{h}(1)}(\mathbf{h}(1))P\{R(1) =(0,0) \mid \mathbf{h}(1)\} w_0\{\mathbf{h}(1)\}$, which is equal to $P\{R_W(1)=(1,0)\} =\frac{1}{l}\int_{\mathbf{h}(1)} f_{\mathbf{h}(1)}(\mathbf{h}(1))P\{R(1) =(1,0) \mid \mathbf{h}(1)\}w_1\{\mathbf{h}(1)\}= \frac{1}{l}\int_{\mathbf{h}(1)} f_{\mathbf{h}(1)}(\mathbf{h}(1))P\{R(1) =(0,0) \mid \mathbf{h}(1)\}w_0\{\mathbf{h}(1)\}$, and similarly for comparisons with the third treatment strategy.
 \vspace{0.2cm}\\

\noindent \textbf{References}\vspace{0.05cm}\\

\noindent Simoneau, Gabrielle, Erica EM Moodie, Jagtar S. Nijjar, Robert W. Platt, and Scottish Early Rheumatoid Arthritis Inception Cohort Investigators (2020). ``Estimating optimal dynamic treatment regimes with survival outcomes." \textit{Journal of the American Statistical Association}, 115(531), pp. 1531-1539.\vspace{0.02cm}\\

\noindent Wallace, Michael P. and Moodie, Erica EM. (2015). ``Doubly-robust dynamic treatment regimen estimation via weighted least squares." \textit{Biometrics}, 71(3), pp. 636-644.\vspace{0.02cm}\\

\noindent Li, F., Thomas, L. E., and Li, F. (2019). ``Addressing extreme propensity scores via the overlap weights." \textit{American Journal of Epidemiology}, 188(1), pp. 250-257.\vspace{0.02cm}\\

\noindent Li, F. (2023). ``Overlap Weighting." In Handbook of Matching and Weighting Adjustments for Causal Inference (pp. 263-282). Chapman and Hall/CRC.

\newpage

 \textbf{Web Appendix D: Variance estimator for a one-stage rule }

Each optimal decision rule is based off estimating equations, and asymptotic theory for each set of estimating equations can be derived using generalized estimating equations theory with the sandwich estimator (Huber, 1967; Liang and Zeger, 1986; Stenfaski and Boos, 2002) or a two-step estimator to account for the estimation of the nuisance models parameters (see e.g., Newey 1994, Chapter 36). The derivation of a variance estimator for an optimal (one-stage) DMAR is shown below. However, for a sequence of decision rules, i.e., a multiple-stage rule, it is less straightforward to compute the cumulative variance of the DMAR estimator. The pseudo-outcomes that are calculated at each stage are not regular and depend on all the previous decision rules, corresponding to all time points that come after the current stage. Thus, they depend on a series of indicator functions, such that, in earlier stages, i.e., all stages except the last decision stage, the finite sample cumulative variance of the estimator could be poorly estimated by the asymptotic variance formula (Chakraborty 2014). Simoneau et al. (2020) proposed the use of \textit{m-out-of-n} bootstrap (Bickel, 2012) as a variance estimator for dynamic weighted least squares-type of approach. This approach has not been tested in a setting with optimal dynamic monitoring strategies but could possibly be used to derive confidence intervals for the blip parameters or the blip functions. \vspace{0.2cm}\\
 
\noindent Under the causal and missingness assumptions in the main manuscript, the estimators for the blip parameters correspond to a two-step M-estimator (Stefanski and Boos, 2002).
Two-step estimators are defined by Newey and McFadden (1994) as estimators that are based on some preliminary, first-step estimator of a parameter vector. Often, a first-step estimator is used to estimate the nuisance parameters (e.g., parameters used in a weight, which will further be incorporated into the two-step estimator). M-estimators, on the other hand, are obtained by solving a sample average equation and often consist of the zero roots of an estimating equation. 

In the case of a one-stage DMAR, our estimator corresponds to a weighted least squares estimator in which estimates for the weight parameters can be plugged in. Let $\bm{\hat{\gamma}}$ be our two-step, doubly-robust estimator and let $\bm{\gamma_0}$ the vector of true parameters. In this derivation, we assume that parameters of the treatment-free model ($\beta$) are included in the vector $\bm{\gamma}$. Because the equations are sums of i.i.d. terms, we have $\sqrt{n}(\bm{\hat{\gamma}-\gamma_0)} \rightarrow N(\mathbf{0}, \mathbf{\Sigma})$ (Newey (1994))  with
\begin{align}
\mathbf{\Sigma}=\mathbf{G_{\gamma}^{-1}} \mathbb{E}\left[ \mathbf{ \left\{ g(o; \gamma_0, \phi_0) - G_{\phi} M^{-1} m(o; \phi_0)\right\}}^{\otimes 2} \right]\mathbf{ G_{\gamma}^{-1}}
\end{align}
where
$
\mathbf{G_{\gamma}}=\mathbf{\mathbb{E}(\triangledown_{\gamma} g(o; \gamma_0, \phi_0))},
\mathbf{G_{\phi}}= \mathbf{\mathbb{E}(\triangledown_{\phi} g(o; \gamma_0, \phi_0))},
\mathbf{M} =\mathbf{ \mathbb{E}(\triangledown_{\phi} m(o; \phi_0)) }
$
for $\mathbf{o}$ the data, and $\mathbf{m(o; \phi_0)}$ and $\mathbf{g(o; \gamma_0, \phi_0)}$ the estimating equations for the nuisance parameters $\bm{\phi}$ (overlap weights model) and the parameters of interest $\bm{\gamma}$, respectively. The symbol $\triangledown_{\delta}$ refers to the partial derivative w.r.t. the parameter(s) $\delta$.\vspace{0.2cm}\\
We can plug in the estimating equations $\mathbf{m(o; \phi_0)}$ and $\mathbf{g(o; \gamma_0, \phi_0)}$ the corresponding parameter estimates to obtain $\mathbf{m(o; \hat{\phi}_0)}$ and $\mathbf{g(o; \hat{\gamma}_0, \hat{\phi}_0)}$. These parameters can be estimated using e.g., the \texttt{glm} function in R or, for the multinomial model, the \texttt{nnet} package. Other functions used in the sandwich formula above can be derived first (some of these functions are derivatives of the estimating equations w.r.t. different sets of parameters), and then parameter estimates be plugged in to obtain variance estimates. \vspace{0.3cm}\\

\noindent \textbf{References}\vspace{0.05cm}\\

\noindent Newey, W. K., and McFadden, D. (1994). ``Large sample estimation and hypothesis testing." \textit{Handbook of Econometrics}, 4, pp. 2111-2245.\vspace{0.1cm}\\

 \noindent Stefanski, L. A., and Boos, D. D. (2002). ``The calculus of M-estimation." \textit{The American Statistician,} 56(1), pp. 29-38.
\newpage

 \textbf{Web Appendix E: More details on the simulation studies and causal diagram}

 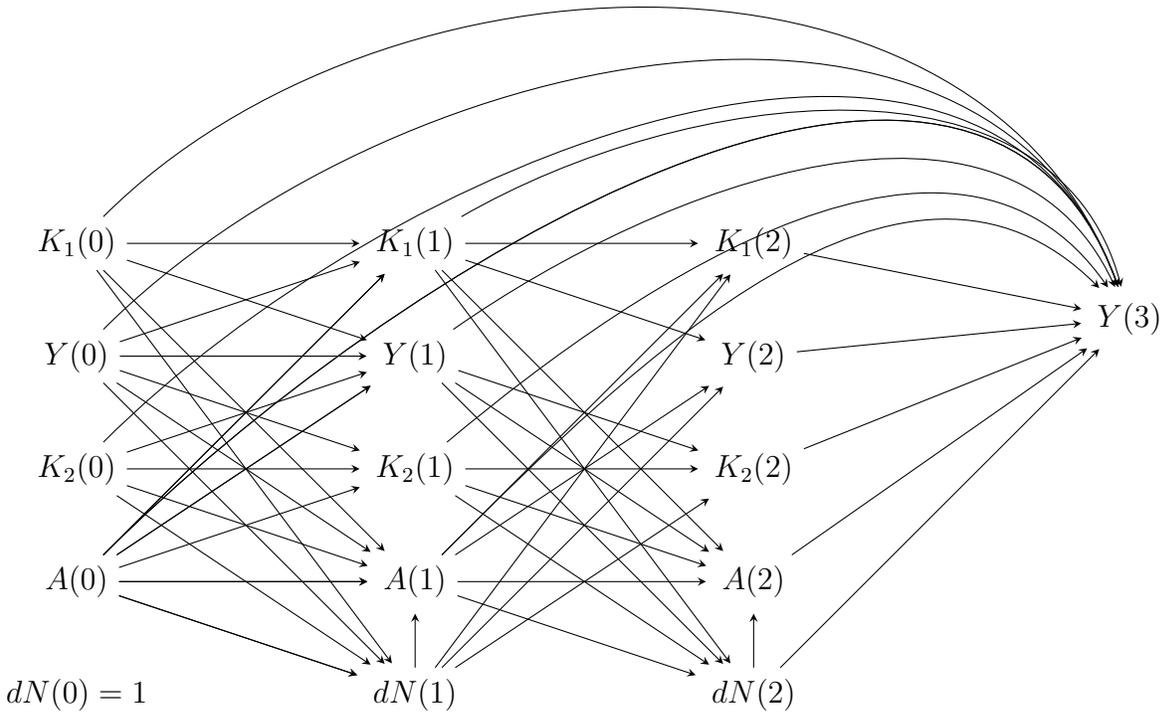
\begin{figure}[hbt!]
 
\begin{tikzpicture}[%
->,
shorten >=2pt,
>=stealth,
node distance=1cm,
pil/.style={
->,
thick,
shorten =2pt,}
]

\node(K0) at (0,0){$K_1(0)$};
\node(Q0) at (0,-1.5){$Y(0)$};
\node(V0) at (0,-3){$K_2(0)$};
\node(A0) at (0,-4.5){$A(0)$};
\node(dN0) at (0,-6){$dN(0)=1$};

\node(K1) at (4.5,0){$K_1(1)$};
\node(Q1) at (4.5,-1.5){$Y(1)$};
\node(V1) at (4.5,-3){$K_2(1)$};
\node(A1) at (4.5,-4.5){$A(1)$};
\node(dN1) at (4.5,-6){$dN(1)$};

\node(K2) at (9,0){$K_1(2)$};
\node(Q2) at (9,-1.5){$Y(2)$};
\node(V2) at (9,-3){$K_2(2)$};
\node(A2) at (9,-4.5){$A(2)$};
\node(dN2) at (9,-6){$dN(2)$};




 \node(yt) at (14,-1){$Y(3)$};
 
\draw[->](dN1) to (A1);
\draw[->](dN2) to (A2);

 \draw[->] (K0) to (K1);
 \draw[->] (K0) to [out=45, in=110] (yt);
\draw[->] (K0) to (Q1);
 \draw[->] (K0) to (A1);
 \draw[->] (K0) to (dN1);

 \draw[->] (Q0) to (K1);
 \draw[->] (Q0) to [out=45, in=110] (yt);
 \draw[->] (Q0) to (Q1);
 \draw[->] (Q0) to (V1);
 \draw[->] (Q0) to (A1);
 \draw[->] (Q0) to (dN1);

 \draw[->] (V0) to [out=45, in=110] (yt);
 \draw[->] (V0) to (Q1);
 \draw[->] (V0) to (A1);
 \draw[->] (V0) to (dN1);

  \draw[->] (A0) to (K1);
 \draw[->] (A0) to [out=45, in=110] (yt);
 \draw[->] (A0) to (Q1);
 \draw[->] (A0) to (A1);
 \draw[->] (A0) to (dN1);

 \draw[->] (V0) to (V1);

  \draw[->] (A1) to [out=45, in=135] (yt);
 \draw[->] (K1) to [out=30, in=105] (yt);
 \draw[->] (Q1) to [out=35, in=115](yt);
 \draw[->] (V1) to [out=40, in=125](yt);
 

\draw[->] (Q2) to (yt);

\draw[->] (V2) to (yt);

\draw[->] (A2) to (yt);
\draw[->] (dN2) to (yt);

 \draw[->] (K1) to (K2);
  \draw[->] (K1) to (Q2);
 \draw[->] (K1) to (A2);
 \draw[->] (K1) to (dN2);

 \draw[->] (Q1) to (V2);
 \draw[->] (Q1) to (A2);
 \draw[->] (Q1) to (dN2);

 \draw[->] (V1) to (V2);
 \draw[->] (V1) to (A2);
 \draw[->] (V1) to (dN2);

 \draw[->] (K2) to (yt);

  \draw[->] (A1) to (K2);
  \draw[->] (A1) to (Q2);
 \draw[->] (A1) to (A2);
 \draw[->] (A1) to (dN2);

  \draw[->] (dN1) to (K2);
  \draw[->] (dN1) to (Q2);
 \draw[->] (dN1) to (V2);


  \draw[->] (A0) to (K1);
 \draw[->] (A0) to [out=45, in=110] (yt);
 \draw[->] (A0) to (Q1);
 \draw[->] (A0) to (V1);
 \draw[->] (A0) to (A1);
 \draw[->] (A0) to (dN1);
\end{tikzpicture}

\caption{
The causal diagram depicting the data generating mechanism used in the main simulation study. In simulation studies, the final outcome to optimize is $Y(3)$ and the set of effect modifiers $\mathbf{Q}(t)$ includes the variables $K_1(t)$ and $Y(t)$.}\label{DAGfig}
\end{figure}
  
We now describe the DGM, removing the patient index $i$ to simplify notation. Note that the notation $\mathcal{N}(a,b)$ refers to the Normal distribution with mean $a$ and standard deviation $b$. First, we simulated two confounders at baseline as $$K_1(0) \sim \mathcal{N}(4, 3)$$ $$K_2(0) \sim \mathcal{N}(5, 1.4),$$ a blood pressure measurement $$Y(0) \sim \mathcal{N}(120, 13),$$ and a randomized drug treatment at baseline $$A(0)\sim \text{Bernoulli}(0.5).$$ 
\noindent Data at month 1 were simulated as 
$$K_1(1) \sim \mathcal{N}(\mu_{k11}, 3), \mu_{k11}= -23 + 0.8 \hspace{0.1cm} K_1(0) +0.2\hspace{0.1cm} Y(0)+0.1\hspace{0.1cm}A(0)$$
$$K_2(1) \sim \mathcal{N}(\mu_{k21}, 3), \mu_{k21}= -43 + 1 \hspace{0.1cm} K_2(0) +0.2\hspace{0.1cm} Y(0)-0.1\hspace{0.1cm}A(0)$$
$$ Y(1) \sim \mathcal{N}(\mu_{y1}, 3)$$
$$ \mu_{y1}= 118  +  0.05 \hspace{0.1cm} A(0)  -0.005 \hspace{0.1cm}Y(0) + 0.02\hspace{0.1cm} K_1(0) +   0.02 \hspace{0.1cm} A(0) \times K_1(0)   + 0.004  \hspace{0.1cm}A(0) \times Y(0)   +  0.02  \hspace{0.1cm} Y(0) \times K_1(0) $$
$$ dN(1) \sim \text{Bernoulli}(p_{dN1}),\text{logit}(p_{dN1})=-2 + 0.3 \hspace{0.1cm} K_1(0)  -0.8 \hspace{0.1cm} K_2(0) + 0.1 \hspace{0.1cm} A(0) + 0.02 \hspace{0.1cm} Y(0) $$
$$ A(1) \mid (dN(1)=1)  \sim \text{Bernoulli}(p_{A1}), \text{logit}(p_{A1})=0.4\hspace{0.1cm} K_1(0)-0.05 \hspace{0.2cm} K_2(0) +  0.2\hspace{0.1cm} A(0) -0.04 \hspace{0.1cm} Y(0), A(1) \mid (dN(1)=0)  = 0 $$
 
\noindent Data at month 2 were simulated as
$$K_1(2) \sim \mathcal{N}(\mu_{k12}, 3), \mu_{k12}= -26 + 0.8 \hspace{0.1cm} K_1(1) +0.2\hspace{0.1cm} Y(1)+0.1\hspace{0.1cm}A(1)+0.1\hspace{0.1cm}dN(1)$$
$$K_2(2) \sim \mathcal{N}(\mu_{k22}, 3), \mu_{k22}= -43 + 1 \hspace{0.1cm} K_2(1) +0.2\hspace{0.1cm} Y(1)+0.1\hspace{0.1cm}A(1) - 0.1\hspace{0.1cm}dN(1)$$
$$Y(2) \sim \mathcal{N}(\mu_{y2}, 3)$$
$$\mu_{y2}= 122  +  0.05 \hspace{0.1cm} A(0)  -0.005 Y(0) + 0.02\hspace{0.1cm} K_1(0) +   0.02 \hspace{0.1cm} A(0) \times K_1(0)   + 0.004  \hspace{0.1cm}A(0) \times Y(0)   +  0.02  \hspace{0.1cm} Y(0) \times K_1(0) +  0.02 \hspace{.1cm} K_1(1)$$
$$- 1.4 \hspace{.1cm} A(1)\times dN(1) -0.0005 +1\hspace{0.1cm} K_1(1)  \times dN(1) +1 \hspace{0.1cm} Y(1) +  0.002 \hspace{0.1cm} dN(1)\times Y(1) + 0.1\hspace{0.1cm} A(1)\times dN(1)\times K_1(1)$$
$$+ 0.04\hspace{0.1cm} A(1)\times dN(1)\times Y(1)$$
$$dN(2) \sim \text{Bernoulli}(p_{dN2}), \text{logit}(p_{dN2})=-18 + 0.3 \hspace{0.1cm} K_1(1)-0.8 \hspace{0.1cm} K_2(1) + 0.1 \hspace{0.1cm} A(1) + 0.02 \hspace{0.1cm} Y(1)  $$
$$A(2) \mid (dN(2)=1)  \sim \text{Bernoulli}(p_{A2}),\text{logit}(p_{A2})=0.4\hspace{0.1cm} K_1(1)-0.05\hspace{0.1cm} K_2(1) +  0.2\hspace{0.1cm} A(1) -0.04 \hspace{0.1cm} Y(1), A(2) \mid (dN(2)=0)  = 0.$$

At month 3, only the simulation of the outcome mattered. It was simulated as
$$Y(3) \sim \mathcal{N}(\mu_{y3}, 3)$$
$$\mu_{y3}=  134  +  0.05 \hspace{0.1cm} A(0)  -0.005 Y(0) + 0.02\hspace{0.1cm} K_1(0) -0.6 \hspace{0.1cm} K_2(0) +   0.02 \hspace{0.1cm} A(0) \times K_1(0)   + 0.004  \hspace{0.1cm}A(0) \times Y(0)   +  0.02  \hspace{0.1cm} Y(0) \times K_1(0)   $$
$$+  0.02 \hspace{.1cm} K_1(1) -1.5 \hspace{0.1cm} K_2(1)  - 1.4 \hspace{.1cm} A(1)\times dN(1) -0.005 \hspace{0.1cm} Y(1)  + 0.18 \hspace{0.1cm} K_1(1)  \times dN(1)  +  0.002 \hspace{0.1cm} dN(1)\times Y(1) + 0.1\hspace{0.1cm} A(1)\times dN(1)\times K_1(1)$$
$$+ 0.04\hspace{0.1cm} A(1)\times dN(1)\times Y(1)  + 1.5\hspace{0.1cm} A(2)\times dN(2) -0.005 \hspace{0.1cm} Y(2) + 0.02 \hspace{0.1cm} K_1(2)  + 0.02 \hspace{0.1cm} dN(2)    -1.5 \hspace{0.1cm} K_2(2) +                           1 \hspace{0.1cm} dN(2) \times K_1(2) $$
$$  +   0.01\hspace{0.1cm} dN(2)\times Y(2) -1.2 \hspace{0.1cm} A(2)\times dN(2) \times K_1(2) +  0.01 \hspace{0.1cm} dN(2)\times A(2) \times Y(2).$$

 \textcolor{black}{In Scenario (B), we removed data on the tailoring variables on any time $t$ when $dN(t)=0$}. In Scenario (C) and (D) (only shown in Web Appendix), informative censoring was respectively generated as follows: $\xi(2) \sim \text{Bern}(1, p_{c2})$ with $p_{c2}=\text{expit}\{8 + 0.2 \hspace{0.1cm} A(0) - 0.1 \hspace{0.1cm} Y(0) + 0.2\hspace{0.1cm} K_1(0)\}$ and $\xi(1) \sim \text{Bern}(1, p_{c1})$ with $p_{c1}=\text{expit}\{10 - 0.2 \hspace{0.1cm} A(0) - 0.1 \hspace{0.1cm} Y(0) + 0.1\hspace{0.1cm} K_1(0)\}$, which together led to an approximate 25\% rate of censoring; and $\xi(t) \sim \text{Bern}(1, p_{ctd})$ with $p_{ctd}=\text{expit}\{10 - 0.2 \hspace{0.1cm} A(t) - 0.1 \hspace{0.1cm} Y(t) + 0.1\hspace{0.1cm} K_1(t)\}$ in the case of time-dependent censoring, which led to a censoring proportion around 7\% for Scenario D.

\textbf{Wrongly specified nuisance models}: In simulations, the wrong treatment-free model did not include the confounder $K_2$ each time $t$, and the wrong propensity score model did not include the confounder $K_2$ each time $t$. We see in the data generating mechanism above that this variable adjustment is mandatory since $K_2(0), K_2(1)$ ad $K_2(2)$ impact the final outcome, and that each $K_2(t)$ impacts treatment and visit $dN(t+1)$ and $A(t+1)$.

\newpage

 \textbf{Web Appendix F: Results of the simulation study when using LOCF to replace missing values}
 
\begin{figure}[hbt!]
\includegraphics[width=1.\textwidth]{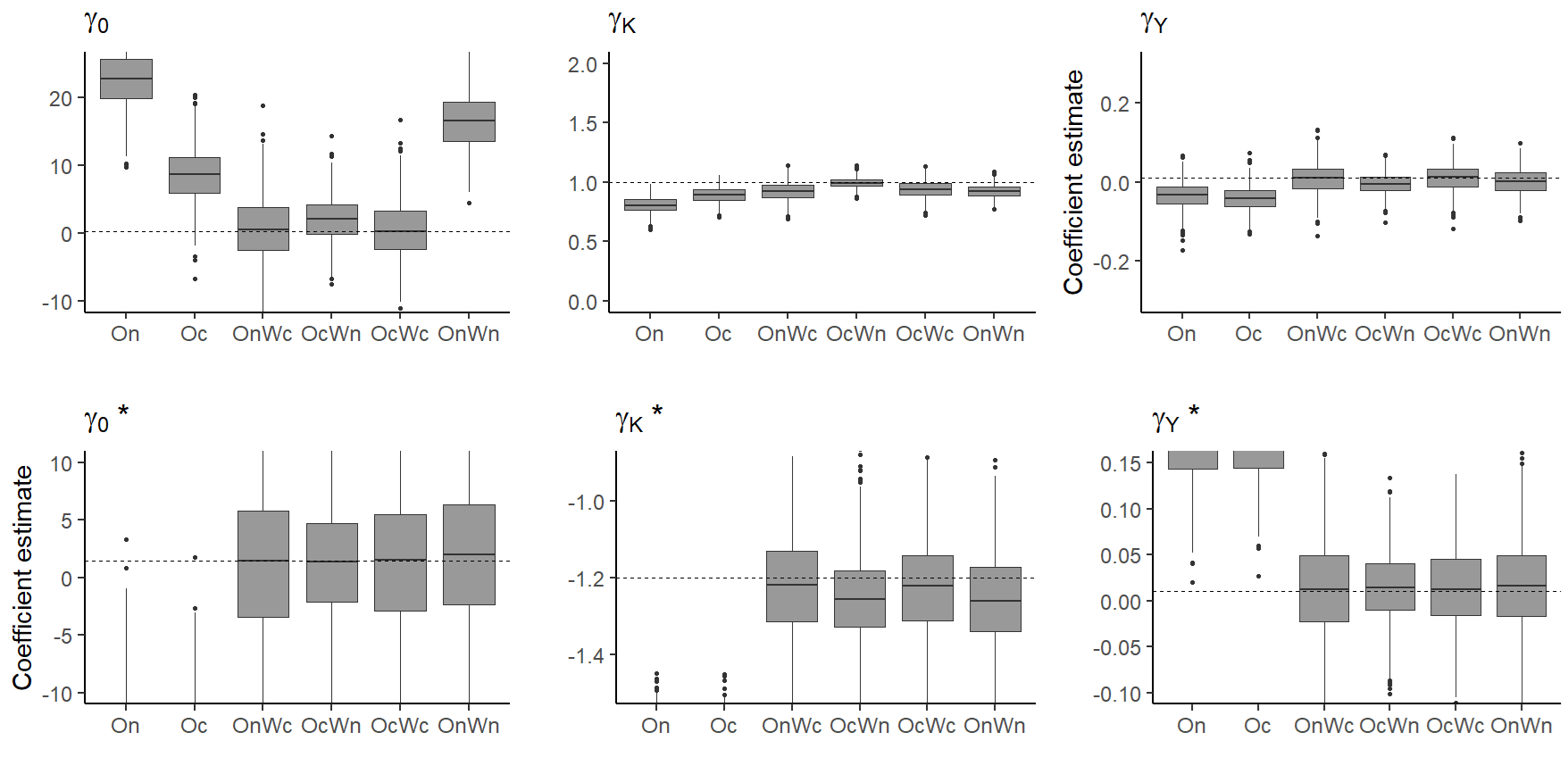}
\caption{Results under \textbf{missing data and LOCF}, with the weighted estimator using \textbf{overlap weights}. Each boxplot represents the distribution of 1000 point estimates. Sample size is 50000. The dashed line represents the true effect. On and Oc: Not correct and correct outcome models; Wn and Wc: Not correct and correct weight models; The top coefficients are those from the visit blip function (intercept, interaction between visit and confounder, and interaction between visit and previous outcome) while the bottom coefficients are the analogs from the treatment blip function. The estimators $O_c$ and $O_n$ correspond to the QLOMA, and all other estimators to the WOMA. }
\end{figure}

 \textbf{Web Appendix G: Results of the simulation studies when using inverse probability of treatment weights instead of overlap weights }

\begin{figure}[hbt!]
\includegraphics[width=1.\textwidth]{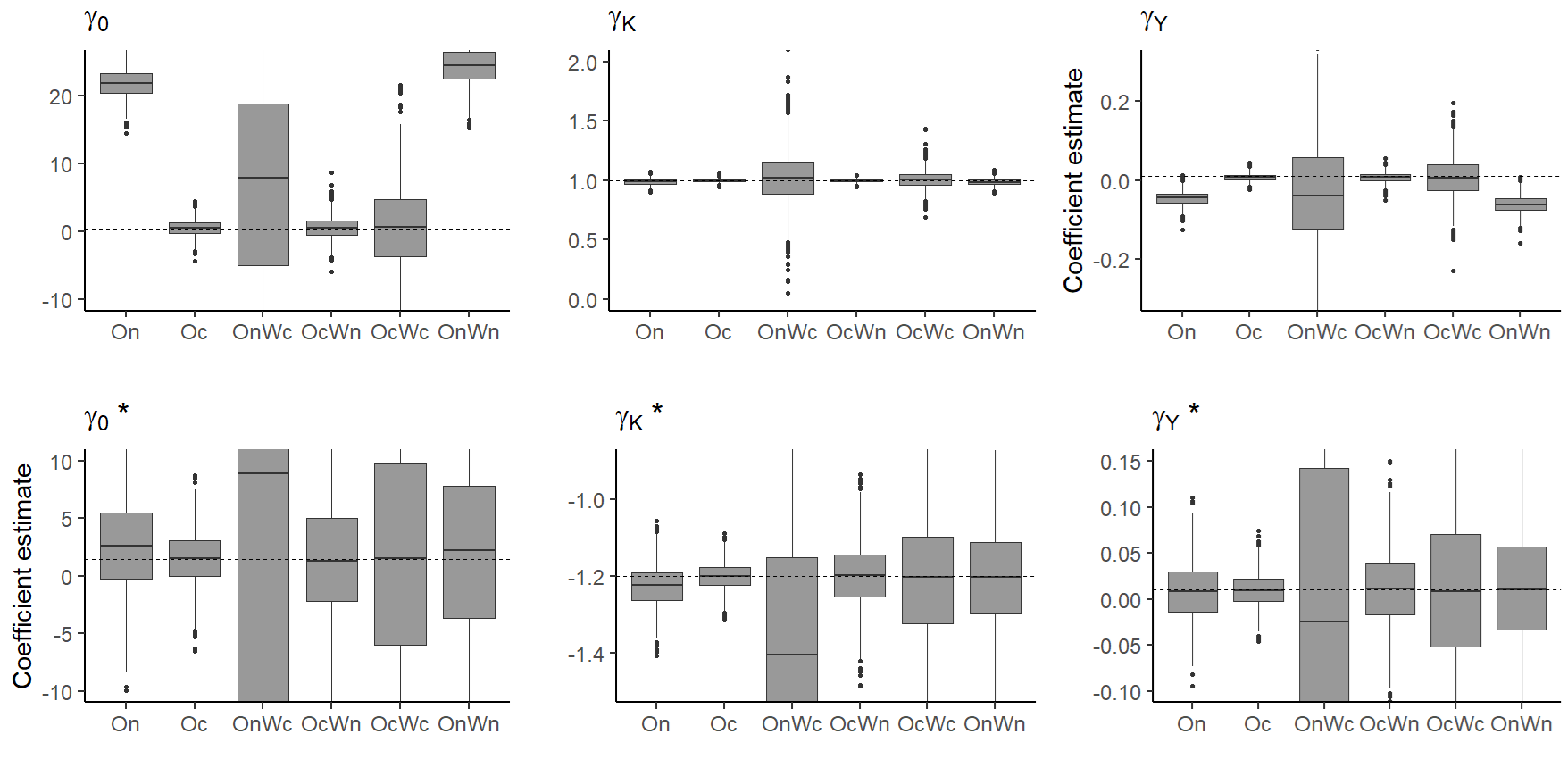}
\caption{Results under \textbf{no missing data} in the effect modifiers, with the weighted estimator using \textbf{inverse probability of treatment weights}. Each boxplot represents the distribution of 1000 point estimates. Sample size is 50000. The dashed line represents the true effect. On and Oc: Not correct and correct outcome models; Wn and Wc: Not correct and correct weight models; The top coefficients are those from the visit blip function (intercept, interaction between visit and confounder, and interaction between visit and previous outcome) while the bottom coefficients are the analogs from the treatment blip function. The estimators $O_c$ and $O_n$ correspond to the QLOMA, and all other estimators to the WOMA. }
\end{figure}

\begin{figure}[hbt!]
\includegraphics[width=1.\textwidth]{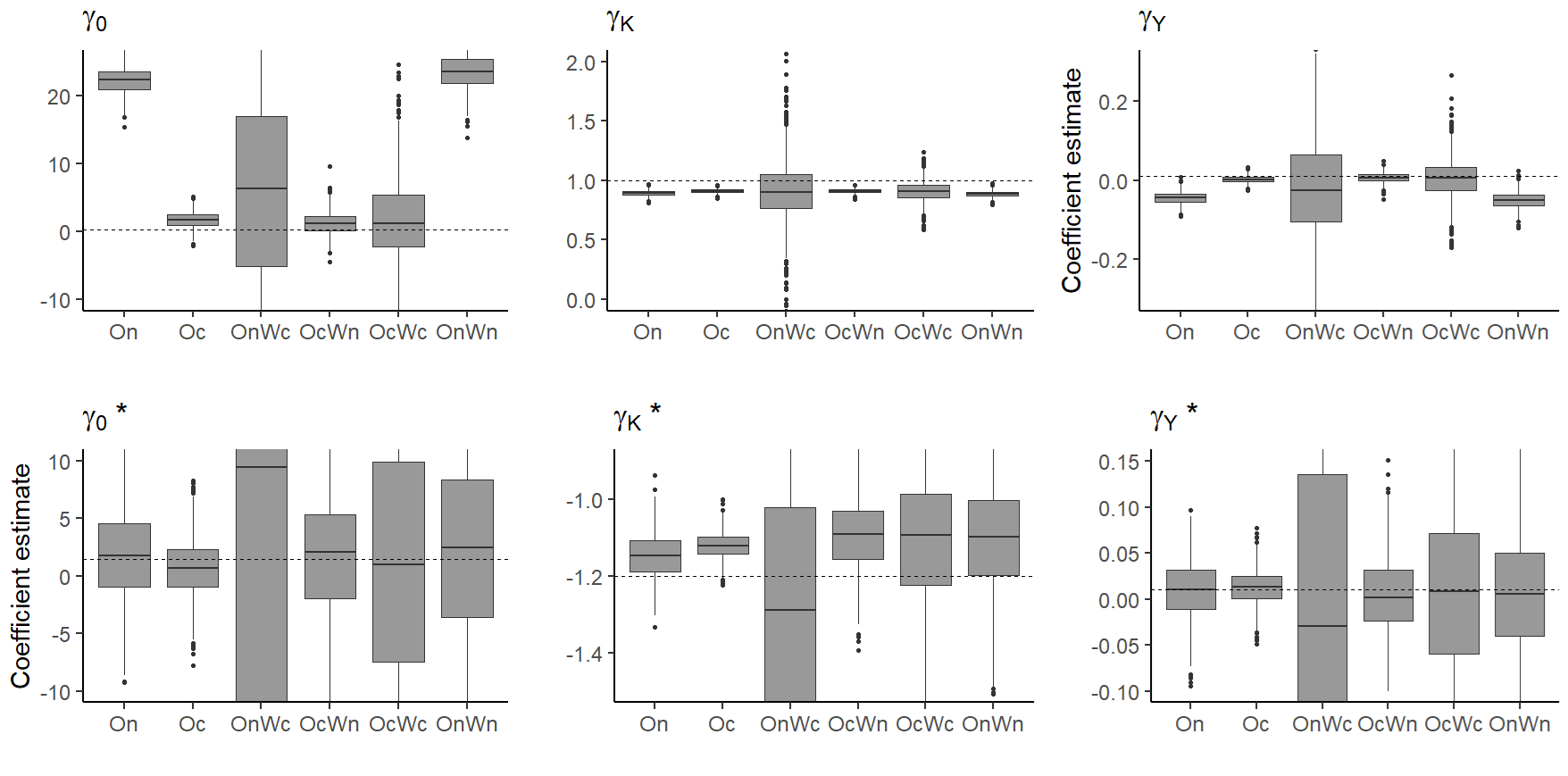}
\caption{Results under \textbf{missing data and multiple imputations using MICE}, with the weighted estimator using \textbf{inverse probability of treatment weights}. Each boxplot represents the distribution of 1000 point estimates. Sample size is 50000. The dashed line represents the true effect. On and Oc: Not correct and correct outcome models; Wn and Wc: Not correct and correct weight models; The top coefficients are those from the visit blip function (intercept, interaction between visit and confounder, and interaction between visit and previous outcome) while the bottom coefficients are the analogs from the treatment blip function. The estimators $O_c$ and $O_n$ correspond to the QLOMA, and all other estimators to the WOMA. }
\end{figure}
\begin{figure}[hbt!]
\includegraphics[width=1.\textwidth]{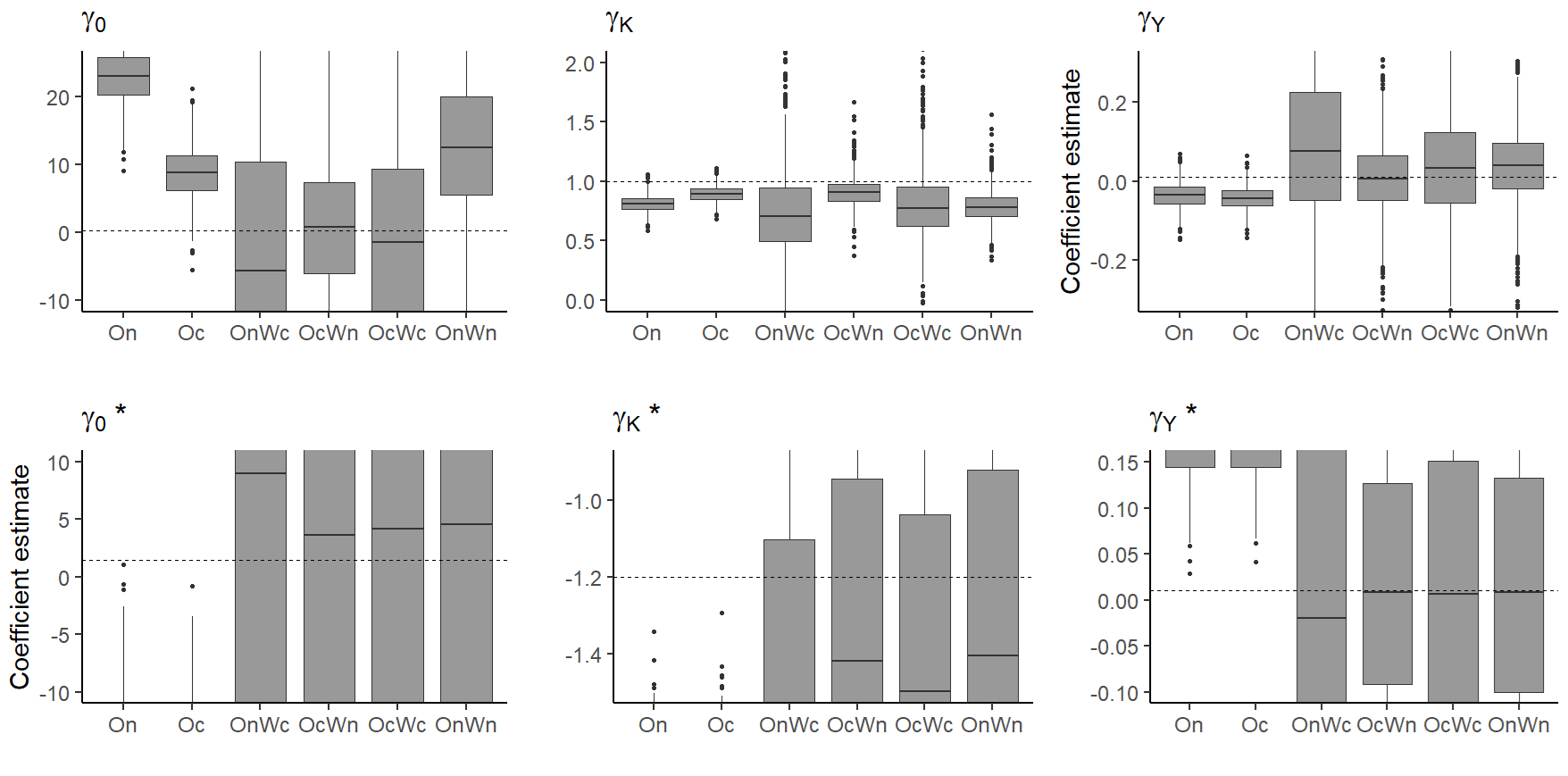}
\caption{Results under \textbf{missing data and LOCF}, with the weighted estimator using \textbf{inverse probability of treatment weights}. Each boxplot represents the distribution of 1000 point estimates. Sample size is 50000. The dashed line represents the true effect. On and Oc: Not correct and correct outcome models; Wn and Wc: Not correct and correct weight models; The top coefficients are those from the visit blip function (intercept, interaction between visit and confounder, and interaction between visit and previous outcome) while the bottom coefficients are the analogs from the treatment blip function. The estimators $O_c$ and $O_n$ correspond to the QLOMA, and all other estimators to the WOMA.}
\end{figure}

\newpage
 \textbf{Web Appendix H: Results of the simulation study under informative censoring, when using inverse probability of censoring weights}
 
\begin{figure}[hbt!]
\includegraphics[width=1.\textwidth]{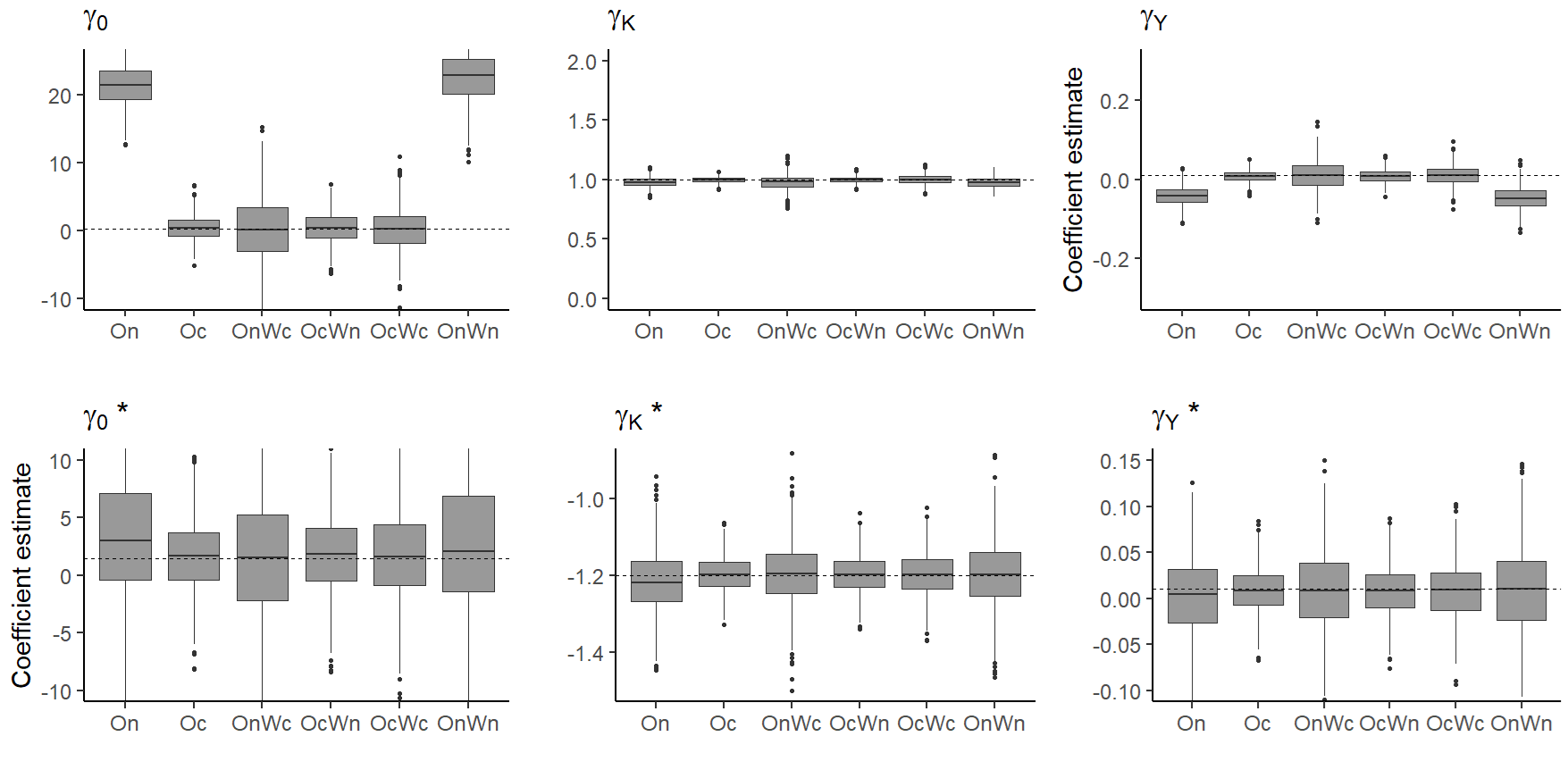}
\caption{Results under \textbf{no missing data} in the effect modifiers, with the weighted estimator using \textbf{overlap weights}, with \textbf{informative censoring due to time-fixed predictors at baseline}. Each boxplot represents the distribution of 1000 point estimates. Sample size is 50000. The dashed line represents the true effect. On and Oc: Not correct and correct outcome models; Wn and Wc: Not correct and correct weight models; The top coefficients are those from the visit blip function (intercept, interaction between visit and confounder, and interaction between visit and previous outcome) while the bottom coefficients are the analogs from the treatment blip function. The estimators $O_c$ and $O_n$ correspond to the QLOMA, and all other estimators to the WOMA.  }
\end{figure}

\begin{figure}[hbt!]
\includegraphics[width=1.\textwidth]{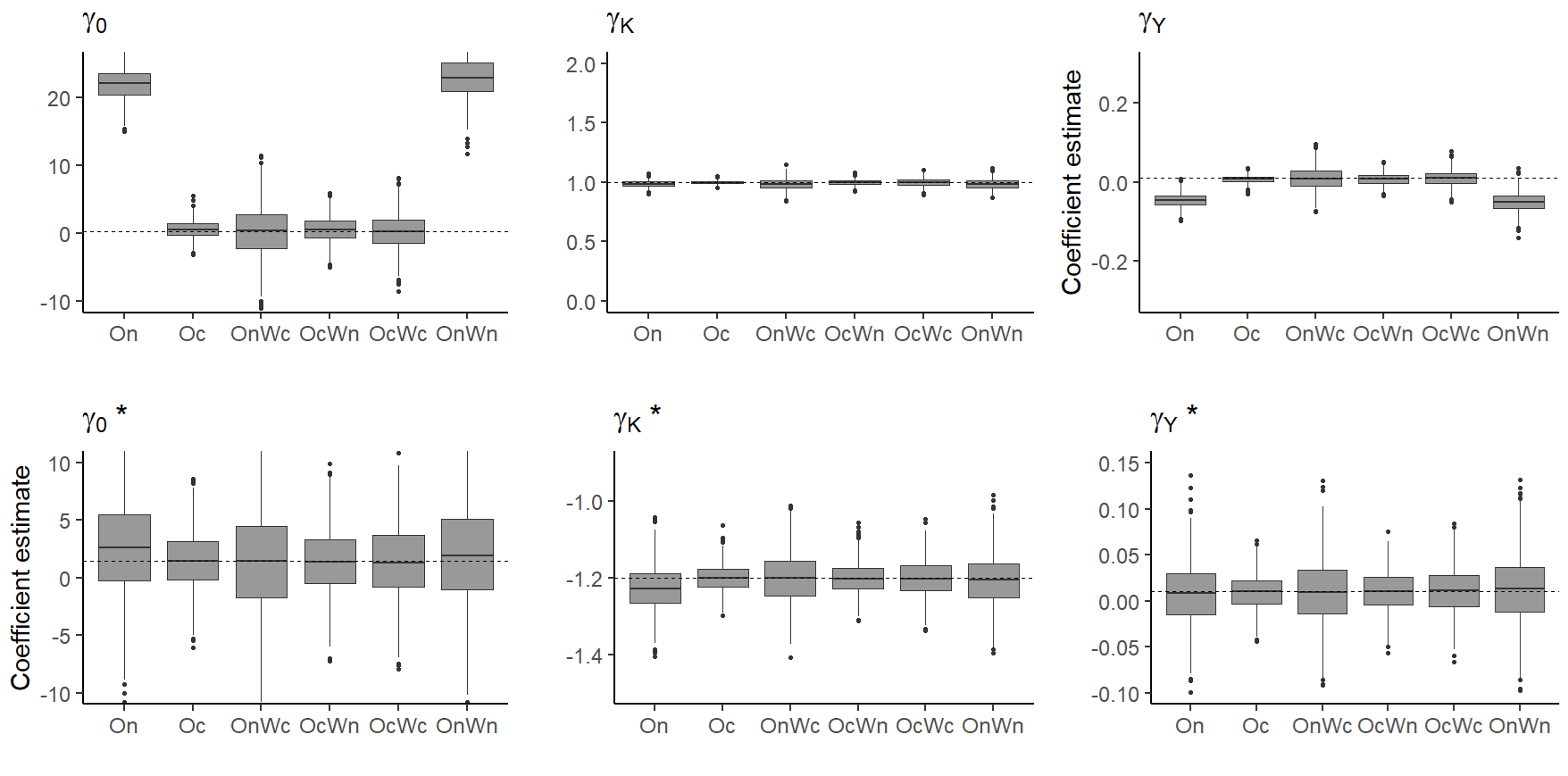}
\caption{Results under \textbf{no missing data} in the effect modifiers, with the weighted estimator using \textbf{overlap weights}, with \textbf{informative censoring due to time-dependent predictors, non-stabilized IPCW}. Each boxplot represents the distribution of 1000 point estimates. Sample size is 50000. The dashed line represents the true effect. On and Oc: Not correct and correct outcome models; Wn and Wc: Not correct and correct weight models; The top coefficients are those from the visit blip function (intercept, interaction between visit and confounder, and interaction between visit and previous outcome) while the bottom coefficients are the analogs from the treatment blip function. The estimators $O_c$ and $O_n$ correspond to the QLOMA, and all other estimators to the WOMA.  }
\end{figure}

\newpage

 \textbf{Web Appendix I: All the simulation study results in terms of empirical bias and SD (tables)}

 \begin{table}[hbt!]
  \begin{center}
    \caption{Blip parameter estimates, analyses using overlap weights \vspace{0.01cm}\\ }  \label{tabvalue}
     \begin{tabular}{|c| c| c| c| c| c| c| c| c| c| c| c| c|}
    \hline 
     \multicolumn{13}{|c|}{\textbf{No missing data, overlap weights}}\\ \hline
& \multicolumn{6}{|c|}{Empirical bias } & \multicolumn{6}{|c|}{Empirical std. deviation } \\ \hline
       Parameter & $O_{n}$ & $O_{c}$ & $O_nW_c$ & $O_cW_n$& $O_cW_c$& $O_nW_n$ & $O_{n}$ & $O_{c}$ & $O_nW_c$ & $O_cW_n$& $O_cW_c$& $O_nW_n$ \\ \hline
   $\gamma_0$ & 21.79 & 0.35 & 0.03 & 0.32 & 0.03 & 22.84 & 2.20 & 1.25 & 3.47 & 1.77 & 2.32 & 3.00 \\ \hline
   $\gamma_K$ & -0.01 & -0.00 & -0.02 & 0.00 & 0.00 & -0.02 & 0.03 & 0.01 & 0.05 & 0.02 & 0.03 & 0.04 \\ \hline
  $\gamma_Y$ & -0.06 & -0.00 & -0.00 & -0.00 & -0.00 & -0.06 & 0.02 & 0.01 & 0.03 & 0.01 & 0.02 & 0.02 \\ \hline
  $\gamma_0^*$ & 1.05 & 0.08 & -0.13 & 0.10 & 0.08 & 0.42 & 4.14 & 2.53 & 4.48 & 2.79 & 3.17 & 4.51 \\ \hline
   $\gamma_K^*$ & -0.03 & -0.00 & 0.00 & -0.00 & 0.00 & -0.00 & 0.06 & 0.03 & 0.06 & 0.04 & 0.04 & 0.06 \\ \hline
   $\gamma_Y^*$ & -0.00 & -0.00 & 0.00 & -0.00 & -0.00 & 0.00 & 0.03 & 0.02 & 0.03 & 0.02 & 0.02 & 0.03 \\ \hline
         \multicolumn{13}{|c|}{\textbf{Missing data, MICE, overlap weights}}\\ \hline
& \multicolumn{6}{|c|}{Empirical bias } & \multicolumn{6}{|c|}{Empirical std. deviation } \\ \hline
       Parameter & $O_{n}$ & $O_{c}$ & $O_nW_c$ & $O_cW_n$& $O_cW_c$& $O_nW_n$ & $O_{n}$ & $O_{c}$ & $O_nW_c$ & $O_cW_n$& $O_cW_c$& $O_nW_n$ \\ \hline
 $\gamma_0$  & 21.95 &  1.50 & 0.46 & 1.12 & 0.78 & 22.65 & 2.08 & 1.20 & 3.33 & 1.66 & 2.26 & 2.76 \\ \hline
  $\gamma_K$ & -0.11 & -0.09 & -0.04 & -0.02 & -0.02 & -0.04 & 0.03 & 0.02 & 0.05 & 0.02 & 0.03 & 0.04 \\ \hline
  $\gamma_Y$ & -0.05 & -0.01 & -0.00 & -0.01 & -0.00 & -0.06 & 0.02 & 0.01 & 0.03 & 0.01 & 0.02 & 0.02 \\ \hline
   $\gamma_0^*$ & 0.48 & -0.85 & -0.15 & -0.07 & -0.11 & 0.71 & 4.25 & 2.44 & 4.51 & 2.77 & 3.15 & 4.71 \\ \hline
   $\gamma_K^*$ & 0.05 & 0.08 & 0.03 & 0.03 & 0.03 & 0.02 & 0.06 & 0.03 & 0.06 & 0.04 & 0.04 & 0.06 \\ \hline
   $\gamma_Y^*$& -0.00 & 0.00 & -0.00 & -0.00 & -0.00 & -0.00 & 0.03 & 0.02 & 0.03 & 0.02 & 0.02 & 0.04 \\ \hline
\end{tabular}
  \end{center}
\end{table} 

   \begin{table}[hbt!]
  \begin{center}

    \begin{tabular}{|c| c| c| c| c| c| c| c| c| c| c| c| c|}
    \hline 
         \multicolumn{13}{|c|}{\textbf{Missing data, LOCF, overlap weights}}\\ \hline
& \multicolumn{6}{|c|}{Empirical bias } & \multicolumn{6}{|c|}{Empirical std. deviation } \\ \hline
       Parameter & $O_{n}$ & $O_{c}$ & $O_nW_c$ & $O_cW_n$& $O_cW_c$& $O_nW_n$ & $O_{n}$ & $O_{c}$ & $O_nW_c$ & $O_cW_n$& $O_cW_c$& $O_nW_n$ \\ \hline
   $\gamma_0$ & 22.67 & 8.36 & 0.43 & 1.91 & 0.18 & 16.30 & 4.38 & 4.06 & 4.97 & 3.23 & 4.26 & 4.09 \\ \hline
  $\gamma_K$ & -0.19 & -0.11 & -0.08 & -0.00 & -0.06 & -0.08 & 0.07 & 0.06 & 0.08 & 0.04 & 0.07 & 0.05 \\ \hline
 $\gamma_Y$ & -0.04 & -0.05 & -0.00 & -0.01 & 0.00 & -0.01 & 0.03 & 0.03 & 0.04 & 0.03 & 0.03 & 0.03 \\ \hline
  $\gamma_0^*$ & -18.68 & -18.92 & -0.20 & -0.20 & -0.20 & 0.49 & 6.17 & 5.28 & 6.92 & 5.23 & 6.24 & 6.46 \\ \hline
 $\gamma_K^*$ & -0.58 & -0.56 & -0.02 & -0.05 & -0.02 & -0.06 & 0.11 & 0.10 & 0.14 & 0.11 & 0.13 & 0.12 \\ \hline
  $\gamma_Y^*$ & 0.16 & 0.16 & 0.00 & 0.01 & 0.00 & 0.01 & 0.05 & 0.04 & 0.05 & 0.04 & 0.05 & 0.05 \\ \hline
             \end{tabular}
  \end{center}
\end{table}

\begin{table}[hbt!]
  \begin{center}
    \caption{Blip parameter estimates, analyses using inverse probability weights \vspace{0.01cm}\\ }  \label{tabvalue}
    \begin{tabular}{|c| c| c| c| c| c| c| c| c| c| c| c| c|}
    \hline 
     \multicolumn{13}{|c|}{\textbf{No missing data, inverse probability weights}}\\ \hline
& \multicolumn{6}{|c|}{Empirical bias } & \multicolumn{6}{|c|}{Empirical std. deviation } \\ \hline
       Parameter & $O_{n}$ & $O_{c}$ & $O_nW_c$ & $O_cW_n$& $O_cW_c$& $O_nW_n$ & $O_{n}$ & $O_{c}$ & $O_nW_c$ & $O_cW_n$& $O_cW_c$& $O_nW_n$ \\ \hline
$\gamma_0$ & 21.63 & 0.29 & 6.16 & 0.31 & 0.34 & 24.25 & 2.22 & 1.24 & 20.29 & 1.72 & 6.87 & 3.01 \\ \hline
  $\gamma_K$  & -0.01 & -0.00 & 0.02 & -0.00 & 0.01 & -0.01 & 0.03 & 0.02 & 0.25 & 0.02 & 0.08 & 0.03 \\ \hline
  $\gamma_Y$  & -0.05 & -0.00 & -0.04 & -0.00 & -0.00 & -0.07 & 0.02 & 0.01 & 0.16 & 0.01 & 0.05 & 0.02 \\ \hline
 $\gamma_0^*$& 1.11 & 0.00 & 5.87 & -0.03 & 0.14 & 0.63 & 4.13 & 2.39 & 34.47 & 5.42 & 13.72 & 8.90 \\ \hline
  $\gamma_K^*$ & -0.03 & -0.00 & -0.20 & 0.00 & -0.01 & -0.00 & 0.06 & 0.04 & 0.43 & 0.09 & 0.18 & 0.14 \\ \hline
 $\gamma_Y^*$ & -0.00 & -0.00 & -0.02 & 0.00 & -0.00 & 0.00 & 0.03 & 0.02 & 0.26 & 0.04 & 0.11 & 0.07 \\ \hline
    \end{tabular}
  \end{center}
\end{table}

\begin{table}[hbt!]
  \begin{center}
      \begin{tabular}{|c| c| c| c| c| c| c| c| c| c| c| c| c|}
    \hline 
    \multicolumn{13}{|c|}{\textbf{Missing data, MICE, inverse probability weights}}\\ \hline
& \multicolumn{6}{|c|}{Empirical bias } & \multicolumn{6}{|c|}{Empirical std. deviation } \\ \hline
       Parameter & $O_{n}$ & $O_{c}$ & $O_nW_c$ & $O_cW_n$& $O_cW_c$& $O_nW_n$ & $O_{n}$ & $O_{c}$ & $O_nW_c$ & $O_cW_n$& $O_cW_c$& $O_nW_n$ \\ \hline
  $\gamma_0$& 22.07 & 1.50 & 4.54 & 0.98 & 1.15 & 23.39 & 1.99 & 1.17 & 20.43 & 1.58 & 6.73 & 2.62 \\ \hline
  $\gamma_K$ & -0.11 & -0.09 & -0.09 & -0.09 & -0.09 & -0.11 & 0.03 & 0.02 & 0.26 & 0.02 & 0.09 & 0.03 \\ \hline
 $\gamma_Y$ & -0.05 & -0.01 & -0.02 & -0.00 & -0.00 & -0.06 & 0.02 & 0.01 & 0.16 & 0.01 & 0.05 & 0.02 \\ \hline
  $\gamma_0^*$ & 0.42 & -0.83 & 6.17 & 0.17 & -0.26 & 1.01 & 4.12 & 2.46 & 34.29 & 5.37 & 13.66 & 9.23 \\ \hline
  $\gamma_K^*$ & 0.05 & 0.08 & -0.08 & 0.11 & 0.10 & 0.10 & 0.06 & 0.03 & 0.44 & 0.10 & 0.19 & 0.15 \\ \hline
  $\gamma_Y^*$& -0.00 & 0.00 & -0.03 & -0.01 & -0.00 & -0.01 & 0.03 & 0.02 & 0.26 & 0.04 & 0.10 & 0.07 \\ \hline

         \multicolumn{13}{|c|}{\textbf{Missing data, LOCF, inverse probability weights}}\\ \hline
& \multicolumn{6}{|c|}{Empirical bias } & \multicolumn{6}{|c|}{Empirical std. deviation } \\ \hline
       Parameter & $O_{n}$ & $O_{c}$ & $O_nW_c$ & $O_cW_n$& $O_cW_c$& $O_nW_n$ & $O_{n}$ & $O_{c}$ & $O_nW_c$ & $O_cW_n$& $O_cW_c$& $O_nW_n$ \\ \hline
  $\gamma_0$ & 22.80 & 8.46 & -7.50 & 0.71 & -2.50 & 12.69 & 4.13 & 3.80 & 32.04 & 13.08 & 23.60 & 13.42 \\ \hline
  $\gamma_K$ & -0.19 & -0.11 & -0.26 & -0.09 & -0.19 & -0.21 & 0.07 & 0.07 & 0.41 & 0.13 & 0.31 & 0.13 \\ \hline
  $\gamma_Y$ & -0.05 & -0.05 & 0.07 & -0.00 & 0.02 & 0.03 & 0.03 & 0.03 & 0.25 & 0.11 & 0.19 & 0.11 \\ \hline
  $\gamma_0^*$ & -18.49 & -18.82 & 6.66 & -2.10 & -1.69 & -1.39 & 5.78 & 5.13 & 55.78 & 29.72 & 46.00 & 30.66 \\ \hline
  $\gamma_K^*$ & -0.58 & -0.57 & -0.30 & -0.05 & -0.13 & -0.05 & 0.11 & 0.10 & 0.89 & 0.65 & 0.77 & 0.67 \\ \hline
   $\gamma_Y^*$ & 0.16 & 0.16 & -0.03 & 0.02 & 0.02 & 0.02 & 0.04 & 0.04 & 0.42 & 0.22 & 0.35 & 0.23 \\ \hline

             \end{tabular}
  \end{center}
\end{table}

\begin{table}[hbt!]
  \begin{center}
    \caption{Blip parameter estimates, analyses with informative censoring using overlap weights }  \label{tabvalue}
    \begin{tabular}{|c| c| c| c| c| c| c| c| c| c| c| c| c|}
    \hline 
     \multicolumn{13}{|c|}{\textbf{No missing data, overlap weights, censoring depending on time-fixed covariates}}\\ \hline
& \multicolumn{6}{|c|}{Empirical bias } & \multicolumn{6}{|c|}{Empirical std. deviation } \\ \hline
       Parameter & $O_{n}$ & $O_{c}$ & $O_nW_c$ & $O_cW_n$& $O_cW_c$& $O_nW_n$ & $O_{n}$ & $O_{c}$ & $O_nW_c$ & $O_cW_n$& $O_cW_c$& $O_nW_n$ \\ \hline
   $\gamma_0$ & 21.27 & 0.19 & -0.09 & 0.16 & -0.07 & 22.56 & 3.11 & 1.76 & 4.52 & 2.17 & 3.04 & 3.82 \\ \hline
   $\gamma_K$ & -0.02 & -0.00 & -0.02 & -0.00 & -0.00 & -0.02 & 0.04 & 0.02 & 0.06 & 0.03 & 0.04 & 0.05 \\ \hline
   $\gamma_Y$ & -0.05 & -0.00 & 0.00 & -0.00 & 0.00 & -0.06 & 0.02 & 0.01 & 0.03 & 0.02 & 0.02 & 0.03 \\ \hline
   $\gamma_0^*$ & 1.72 & 0.19 & 0.09 & 0.23 & 0.28 & 1.01 & 5.44 & 3.09 & 5.64 & 3.48 & 3.85 & 6.13 \\ \hline
   $\gamma_K^*$ & -0.02 & 0.00 & 0.00 & 0.00 & 0.00 & 0.00 & 0.08 & 0.04 & 0.08 & 0.05 & 0.06 & 0.09 \\ \hline
    $\gamma_Y^*$ & -0.01 & -0.00 & -0.00 & -0.00 & -0.00 & -0.00 & 0.04 & 0.02 & 0.04 & 0.03 & 0.03 & 0.05 \\ \hline
     \end{tabular}
  \end{center}
\end{table}

\begin{table}[hbt!]
  \begin{center}
      \begin{tabular}{|c| c| c| c| c| c| c| c| c| c| c| c| c|}
    \hline 
         \multicolumn{13}{|c|}{\textbf{No missing data, overlap weights, censoring depending on time-dependent covariates}}\\ \hline
& \multicolumn{6}{|c|}{Empirical bias } & \multicolumn{6}{|c|}{Empirical std. deviation } \\ \hline
       Parameter & $O_{n}$ & $O_{c}$ & $O_nW_c$ & $O_cW_n$& $O_cW_c$& $O_nW_n$ & $O_{n}$ & $O_{c}$ & $O_nW_c$ & $O_cW_n$& $O_cW_c$& $O_nW_n$ \\ \hline
 $\gamma_0$ & 21.78 & 0.34 & 0.07 & 0.36 & 0.04 & 22.84 & 2.29 & 1.28 & 3.80 & 1.80 & 2.47 & 3.17 \\ \hline
  $\gamma_K$ & -0.01 & -0.00 & -0.02 & -0.00 & -0.00 & -0.02 & 0.03 & 0.02 & 0.05 & 0.02 & 0.03 & 0.04 \\ \hline
  $\gamma_Y$ & -0.06 & -0.00 & -0.00 & -0.00 & -0.00 & -0.06 & 0.02 & 0.01 & 0.03 & 0.01 & 0.02 & 0.02 \\ \hline
 $\gamma_0^*$ & 1.15 & 0.00 & -0.11 & -0.06 & -0.12 & 0.44 & 4.36 & 2.53 & 4.69 & 2.81 & 3.25 & 4.81 \\ \hline
 $\gamma_K^*$  & -0.03 & -0.00 & -0.00 & -0.00 & -0.00 & -0.01 & 0.06 & 0.03 & 0.07 & 0.04 & 0.05 & 0.07 \\ \hline
 $\gamma_Y^*$  & -0.00 & 0.00 & 0.00 & 0.00 & 0.00 & 0.00 & 0.03 & 0.02 & 0.04 & 0.02 & 0.02 & 0.04 \\ \hline
             \end{tabular}
  \end{center}
\end{table}

\clearpage
 \textbf{Web Appendix J: Results from the application to the SPRINT  } 

In this Appendix, we show 1) the list of antihypertensive drug classes used in our add-on definition; 2) the table of characteristics stratified by treatment strategy before and after overlap weighting; 3) the blip coefficient estimates for all months 2 to 11 and the frequency of statistically significant effect modifiers; and 4) the contingency tables of the received vs optimal treatment strategies in the first imputed dataset.

List of antihypertensive drug classes considered in the application to the SPRINT trial: ACE inhibitors; Alpha 1 blockers; Alpha 2 agonists; ARA; ARB; Beta blockers, Cardioselective beta blockers; Beta-Blocker with intrinsic sympathomimetic activity; Beta blockers with Alpha-blocking activity; Nonselective Beta blockers; Cardioselective and vasodilator Beta blockers; Calcium channel blockers, Dihydropyridine; Nondihydropyridine Calcium channel blockers; Potassium sparing diuretics; Loop diuretics; Thiazide-like diuretics; Thiazide-type diuretics; Direct arterial vasodilators; Direct Renin inhibitors; and Peripheral Adrenergic antagonists.\vspace{0.2cm}\\

\begin{table}[hbt!]
  \begin{center}
    \caption{Balance in patient characteristics on month 11, before and after overlap weighting, computed using one imputed dataset. The strategies (0,0), (1,0) and (1,1) respectively refer to no visit, a visit with no add-on and a visit with an add-on drug (corresponding to either an add-on or a switch of antihypertensive drug class), SPRINT trial, 2010-2013. }  \label{tabbal}
   \scriptsize
    \begin{tabular}{|l|c|c|c|c|c|c|c|c|}
    \hline 
         &  \multicolumn{4}{|c|}{No weights} &  \multicolumn{4}{|c|}{With overlap weights} \\ \hline
        Variable & (0,0) & (1,0)& (1,1) & All &  (0,0) & (1,0)& (1,1) & All   \\ \hline  \hline  
      Intensive group &49 & 57& 59& 50& 58& 58& 58 &58 \\  \hline  
Visit Month 10 & 10& 30& 17&13 &18 &19 &18 &19 \\\hline

Add-On Month 10 &5 &18 & 8& 7& 9& 9& 9& 9\\\hline
Age$^*$    &  67.7 (9.4)& 69.3 (9.4)& 68.1 (9.7)& 67.9 (9.4) & 68.4 (9.5) & 68.5 (9.5) & 68.4 (9.7)  & 68.5 (9.6) \\  \hline   
       Female, \%    & 35 & 39 &36 & 36&37 &37 &37  &37 \\  \hline   
          Race, \%   & & & & & & &  & \\  \hline   
           \hspace{0.2cm} Black  & 30& 28 & 33 &29 & 22&22 &21  &22 \\  \hline   
            \hspace{0.2cm} Hispanic    & 11&7 &9 &11 & 8& 8&9  &8 \\  \hline\hspace{0.2cm} White   &57 & 63& 54& 58& 57&57 &57  &57 \\  \hline   
             
               \hspace{0.2cm} Other    & 2&2 &4 & 2&3 &3 &3  &3\\ \hline 
              BMI$^*$,  $kg/m^2$    & 29.9 (5.7)&29.8 (6.0) &29.6 (5.8) &29.9 (5.8) & 29.7 (5.8)&29.6 (6.0) &29.7 (5.7)  &29.7 (5.8) \\  \hline  
              CVD & 20& 18&20 &20 & 19&19 &19  &19 \\  \hline  
             Smoking (ever)  &13 &12 &16 &13 &15 &15 &15  &15 \\  \hline 
             SBP$^*$, mmHg &&&&&&&&\\ \hline
              \hspace{0.2cm}  Baseline    & 139.3 (15.5)& 141.7 (15.7)& 141.9 (16.4)& 139.7 (15.6)& 141.8 (16.1)&142.0 (15.5) &141.7 (16.3)  &141.8 (16.0) \\ 
             \hline
             \hspace{0.2cm}
             Month 10   & 128.1 (15.7)&129.7 (15.6) &130.7 (16.4) &128.4 (15.7)& 130.1 (16.2)& 130.5 (15.7)& 130.2 (16.5) & 130.3 (16.1)\\ \hline   
                    Aspirin use  & 51& 53&48 &51 & 49& 50& 50 &49 \\  \hline 
Statin use &44 &43 &41 &44 &42 &42 &42  &42 \\  \hline 
HDL$^*$, mg/dl &52.7 (14.3) &54.1 (15.5) &53.7 (15.7) &52.9 (14.5) &53.7 (14.8) &53.9 (15.7) &53.7 (15.5) &53.8 (15.3) \\\hline
                          \end{tabular}

                          \scriptsize{$*$ Mean (SD). Acronyms: BMI, Body Mass index; CVD, Cardiovascular disease; HDL, High-Density Lipoprotein; SBP, 
                          Systolic Blood Pressure. }
  \end{center}  
\end{table}

\noindent \textbf{Blip estimates for each imputed dataset and each monthly decision rule: Each Table  from Web Table 5 to 14 contains one monthly decision rule.}\vspace{1cm}\\

 \begin{table}[hbt!]
  \begin{center}
    \caption{ Decision rule at month 2, mean of the blip estimates over 25 imputed datasets, SPRINT trial, 2010-2013.   }  \label{tabbal} 
    
  \begin{tabular}{|l|c|c|c|}
    \hline
    Variable & QLOMA & WOMA & No. of times stat. significant$^*$ \\ \hline
    Intercept visit          & -2.5 & -8.5 & 11 \\ \hline
    Intercept add-on          & 4.4 & 7.6 & 8 \\ \hline
    \multicolumn{4}{|c|}{Visit interaction with:} \\ \hline
    Intensive group           & 0.6 & 1.6 & 16 \\ \hline
    Age                       & 0.0 & 0.0 & 9 \\ \hline
    Female sex                & 0.4 & 0.5 & 8 \\ \hline
    Race Black                & \multicolumn{3}{|c|}{Reference}  \\ \hline
    Race Hispanic             & 0.6 & 0.9 & 10 \\ \hline
    Race White                & 0.5 & -0.5 & 4 \\ \hline
    \hspace{0.2cm} Other      & 0.2 & -1.3 & 5 \\ \hline
    Smoking (ever)            & -0.8 & -1.0 & 9 \\ \hline
    BMI                       & 0.0 & 0.0 & 7 \\ \hline
    HDL                       & 0.0 & 0.0 & 7 \\ \hline
    SBP Baseline              & 0.0 & 0.1 & 17 \\ \hline
    SBP Current month         & -0.0 & -0.0 & 15 \\ \hline
    CVD                       & 0.1 & 0.9 & 10 \\ \hline
    Aspirin use               & -0.0 & -0.3 & 5 \\ \hline
    Statin use                & -1.0 & -1.2 & 11 \\ \hline
 \end{tabular}
   \end{center}  
\end{table}
 
  \begin{table}[hbt!]
  \begin{center}
     \begin{tabular}{|l|c|c|c|}
    \hline
    Variable & QLOMA & WOMA & No. of times stat. significant$^*$ \\ \hline
 \multicolumn{4}{|c|}{Add-on and visit interaction with:} \\ \hline
    Intensive group           & -0.1 & -0.4 & 1 \\ \hline
    Age                       & -0.0 & -0.1 & 7 \\ \hline
    Female sex                & 0.3 & -0.5 & 3 \\ \hline
    Race Black                & \multicolumn{3}{|c|}{Reference}  \\ \hline
    Race Hispanic             & 0.5 & 0.6 & 1 \\ \hline
    Race White                & -1.4 & -1.8 & 17 \\ \hline
    \hspace{0.2cm} Other      & 1.3 & 2.5 & 5 \\ \hline
    Smoking (ever)            & 0.7 & -0.7 & 1 \\ \hline
    BMI                       & 0.0 & 0.1 & 5 \\ \hline
    HDL                       & -0.0 & -0.0 & 2 \\ \hline
    SBP Baseline              & -0.0 & -0.0 & 6 \\ \hline
    SBP Current month         & 0.0 & 0.0 & 5 \\ \hline
    CVD                       & 0.3 & 0.1 & 0 \\ \hline
    Aspirin use               & 0.6 & 0.3 & 2 \\ \hline
    Statin use                & -0.1 & -0.3 & 3 \\ \hline
\end{tabular}

                          \scriptsize{$*$For the analysis using the dynamic weighted approach only, number of times over 25 imputed datasets for which the variable was a statistically significant effect modifier in the outcome model.  Acronyms: BMI, Body Mass index; CVD, Cardiovascular disease; HDL, High-Density Lipoprotein; SBP, 
                          Systolic Blood Pressure.}
  \end{center}  
\end{table}

 \begin{table}[hbt!]
  \begin{center}
    \caption{ Decision rule at month 3, mean of the blip estimates over 25 imputed datasets, SPRINT trial, 2010-2013.  }  \label{tabbal}

  \begin{tabular}{|l|c|c|c|}
    \hline
    Variable & QLOMA & WOMA & No. of times stat. significant$^*$ \\ \hline
    Intercept visit          & -8.3 & -2.4 & 2 \\ \hline
    Intercept add-on          & -3.1 & -5.5 & 0 \\ \hline
    \multicolumn{4}{|c|}{Visit interaction with:} \\ \hline
    Intensive group           & -0.5 & -1.3 & 0 \\ \hline
    Age                       & 0.0 & -0.0 & 0 \\ \hline
    Female sex                & 0.2 & 0.8 & 2 \\ \hline
    Race Black                & \multicolumn{3}{|c|}{Reference}  \\ \hline
    Race Hispanic             & 0.6 & -0.2 & 0 \\ \hline
    Race White                & 0.7 & 1.9 & 0 \\ \hline
    \hspace{0.2cm} Other      & -0.1 & 0.4 & 0 \\ \hline
    Smoking (ever)            & -1.4 & -1.8 & 1 \\ \hline
    BMI                       & 0.0 & -0.1 & 0 \\ \hline
    HDL                       & 0.0 & -0.0 & 1 \\ \hline
    SBP Baseline              & 0.0 & 0.0 & 2 \\ \hline
    SBP Current month         & 0.0 & 0.0 & 1 \\ \hline
    CVD                       & 0.9 & 1.6 & 2 \\ \hline
    Aspirin use               & 0.0 & -1.1 & 0 \\ \hline
    Statin use                & 0.3 & 0.6 & 0 \\ \hline
   \end{tabular}
   \end{center}  
\end{table}
 
  \begin{table}[hbt!]
  \begin{center}
     \begin{tabular}{|l|c|c|c|}
    \hline
    Variable & QLOMA & WOMA & No. of times stat. significant$^*$ \\ \hline
    \multicolumn{4}{|c|}{Add-on and visit interaction with:} \\ \hline
    Intensive group           & -0.2 & -0.9 & 0 \\ \hline
    Age                       & -0.0 & 0.0 & 0 \\ \hline
    Female sex                & -0.1 & 0.3 & 1 \\ \hline
    Race Black                & \multicolumn{3}{|c|}{Reference}  \\ \hline
    Race Hispanic             & -0.6 & -2.9 & 0 \\ \hline
    Race White                & 0.4 & 0.8 & 0 \\ \hline
    \hspace{0.2cm} Other      & 0.7 & 1.9 & 0 \\ \hline
    Smoking (ever)            & 1.6 & 1.4 & 3 \\ \hline
    BMI                       & 0.1 & 0.2 & 12 \\ \hline
    HDL                       & 0.0 & 0.0 & 11 \\ \hline
    SBP Baseline              & -0.0 & 0.0 & 0 \\ \hline
    SBP Current month         & -0.0 & -0.0 & 1 \\ \hline
    CVD                       & 0.3 & -0.4 & 0 \\ \hline
    Aspirin use               & -0.7 & -1.4 & 1 \\ \hline
    Statin use                & 0.6 & 0.3 & 0 \\ \hline
\end{tabular}

                          \scriptsize{$*$For the analysis using the dynamic weighted approach only, number of times over 25 imputed datasets for which the variable was a statistically significant effect modifier in the outcome model.  Acronyms: BMI, Body Mass index; CVD, Cardiovascular disease; HDL, High-Density Lipoprotein; SBP, 
                          Systolic Blood Pressure.}
  \end{center}  
\end{table}

 \begin{table}[hbt!]
  \begin{center}
    \caption{ Decision rule at month 4, mean of the blip estimates over 25 imputed datasets, SPRINT trial, 2010-2013.   }  \label{tabbal}

 \begin{tabular}{|l|c|c|c|}
    \hline
    Variable &QLOMA & WOMA & No. of times stat. significant$^*$ \\ \hline
    Intercept visit          & 8.8 & 9.8 & 2 \\ \hline
    Intercept add-on          & -10.3 & -7.0 & 2 \\ \hline
    \multicolumn{4}{|c|}{Visit interaction with:} \\ \hline
    Intensive group           & -0.5 & -0.0 & 1 \\ \hline
    Age                       & -0.0 & -0.1 & 1 \\ \hline
    Female sex                & 0.6 & 0.0 & 0 \\ \hline
    Race Black                & \multicolumn{3}{|c|}{Reference}  \\ \hline
    Race Hispanic             & -0.7 & 0.0 & 0 \\ \hline
    Race White                & 0.0 & -0.1 & 0 \\ \hline
    \hspace{0.2cm} Other      & -0.8 & 0.4 & 0 \\ \hline
    Smoking (ever)            & -0.2 & -0.7 & 0 \\ \hline
    BMI                       & -0.0 & -0.1 & 0 \\ \hline
    HDL                       & -0.1 & -0.0 & 17 \\ \hline
    SBP Baseline              & 0.0 & -0.0 & 0 \\ \hline
    SBP Current month         & -0.0 & -0.0 & 0 \\ \hline
    CVD                       & -0.7 & -1.2 & 0 \\ \hline
    Aspirin use               & 0.4 & -0.1 & 0 \\ \hline
    Statin use                & 0.6 & 0.7 & 0 \\ \hline
  \end{tabular}
   \end{center}  
\end{table}
 
  \begin{table}[hbt!]
  \begin{center}
     \begin{tabular}{|l|c|c|c|}
    \hline
    Variable & QLOMA & WOMA & No. of times stat. significant$^*$ \\ \hline
    \multicolumn{4}{|c|}{Add-on and visit interaction with:} \\ \hline
    Intensive group           & 0.9 & 0.5 & 0 \\ \hline
    Age                       & 0.0 & 0.1 & 0 \\ \hline
    Female sex                & -1.4 & -0.9 & 1 \\ \hline
    Race Black                & \multicolumn{3}{|c|}{Reference}  \\ \hline
    Race Hispanic             & 2.3 & 0.8 & 0 \\ \hline
    Race White                & 1.0 & 0.4 & 0 \\ \hline
    \hspace{0.2cm} Other      & 2.0 & 2.1 & 0 \\ \hline
    Smoking (ever)            & 2.7 & 3.3 & 5 \\ \hline
    BMI                       & 0.1 & 0.1 & 0 \\ \hline
    HDL                       & 0.1 & 0.0 & 0 \\ \hline
    SBP Baseline              & -0.0 & -0.0 & 0 \\ \hline
    SBP Current month         & 0.0 & 0.0 & 0 \\ \hline
    CVD                       & -0.6 & -0.5 & 0 \\ \hline
    Aspirin use               & 0.5 & 0.5 & 0 \\ \hline
    Statin use                & -0.3 & -0.3 & 0 \\ \hline
\end{tabular}
                      
                          \scriptsize{$*$For the analysis using the dynamic weighted approach only, number of times over 25 imputed datasets for which the variable was a statistically significant effect modifier in the outcome model.  Acronyms: BMI, Body Mass index; CVD, Cardiovascular disease; HDL, High-Density Lipoprotein; SBP, 
                          Systolic Blood Pressure.}
  \end{center}  
\end{table}

 \begin{table}[hbt!]
  \begin{center}
    \caption{ Decision rule at month 5, mean of the blip estimates over 25 imputed datasets, SPRINT trial, 2010-2013.   }  \label{tabbal}

  \begin{tabular}{|l|c|c|c|}
    \hline
    Variable &QLOMA & WOMA & No. of times stat. significant$^*$ \\ \hline
    Intercept visit          & 9.1 & 9.7 & 11 \\ \hline
    Intercept add-on          & -3.2 & -4.8 & 0 \\ \hline
    \multicolumn{4}{|c|}{Visit interaction with:} \\ \hline
    Intensive group           & -1.7 & -1.7 & 20 \\ \hline
    Age                       & -0.0 & -0.0 & 4 \\ \hline
    Female sex                & 0.7 & 0.5 & 2 \\ \hline
    Race Black                & \multicolumn{3}{|c|}{Reference}  \\ \hline
    Race Hispanic             & -1.9 & -1.5 & 2 \\ \hline
    Race White                & 0.3 & 0.4 & 1 \\ \hline
    \hspace{0.2cm} Other      & -1.0 & -0.5 & 0 \\ \hline
    Smoking (ever)            & 0.0 & -0.4 & 0 \\ \hline
    BMI                       & -0.0 & -0.0 & 0 \\ \hline
    HDL                       & 0.0 & 0.0 & 0 \\ \hline
    SBP Baseline              & 0.0 & 0.0 & 0 \\ \hline
    SBP Current month         & -0.1 & -0.1 & 20 \\ \hline
    CVD                       & -1.2 & -1.8 & 7 \\ \hline
    Aspirin use               & -0.4 & -0.3 & 0 \\ \hline
    Statin use                & 0.5 & 0.3 & 1 \\ \hline
   \end{tabular}
   \end{center}  
\end{table}
 
  \begin{table}[hbt!]
  \begin{center}
     \begin{tabular}{|l|c|c|c|}
    \hline
    Variable & QLOMA & WOMA & No. of times stat. significant$^*$ \\ \hline
    \multicolumn{4}{|c|}{Add-on and visit interaction with:} \\ \hline
    Intensive group           & 0.6 & 1.0 & 0 \\ \hline
    Age                       & -0.1 & -0.1 & 0 \\ \hline
    Female sex                & -0.8 & -0.8 & 0 \\ \hline
    Race Black                & \multicolumn{3}{|c|}{Reference}  \\ \hline
    Race Hispanic             & 3.4 & 3.1 & 4 \\ \hline
    Race White                & -0.9 & -1.1 & 0 \\ \hline
    \hspace{0.2cm} Other      & 1.3 & 1.3 & 0 \\ \hline
    Smoking (ever)            & 0.2 & 0.2 & 0 \\ \hline
    BMI                       & 0.1 & 0.1 & 0 \\ \hline
    HDL                       & 0.1 & 0.1 & 1 \\ \hline
    SBP Baseline              & -0.0 & -0.0 & 0 \\ \hline
    SBP Current month         & 0.0 & 0.1 & 1 \\ \hline
    CVD                       & 0.4 & 1.0 & 0 \\ \hline
    Aspirin use               & 1.9 & 1.8 & 3 \\ \hline
    Statin use                & -0.6 & -0.7 & 0 \\ \hline
\end{tabular}
                       
                          \scriptsize{$*$For the analysis using the dynamic weighted approach only, number of times over 25 imputed datasets for which the variable was a statistically significant effect modifier in the outcome model.  Acronyms: BMI, Body Mass index; CVD, Cardiovascular disease; HDL, High-Density Lipoprotein; SBP, 
                          Systolic Blood Pressure.}
  \end{center}  
\end{table}

 \begin{table}[hbt!]
  \begin{center}
    \caption{ Decision rule at month 6, mean of the blip estimates over 25 imputed datasets, SPRINT trial, 2010-2013.  }  \label{tabbal}

    \begin{tabular}{|l|c|c|c|}
    \hline
    Variable & QLOMA & WOMA & No. of times stat. significant$^*$ \\ \hline
    Intercept visit          & 3.5 & 1.6 & 0 \\ \hline
    Intercept add-on          & -16.4 & -13.5 & 5 \\ \hline
    \multicolumn{4}{|c|}{Visit interaction with:} \\ \hline
    Intensive group           & 0.7 & 1.2 & 3 \\ \hline
    Age                       & 0.0 & 0.1 & 0 \\ \hline
    Female sex                & 0.6 & 0.6 & 0 \\ \hline
    Race Black                & \multicolumn{3}{|c|}{Reference}  \\ \hline
    Race Hispanic             & 0.3 & 0.4 & 1 \\ \hline
    Race White                & -0.5 & -0.3 & 0 \\ \hline
    \hspace{0.2cm} Other      & -2.2 & -2.7 & 0 \\ \hline
    Smoking (ever)            & -0.5 & -0.6 & 2 \\ \hline
    BMI                       & -0.0 & -0.0 & 1 \\ \hline
    HDL                       & -0.0 & 0.0 & 0 \\ \hline
    SBP Baseline              & -0.0 & -0.0 & 1 \\ \hline
    SBP Current month         & -0.0 & -0.0 & 2 \\ \hline
    CVD                       & -0.2 & -0.5 & 1 \\ \hline
    Aspirin use               & 0.2 & -0.0 & 0 \\ \hline
    Statin use                & -0.7 & -0.4 & 1 \\ \hline
   \end{tabular}
   \end{center}  
\end{table}
 
  \begin{table}[hbt!]
  \begin{center}
     \begin{tabular}{|l|c|c|c|}
    \hline
    Variable & QLOMA & WOMA & No. of times stat. significant$^*$ \\ \hline
    \multicolumn{4}{|c|}{Add-on and visit interaction with:} \\ \hline
    Intensive group           & -1.3 & -1.0 & 0 \\ \hline
    Age                       & 0.0 & 0.1 & 0 \\ \hline
    Female sex                & 0.0 & 0.8 & 0 \\ \hline
    Race Black                & \multicolumn{3}{|c|}{Reference}  \\ \hline
    Race Hispanic             & -0.7 & -0.9 & 0 \\ \hline
    Race White                & 0.2 & -0.2 & 0 \\ \hline
    \hspace{0.2cm} Other      & 4.4 & 6.3 & 0 \\ \hline
    Smoking (ever)            & 1.7 & 2.6 & 0 \\ \hline
    BMI                       & 0.2 & 0.1 & 7 \\ \hline
    HDL                       & 0.1 & 0.1 & 8 \\ \hline
    SBP Baseline              & -0.0 & -0.0 & 0 \\ \hline
    SBP Current month         & 0.1 & 0.1 & 4 \\ \hline
    CVD                       & -1.3 & -0.8 & 1 \\ \hline
    Aspirin use               & -1.4 & -0.9 & 1 \\ \hline
    Statin use                & -2.2 & -2.6 & 8 \\ \hline
\end{tabular}

                          \scriptsize{$*$For the analysis using the dynamic weighted approach only, number of times over 25 imputed datasets for which the variable was a statistically significant effect modifier in the outcome model.  Acronyms: BMI, Body Mass index; CVD, Cardiovascular disease; HDL, High-Density Lipoprotein; SBP, 
                          Systolic Blood Pressure.}
  \end{center}  
\end{table}

 \begin{table}[hbt!]
  \begin{center}
    \caption{ Decision rule at month 7, mean of the blip estimates over 25 imputed datasets, SPRINT trial, 2010-2013.   }  \label{tabbal}

    \begin{tabular}{|l|c|c|c|}
    \hline
    Variable &  QLOMA & WOMA  & No. of times stat. significant$^*$ \\ \hline
    Intercept visit          & 18.4 & 23.1 & 20 \\ \hline
    Intercept add-on          & -20.3 & -27.0 & 16 \\ \hline
    \multicolumn{4}{|c|}{Visit interaction with:} \\ \hline
    Intensive group           & -1.7 & -1.7 & 10 \\ \hline
    Age                       & -0.1 & -0.1 & 1 \\ \hline
    Female sex                & -0.4 & -0.2 & 1 \\ \hline
    Race Black                & \multicolumn{3}{|c|}{Reference}  \\ \hline
    Race Hispanic             & -0.5 & -1.0 & 0 \\ \hline
    Race White                & 1.2 & 1.5 & 1 \\ \hline
    \hspace{0.2cm} Other      & 7.6 & 7.0 & 12 \\ \hline
    Smoking (ever)            & 2.1 & 1.0 & 5 \\ \hline
    BMI                       & -0.2 & -0.3 & 17 \\ \hline
    HDL                       & -0.0 & -0.0 & 0 \\ \hline
    SBP Baseline              & 0.0 & -0.0 & 0 \\ \hline
    SBP Current month         & -0.1 & -0.1 & 21 \\ \hline
    CVD                       & 0.7 & 0.9 & 0 \\ \hline
    Aspirin use               & -0.3 & -0.1 & 0 \\ \hline
    Statin use                & -0.2 & 0.4 & 0 \\ \hline
   \end{tabular}
   \end{center}  
\end{table}
 
  \begin{table}[hbt!]
  \begin{center}
     \begin{tabular}{|l|c|c|c|}
    \hline
    Variable & QLOMA & WOMA & No. of times stat. significant$^*$ \\ \hline
    \multicolumn{4}{|c|}{Add-on and visit interaction with:} \\ \hline
    Intensive group           & 0.6 & 0.7 & 1 \\ \hline
    Age                       & 0.1 & 0.1 & 0 \\ \hline
    Female sex                & 1.5 & 1.3 & 1 \\ \hline
    Race Black                & \multicolumn{3}{|c|}{Reference}  \\ \hline
    Race Hispanic             & -0.9 & -1.1 & 0 \\ \hline
    Race White                & -0.5 & -1.5 & 0 \\ \hline
    \hspace{0.2cm} Other      & -6.2 & -4.8 & 4 \\ \hline
    Smoking (ever)            & -1.0 & 0.0 & 0 \\ \hline
    BMI                       & 0.4 & 0.4 & 24 \\ \hline
    HDL                       & -0.0 & -0.0 & 2 \\ \hline
    SBP Baseline              & -0.0 & 0.0 & 0 \\ \hline
    SBP Current month         & 0.1 & 0.1 & 11 \\ \hline
    CVD                       & 0.2 & 0.5 & 0 \\ \hline
    Aspirin use               & 0.3 & 0.6 & 0 \\ \hline
    Statin use                & -0.3 & -1.0 & 0 \\ \hline
\end{tabular}

                          \scriptsize{$*$For the analysis using the dynamic weighted approach only, number of times over 25 imputed datasets for which the variable was a statistically significant effect modifier in the outcome model.  Acronyms: BMI, Body Mass index; CVD, Cardiovascular disease; HDL, High-Density Lipoprotein; SBP, 
                          Systolic Blood Pressure.}
  \end{center}  
\end{table}

 \begin{table}[hbt!]
  \begin{center}
    \caption{ Decision rule at month 8, mean of the blip estimates over 25 imputed datasets, SPRINT trial, 2010-2013.  }  \label{tabbal}

 \begin{tabular}{|l|c|c|c|}
    \hline
    Variable &  QLOMA & WOMA  & No. of times stat. significant$^*$ \\ \hline
    Intercept visit          & 10.4 & 9.6 & 12 \\ \hline
    Intercept add-on          & -3.3 & -4.1 & 0 \\ \hline
    \multicolumn{4}{|c|}{Visit interaction with:} \\ \hline
    Intensive group           & -2.3 & -1.8 & 22 \\ \hline
    Age                       & -0.0 & 0.0 & 0 \\ \hline
    Female sex                & 0.3 & -0.0 & 0 \\ \hline
    Race Black                & \multicolumn{3}{|c|}{Reference}  \\ \hline
    Race Hispanic             & 0.8 & 1.2 & 1 \\ \hline
    Race White                & -0.2 & -0.5 & 0 \\ \hline
    \hspace{0.2cm} Other      & 1.9 & 0.7 & 1 \\ \hline
    Smoking (ever)            & -0.0 & -0.2 & 0 \\ \hline
    BMI                       & 0.0 & 0.0 & 0 \\ \hline
    HDL                       & 0.0 & 0.1 & 3 \\ \hline
    SBP Baseline              & -0.0 & -0.0 & 0 \\ \hline
    SBP Current month         & -0.1 & -0.1 & 25 \\ \hline
    CVD                       & 1.3 & 1.1 & 6 \\ \hline
    Aspirin use               & -0.4 & -0.7 & 0 \\ \hline
    Statin use                & 1.2 & 1.6 & 5 \\ \hline
     \end{tabular}
   \end{center}  
\end{table}
 
  \begin{table}[hbt!]
  \begin{center}
     \begin{tabular}{|l|c|c|c|}
    \hline
    Variable & QLOMA & WOMA & No. of times stat. significant$^*$ \\ \hline
    \multicolumn{4}{|c|}{Add-on and visit interaction with:} \\ \hline
    Intensive group           & 2.4 & 2.3 & 4 \\ \hline
    Age                       & 0.1 & 0.1 & 4 \\ \hline
    Female sex                & -0.9 & -0.9 & 1 \\ \hline
    Race Black                & \multicolumn{3}{|c|}{Reference}  \\ \hline
    Race Hispanic             & 2.3 & 2.2 & 1 \\ \hline
    Race White                & -0.5 & -0.3 & 0 \\ \hline
    \hspace{0.2cm} Other      & -6.6 & -4.4 & 1 \\ \hline
    Smoking (ever)            & 4.3 & 4.2 & 9 \\ \hline
    BMI                       & -0.0 & -0.0 & 0 \\ \hline
    HDL                       & -0.0 & -0.0 & 0 \\ \hline
    SBP Baseline              & -0.1 & -0.0 & 1 \\ \hline
    SBP Current month         & 0.0 & 0.0 & 2 \\ \hline
    CVD                       & -3.1 & -3.1 & 8 \\ \hline
    Aspirin use               & -0.1 & 0.1 & 0 \\ \hline
    Statin use                & 0.1 & -0.4 & 1 \\ \hline
\end{tabular}

                          \scriptsize{$*$For the analysis using the dynamic weighted approach only, number of times over 25 imputed datasets for which the variable was a statistically significant effect modifier in the outcome model.  Acronyms: BMI, Body Mass index; CVD, Cardiovascular disease; HDL, High-Density Lipoprotein; SBP, 
                          Systolic Blood Pressure.}
  \end{center}  
\end{table}

 \begin{table}[hbt!]
  \begin{center}
    \caption{ Decision rule at month 9, mean of the blip estimates over 25 imputed datasets, SPRINT trial, 2010-2013.   }  \label{tabbal}

 \begin{tabular}{|l|c|c|c|}
    \hline
    Variable &  QLOMA & WOMA  & No. of times stat. significant$^*$ \\ \hline
    Intercept visit          & -0.7 & 0.8 & 0 \\ \hline
    Intercept add-on          & -9.0 & -9.0 & 0 \\ \hline
    \multicolumn{4}{|c|}{Visit interaction with:} \\ \hline
    Intensive group           & 1.4 & 0.8 & 6 \\ \hline
    Age                       & -0.0 & -0.1 & 3 \\ \hline
    Female sex                & -0.6 & -0.5 & 2 \\ \hline
    Race Black                & \multicolumn{3}{|c|}{Reference}  \\ \hline
    Race Hispanic             & -0.6 & -0.6 & 1 \\ \hline
    Race White                & -1.0 & -1.4 & 1 \\ \hline
    \hspace{0.2cm} Other      & -1.2 & -1.7 & 1 \\ \hline
    Smoking (ever)            & -0.2 & -0.3 & 0 \\ \hline
    BMI                       & 0.0 & -0.0 & 1 \\ \hline
    HDL                       & 0.0 & 0.0 & 2 \\ \hline
    SBP Baseline              & 0.0 & 0.0 & 3 \\ \hline
    SBP Current month         & -0.0 & -0.0 & 1 \\ \hline
    CVD                       & 0.6 & 0.6 & 0 \\ \hline
    Aspirin use               & 0.6 & 0.9 & 1 \\ \hline
    Statin use                & 0.3 & 0.2 & 0 \\ \hline
    \end{tabular}
   \end{center}  
\end{table}
 
  \begin{table}[hbt!]
  \begin{center}
     \begin{tabular}{|l|c|c|c|}
    \hline
    Variable & QLOMA & WOMA & No. of times stat. significant$^*$ \\ \hline
    \multicolumn{4}{|c|}{Add-on and visit interaction with:} \\ \hline
    Intensive group           & -3.3 & -3.5 & 16 \\ \hline
    Age                       & 0.1 & 0.0 & 0 \\ \hline
    Female sex                & 1.2 & 1.5 & 0 \\ \hline
    Race Black                & \multicolumn{3}{|c|}{Reference}  \\ \hline
    Race Hispanic             & 1.2 & 0.8 & 0 \\ \hline
    Race White                & 0.1 & 0.2 & 0 \\ \hline
    \hspace{0.2cm} Other      & 17.1 & 20.7 & 24 \\ \hline
    Smoking (ever)            & -1.7 & -2.2 & 0 \\ \hline
    BMI                       & -0.1 & -0.1 & 2 \\ \hline
    HDL                       & -0.0 & -0.0 & 0 \\ \hline
    SBP Baseline              & 0.1 & 0.1 & 3 \\ \hline
    SBP Current month         & 0.0 & 0.0 & 0 \\ \hline
    CVD                       & -2.5 & -2.6 & 5 \\ \hline
    Aspirin use               & 0.9 & 1.6 & 1 \\ \hline
    Statin use                & -0.2 & -0.7 & 0 \\ \hline
\end{tabular}
                     
                          \scriptsize{$*$For the analysis using the dynamic weighted approach only, number of times over 25 imputed datasets for which the variable was a statistically significant effect modifier in the outcome model.  Acronyms: BMI, Body Mass index; CVD, Cardiovascular disease; HDL, High-Density Lipoprotein; SBP, 
                          Systolic Blood Pressure.}
  \end{center}  
\end{table}

 \begin{table}[hbt!]
  \begin{center}
    \caption{ Decision rule at month 10, mean of the blip estimates over 25 imputed datasets, SPRINT trial, 2010-2013.  }  \label{tabbal}

  \begin{tabular}{|l|c|c|c|}
    \hline
    Variable & QLOMA & WOMA  & No. of times stat. significant$^*$ \\ \hline
    Intercept visit          & 11.4 & 4.7 & 5 \\ \hline
    Intercept add-on          & -5.2 & 2.0 & 0 \\ \hline
    \multicolumn{4}{|c|}{Visit interaction with:} \\ \hline
    Intensive group           & -2.2 & -1.7 & 13 \\ \hline
    Age                       & -0.0 & 0.0 & 1 \\ \hline
    Female sex                & -1.4 & -1.9 & 3 \\ \hline
    Race Black                & \multicolumn{3}{|c|}{Reference}  \\ \hline
    Race Hispanic             & -1.1 & -0.9 & 0 \\ \hline
    Race White                & 0.8 & 0.9 & 1 \\ \hline
    \hspace{0.2cm} Other      & 6.3 & 3.1 & 6 \\ \hline
    Smoking (ever)            & -2.9 & -3.0 & 7 \\ \hline
    BMI                       & 0.1 & 0.1 & 0 \\ \hline
    HDL                       & 0.0 & 0.0 & 0 \\ \hline
    SBP Baseline              & 0.0 & 0.0 & 0 \\ \hline
    SBP Current month         & -0.1 & -0.1 & 24 \\ \hline
    CVD                       & 0.3 & 0.3 & 0 \\ \hline
    Aspirin use               & 0.4 & -0.0 & 0 \\ \hline
    Statin use                & -2.1 & -0.8 & 10 \\ \hline
    \end{tabular}
   \end{center}  
\end{table}
 
  \begin{table}[hbt!]
  \begin{center}
     \begin{tabular}{|l|c|c|c|}
    \hline
    Variable & QLOMA & WOMA & No. of times stat. significant$^*$ \\ \hline
    \multicolumn{4}{|c|}{Add-on and visit interaction with:} \\ \hline
    Intensive group           & 1.1 & 1.9 & 1 \\ \hline
    Age                       & 0.1 & 0.1 & 4 \\ \hline
    Female sex                & 0.7 & 1.3 & 2 \\ \hline
    Race Black                & \multicolumn{3}{|c|}{Reference}  \\ \hline
    Race Hispanic             & 2.6 & 1.5 & 1 \\ \hline
    Race White                & -1.8 & -2.2 & 2 \\ \hline
    \hspace{0.2cm} Other      & -15.3 & -14.6 & 21 \\ \hline
    Smoking (ever)            & 1.3 & 0.9 & 2 \\ \hline
    BMI                       & -0.2 & -0.3 & 3 \\ \hline
    HDL                       & -0.0 & -0.1 & 1 \\ \hline
    SBP Baseline              & -0.1 & -0.1 & 9 \\ \hline
    SBP Current month         & 0.1 & 0.1 & 15 \\ \hline
    CVD                       & -0.4 & 0.2 & 0 \\ \hline
    Aspirin use               & -0.0 & 0.1 & 0 \\ \hline
    Statin use                & 2.3 & 0.9 & 6 \\ \hline
\end{tabular}

                          \scriptsize{$*$For the analysis using the dynamic weighted approach only, number of times over 25 imputed datasets for which the variable was a statistically significant effect modifier in the outcome model.  Acronyms: BMI, Body Mass index; CVD, Cardiovascular disease; HDL, High-Density Lipoprotein; SBP, 
                          Systolic Blood Pressure.}
  \end{center}  
\end{table}

 \begin{table}[hbt!]
  \begin{center}
    \caption{ Decision rule at month 11, mean of the blip estimates over 25 imputed datasets, SPRINT trial, 2010-2013.  }  \label{tabbal}

   \begin{tabular}{|l|c|c|c|}
    \hline
    Variable & QLOMA & WOMA  & No. of times stat. significant$^*$ \\ \hline
    Intercept visit          & 12.9 & 5.1 & 10 \\ \hline
    Intercept add-on          & 2.3 & 5.4 & 0 \\ \hline
    \multicolumn{4}{|c|}{Visit interaction with:} \\ \hline
    Intensive group           & -2.4 & -1.4 & 21 \\ \hline
    Age                       & -0.0 & -0.0 & 1 \\ \hline
    Female sex                & 0.4 & 0.5 & 0 \\ \hline
    Race Black                & \multicolumn{3}{|c|}{Reference}  \\ \hline
    Race Hispanic             & -0.4 & -1.1 & 0 \\ \hline
    Race White                & 0.9 & 1.0 & 2 \\ \hline
    \hspace{0.2cm} Other      & 0.6 & 1.0 & 0 \\ \hline
    Smoking (ever)            & -1.3 & -1.3 & 0 \\ \hline
    BMI                       & 0.0 & 0.0 & 0 \\ \hline
    HDL                       & -0.0 & -0.0 & 0 \\ \hline
    SBP Baseline              & 0.0 & 0.0 & 0 \\ \hline
    SBP Current month         & -0.1 & -0.1 & 23 \\ \hline
    CVD                       & 0.7 & 0.2 & 0 \\ \hline
    Aspirin use               & -0.6 & -0.8 & 1 \\ \hline
    Statin use                & -0.1 & 0.3 & 0 \\ \hline
    \end{tabular}
   \end{center}  
\end{table}
 
  \begin{table}[hbt!]
  \begin{center}
     \begin{tabular}{|l|c|c|c|}
    \hline
    Variable & QLOMA & WOMA & No. of times stat. significant$^*$ \\ \hline
    \multicolumn{4}{|c|}{Add-on and visit interaction with:} \\ \hline
    Intensive group           & -1.7 & -1.3 & 1 \\ \hline
    Age                       & 0.1 & 0.1 & 3 \\ \hline
    Female sex                & -2.0 & -2.1 & 0 \\ \hline
    Race Black                & \multicolumn{3}{|c|}{Reference}  \\ \hline
    Race Hispanic             & 1.2 & 1.6 & 0 \\ \hline
    Race White                & -0.5 & -0.8 & 0 \\ \hline
    \hspace{0.2cm} Other      & 4.0 & 3.6 & 0 \\ \hline
    Smoking (ever)            & -1.2 & -1.2 & 0 \\ \hline
    BMI                       & 0.1 & 0.1 & 1 \\ \hline
    HDL                       & -0.0 & -0.0 & 0 \\ \hline
    SBP Baseline              & -0.1 & -0.1 & 2 \\ \hline
    SBP Current month         & -0.0 & -0.0 & 0 \\ \hline
    CVD                       & -1.9 & -1.1 & 0 \\ \hline
    Aspirin use               & -0.2 & -0.1 & 0 \\ \hline
    Statin use                & 0.5 & 0.1 & 0 \\ \hline
\end{tabular}
                    
                          \scriptsize{$*$For the analysis using the dynamic weighted approach only, number of times over 25 imputed datasets for which the variable was a statistically significant effect modifier in the outcome model.  Acronyms: BMI, Body Mass index; CVD, Cardiovascular disease; HDL, High-Density Lipoprotein; SBP, 
                          Systolic Blood Pressure.}
  \end{center}  
\end{table}

\clearpage

\begin{table}[htb!]
\caption{Monthly contingency tables of the received vs the optimal strategy in the first imputed dataset; 1 is no visit, 2 is a visit and no add-on, and 3 is a visit and an add-on. The numbers in each cell are the frequency of patients on a total of 9361 patients, SPRINT trial, 2010-2013.}
\begin{tabular}{|l|c|c|c|c|}
\hline
 & & \multicolumn{3}{|c|}{Optimal} \\ \hline
Month &Received & 1 & 2 & 3 \\ \hline
    & 1 & 132& 105&280 \\
   2 & 2 &2324 &1713 &3437 \\
    & 3 & 265& 439& 666\\ \hline  
        & 1 &133 &47 &227 \\
   3 & 2 &2339 & 1017&4508 \\
    & 3 & 359&147 &584 \\ \hline 
            & 1 & 3765& 1345&2339 \\
   4 & 2 & 476&191 &354 \\
    & 3 & 347&190 &354 \\ \hline 
            & 1 & 2985&2824 &1239 \\
   5 & 2 &843 &671 &347 \\
    & 3 & 274&60 &118 \\ \hline 
            & 1 &166 & 172&283 \\
   6 & 2 &2291 &3012 &3073 \\
    & 3 & 91& 130& 143\\ \hline 
            & 1 & 1116&2564 &4176 \\
   7 & 2 &111 &238 &447 \\
    & 3 &126 & 160&423 \\ \hline 
  \end{tabular}
\end{table}
\begin{table}[htb!]
\begin{tabular}{|l|c|c|c|c|}
\hline
 & & \multicolumn{3}{|c|}{Optimal} \\ \hline
Month &Received & 1 & 2 & 3 \\ \hline
            & 1 &2238 &2078 &3311 \\
            8 & 2 & 454&317 &610 \\
    & 3 & 157&56 &140 \\ \hline 
            & 1 & 245& 408&173 \\
   9 & 2 & 2326& 4439&1466 \\
    & 3 & 79&166 &59 \\ \hline 
            & 1 &1299 &4718 &2131 \\
   10 & 2 & 95&345 &164 \\
    & 3 & 112& 300&197 \\ \hline 
            & 1 & 3198&1964 &2740 \\
   11 & 2 &522 &305 &297 \\
    & 3 & 219&47 &69 \\ \hline 
\end{tabular}
\end{table}

\end{document}